\documentclass[12pt]{article}

\pdfoutput=1

\usepackage{ifpdf}
\ifpdf
\usepackage{graphicx,color}
\usepackage{hyperref}
\else
\usepackage[dvipdfmx]{graphicx,color}
\usepackage[dvipdfmx]{hyperref}
\fi
\usepackage{amssymb,amsfonts,amsmath,cancel,cite,multirow}
\usepackage[capitalise]{cleveref}
\usepackage{slashed}
\usepackage{subcaption}
\usepackage[margin=1in]{geometry}

\newcommand{\eqs}[1]{\begin{equation}\begin{split} #1 \end{split}\end{equation}}

\begin{document}

\begin{titlepage}

\begin{flushright}
\end{flushright}

\vskip 1.35cm
\begin{center}

{\large
\textbf{
Maximally self-interacting dark matter: \\
models and predictions}}
\vskip 1.2cm

Ayuki Kamada$^{a}$,
Hee Jung Kim$^{b}$,
and Takumi Kuwahara$^{a}$

\vskip 0.4cm

\textit{$^a$
Center for Theoretical Physics of the Universe,
Institute for Basic Science (IBS), Daejeon 34126, Korea
}

\textit{$^b$
Department of Physics, KAIST, Daejeon 34141, Korea
}

\vskip 1.5cm

\begin{abstract}
We study self-interacting dark matter (SIDM) scenarios, where the $s$-wave self-scattering cross section almost saturates the Unitarity bound.
Such self-scattering cross sections are singly parameterized by the dark matter mass, and are featured by strong velocity dependence in a wide range of velocities.
They may be indicated by observations of dark matter halos in a wide range of masses, from Milky Way's dwarf spheroidal galaxies to galaxy clusters.
We pin down the model parameters that saturates the Unitarity bound in well-motivated SIDM models: the gauged $L_{\mu} - L_{\tau}$ model and composite asymmetric dark matter model.
We discuss implications and predictions of such model parameters for cosmology like the $H_{0}$ tension and dark-matter direct-detection experiments, and particle phenomenology like the beam-dump experiments.
\end{abstract}

\end{center}
\end{titlepage}

\section{Introduction} \label{sec:intro}
Collisionless cold dark matter (CDM) is a minimal hypothesis for missing mass of the Universe.
The CDM hypothesis works well in explaining the large-scale structure of the Universe from cosmic microwave background (CMB) anisotropies to galaxy clusterings.
On the other hand, there have been tensions between na\"ive CDM predictions and the observed small-scale structure of the Universe.
Such tensions are collectively dubbed as the small-scale issues, and challenge our understanding of how Milky Way (MW)-size and dwarf galaxies form (see Ref.~\cite{Bullock:2017xww} for a review).

The small-scale issues may indicate the nature of DM, e.g., self-interacting dark matter (SIDM)~\cite{Spergel:1999mh} (see Ref.~\cite{Tulin:2017ara} for a review).
It is intriguing that we can probe self-interaction of DM, which is not accessible in terrestrial experiments, in cosmological observations.
Such gravitational probes of DM have been attracting growing interests~\cite{Buckley:2017ijx}, in light of null detection of DM interactions with standard-model (SM) particles in collider searches and DM direct-detection and indirect-detection experiments~\cite{Arcadi:2017kky,Roszkowski:2017nbc}.
Since traditional weakly interacting massive particle (WIMP) DM behaves as CDM on galactic scales, the small-scale issues may call for a new paradigm, i.e., beyond-WIMP DM.

DM self-interaction is not necessarily constant in collision energy.
Ref.~\cite{Kaplinghat:2015aga} illustrates how we can probe velocity dependence of the self-scattering cross section by combining observations of DM halos in a wide range of masses.
The velocity dependence may be already indicated.
To form cores of galaxy clusters~\cite{Newman:2012nw} and evade constraints from bullet clusters~\cite{Randall:2007ph} and merging clusters~\cite{Harvey:2018uwf}, where the collision velocity is $v \sim 1000 \, {\rm km/s}$, the self-scattering cross section per mass is preferred to be $\sigma / m \sim 0.1 \, {\rm cm^{2}/g}$ (see also Ref.~\cite{Sagunski:2020spe} for a recent reanalysis of galaxy clusters and a new constraint from galaxy groups~\cite{Newman:2015kzv}).
On the other hand, to explain the diversity of galactic rotation curves~\cite{Oman:2015xda}, where $v \sim 100 \, {\rm km/s}$, it is preferred to be $\sigma / m \sim 1 \, {\rm cm^{2}/g}$~\cite{Kamada:2016euw,Ren:2018jpt}.
This velocity dependence call for particle-physics model buildings:
DM with a light mediator~\cite{Tulin:2012wi,Tulin:2013teo,Kahlhoefer:2017umn,Ma:2017ucp,Duch:2017khv,Duerr:2018mbd,Kamada:2018zxi,Kamada:2018kmi,Duch:2019vjg}, DM with a resonant mediator~\cite{Duch:2017nbe,Chu:2018fzy}, and composite DM~\cite{CyrRacine:2012fz,Cline:2013pca,Cline:2013zca,Boddy:2014yra,Boddy:2014qxa,Boddy:2016bbu,Ibe:2018juk,Chu:2018faw}.
Hereafter, we focus on an elastic scattering, while an inelastic scattering may also introduce the scale-dependence of core formation~\cite{Loeb:2010gj,Schutz:2014nka,McDermott:2017vyk,Vogelsberger:2018bok,Kamada:2019wjo}.

MW's dwarf spheroidal satellite galaxies, where $v \sim 30 \, {\rm km/s}$, are the subject of active debate.
They show a large diversity in the central density profiles~\cite{Gilmore:2007fy}.
The preferred self-scattering cross section also vary as $\sigma / m \sim 0.1 \text{-} 40 \,{\rm cm^{2}/g}$~\cite{Valli:2017ktb}.
The high central density of Draco actually provides a tight constraint of $\sigma / m < 0.57 \,{\rm cm^{2}/g}$~\cite{Read:2018pft} (see also the recent analysis of ultra-faint dwarf galaxies~\cite{Hayashi:2020syu}).
However, it was recently pointed out that while the above discussions focus on the so-called core expansion (formation) phase, the situation changes in the core contraction phase (but long before the core collapse time).
There, SIDM halo may experience a gravothermal core collapse~\cite{Balberg:2001qg}, leading to a cuspy profile.
The gravothermal core collapse may be further accelerated by the tidal stripping~\cite{Nishikawa:2019lsc}.
In addition, the core collapse may explain the anti-correlation between the central DM densities and their orbital pericenter distances~\cite{Kaplinghat:2019svz}.

If the MW's dwarf spheroidal satellite galaxies are in the core contraction phase, the self-scattering cross section may need to be as large as $\sigma/m \sim 30 \text{-} 200 \, {\rm cm^{2} / g}$ to have a sufficient core collapse~\cite{Correa:2020qam}.
This implies a strong velocity dependence of $\sigma / m$.
Meanwhile, such large $\sigma / m$ seem not compatible with the inferred values of $\sigma / m$ from dwarf galaxies in the field~\cite{Tulin:2017ara}, $\sigma / m \sim 0.1 \text{-} 10 \, {\rm cm^2/g}$, though they have similar collision velocities, $v \sim 20 \text{-} 60 \, {\rm km/s}$.
It might be because Ref.~\cite{Tulin:2017ara} assumes the halos to be in the core expansion phase.
If we repeated their analyses in the core contraction phase, we might find a larger value of $\sigma / m$ for the field dwarf galaxies as well.
On the other hand, the inferred $\sigma / m$ from the MW's satellites changes with different initial conditions of halos and modeling of tidal stripping.
With higher initial concentrations of halos, the inferred cross sections may be appreciably lower than in Ref.~\cite{Correa:2020qam}, $\sigma / m \sim 3 \, {\rm cm^2/g}$~\cite{Sameie:2019zfo}, mitigating the tension between the MW's dwarf spheroidal satellite galaxies and the field dwarf galaxies.
In this paper, we take the inferred values from Ref.~\cite{Correa:2020qam} at face values and study the implications of the strong velocity dependence of $\sigma / m$ (see \cref{sec:sat-vs-field} for the changes in conclusions if the inferred $\sigma / m$ in the MW's satellites was lowered).

Furthermore, the above values of $\sigma / m$ are inferred under the assumption of a constant cross section.
If we reanalyze the astronomical data by taking velocity-dependent cross sections, the tension between the MW's satellites and the field dwarf galaxies could be mitigated.
For velocity-dependent cross sections, we also have to take an average over local velocity distribution (see \cref{sec:v-average} for more discussion).
In this paper, we compare the velocity-dependent cross section without the distribution averaging with the inferred values of $\sigma / m$ mentioned above.
The velocity-dependent cross section considered in this paper may give a benchmark with which the astronomical data will be reanalyzed.

As seen in the next section, this strong velocity dependence is realized when the self-scattering cross section almost saturates the $s$-wave Unitarity bound, i.e., on the quantum (zero-energy) resonance~\cite{Tulin:2012wi,Tulin:2013teo}. 
On the quantum resonanace, the self-scattering cross section is singly parameterized by the DM mass.
We take a closer look at the quantum resonance by using the effective-range theory.
The quantum resonance allows us to pin down the model parameters as well as the DM mass.
In \cref{sec:Lmu-Ltau}, we consider DM with a light mediator based on the gauged $L_{\mu}-L_{\tau}$ model.
We identify the mediator and DM masses that explain the strong velocity dependence of DM self-interaction.
Interestingly, we find that the discrepancy in muon anomalous magnetic moment $g-2$ and tension in $H_{0}$ are mitigated in the very parameter points.
The light $L_{\mu}-L_{\tau}$ gauge boson are also subject to various experimental searches such as non-standard interactions of neutrinos and missing-energy events in colliders.
In \cref{sec:ADM}, we consider composite asymmetric dark matter (ADM) based on the dark quantum chromodynamics (QCD) and electrodynamics (QED).
The dark QCD scale and dark pion masses are constrained so that dark nucleon DM can have the indicated velocity-dependent cross section.
The binding energies of two nucleons and vector resonances are also predicted, which have important implications for the dark nucleosynthesis in the early Universe and dark spectroscopy measurements in lepton colliders, respectively.
The dark photon, which is required to make cosmology viable, is subject to intensive experimental efforts such as beam-dump experiments.
The dark proton may be found in DM direct-detection experiments.
We give concluding remarks in \cref{sec:conclusion}.

\section{Effective-range theory \label{sec:eff-ran}}

The effective-range theory is first developed in the attempt to understand low-energy scatterings of nucleons~\cite{Blatt:1949zz,Bethe:1949yr}.
See Ref.~\cite{Kaplan:2005es} for a review of the effective-range theory in the context of effective-field theory.
Here we refer to Ref.~\cite{Chu:2019awd} that revisits the effective-range theory in the context of SIDM.
The 2-body scattering cross section can be decomposed into the partial-wave (multipole $\ell$) contributions:
\eqs{
\sigma = \sum_{\ell} \sigma_{\ell} = \sum_{\ell} \frac{4 \pi}{k^{2}} (2 \ell + 1) \sin^{2} \delta_{\ell} \,.
\label{eq:pwesigma}
}
Here we ignore a possible quantum interference for the identical particles.%
\footnote{One may also choose to use the transfer cross section $\sigma_T$, which may be a more suitable quantity to parameterize the momentum-transfer effect among DM particles through elastic scattering.
For the scattering of identical particles, $\sigma_T$ is written as~\cite{Kahlhoefer:2013dca}
\begin{equation}
    \sigma_T = \int d \Omega (1 - |\cos{\theta}|) \frac{d \sigma}{d \Omega}\,,
\end{equation}
where $\sigma$ is the standard cross section, $\sigma=\int d\Omega (d\sigma/d\Omega)$.
Since we focus on the $s$-wave scattering, this amounts to an additional factor of $1/2$, $\sigma_T = \sigma/2$.
}.
The momentum $k = \mu v_{\rm rel}$ is given by the reduced mass $\mu$ ($\mu = m / 2$ for the identical particles with the mass of $m$) and relative velocity $v_{\rm rel}$.
The analytic properties of the wave function determine the low-energy behavior of  $\delta_{\ell}$ as
\eqs{
k^{2 \ell + 1} \cot \delta_{\ell} \to - \frac{1}{a_{\ell}^{2 \ell + 1}} + \frac{1}{2 r_{e\ell}^{2 \ell - 1}} k^{2} \quad (k \to 0) \,.
}
Here $a_{\ell}$ and $r_{e \ell}$ are called the scattering length and effective range, respectively.

In the following, we focus on the $s$-wave ($\ell = 0$), which is expected to be dominant at the low energy.%
\footnote{
One prominent counterexample is the Rutherford scattering via the exchange of an (almost) massless mediator.
Hereafter we omit the subscript of $\ell = 0$ for notational simplicity.
}
Then, the low-energy cross section is given by
\eqs{
\sigma = \frac{4 \pi a^{2}}{1 + k^{2} (a^{2} - a r_{e}) + a^{2} r_{e}^{2} k^{4} / 4} \,.
}
When $|a| \gg |r_{e}|$, in the range of $1/|r_{e}| > k > 1/|a|$, the cross section saturates the Unitarity bound,
\eqs{
\sigma_{\rm max} = \frac{4 \pi}{k^{2}} \,,
}
which is singly parameterized by the DM mass, and thus gives a good benchmark for studying halos in velocity-dependent SIDM.
For the DM mass of $m \sim 10 \, {\rm GeV}$, $\sigma_{\rm max} / m$  simultaneously explains the inferred $\sigma / m$'s from MW's dwarf spheroidal satellite galaxies and galaxy clusters, as we will see shortly.
A large $|a|$ indicates the existence of the shallow bound state ($a > 0$) or virtual state ($a < 0$; bound state with a non-normalizable wave function), depending on its sign.
The bound (virtual) state energy is given by the positive (negative) imaginary pole $k_{\rm pole}$:
\eqs{
E_{b} = - \frac{k_{\rm pole}^{2}}{2 \mu} \,, \quad k_{\rm pole} = \frac{i}{r_{e}} \left( 1 - \sqrt{1 - 2 r_{e} / a} \right) \,.
}

\begin{figure}
\centering
\begin{subfigure}{1.\textwidth}
\centering
  \includegraphics[width=0.93\linewidth]{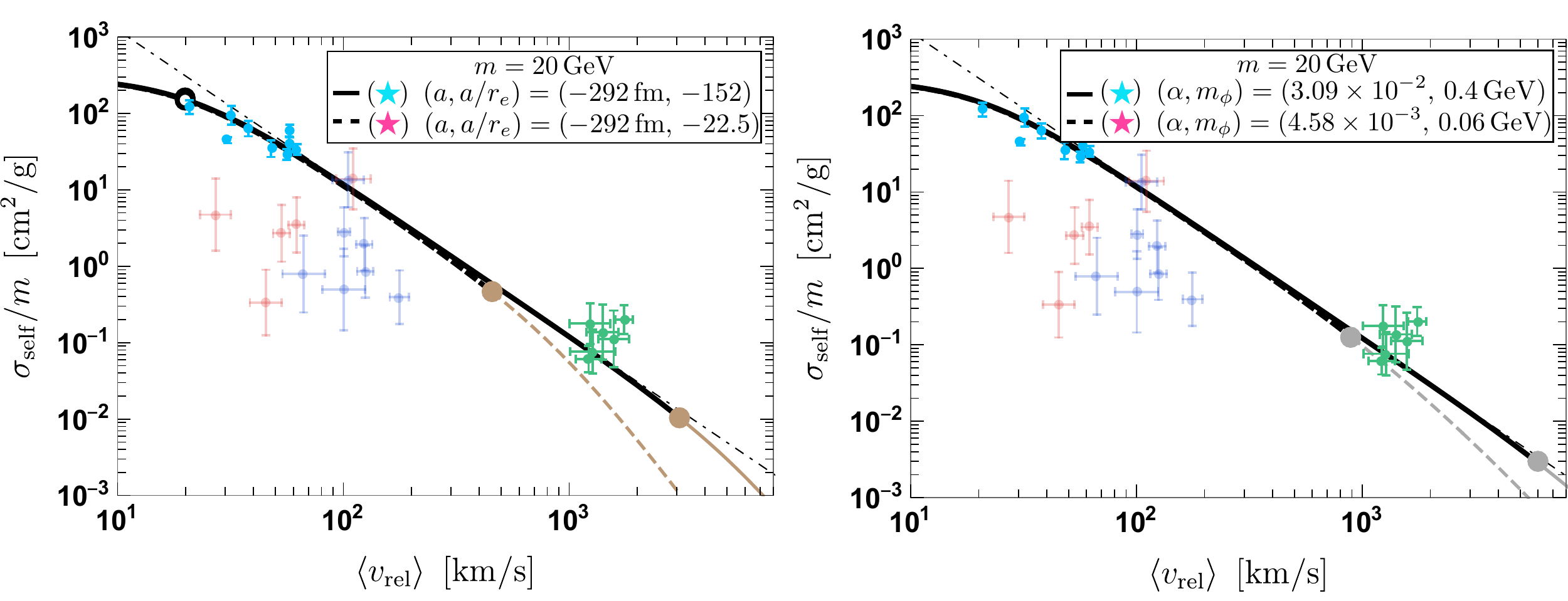}
  \caption{}
  \label{fig:ERTvelocity1}
  \par\bigskip
  \includegraphics[width=0.93\linewidth]{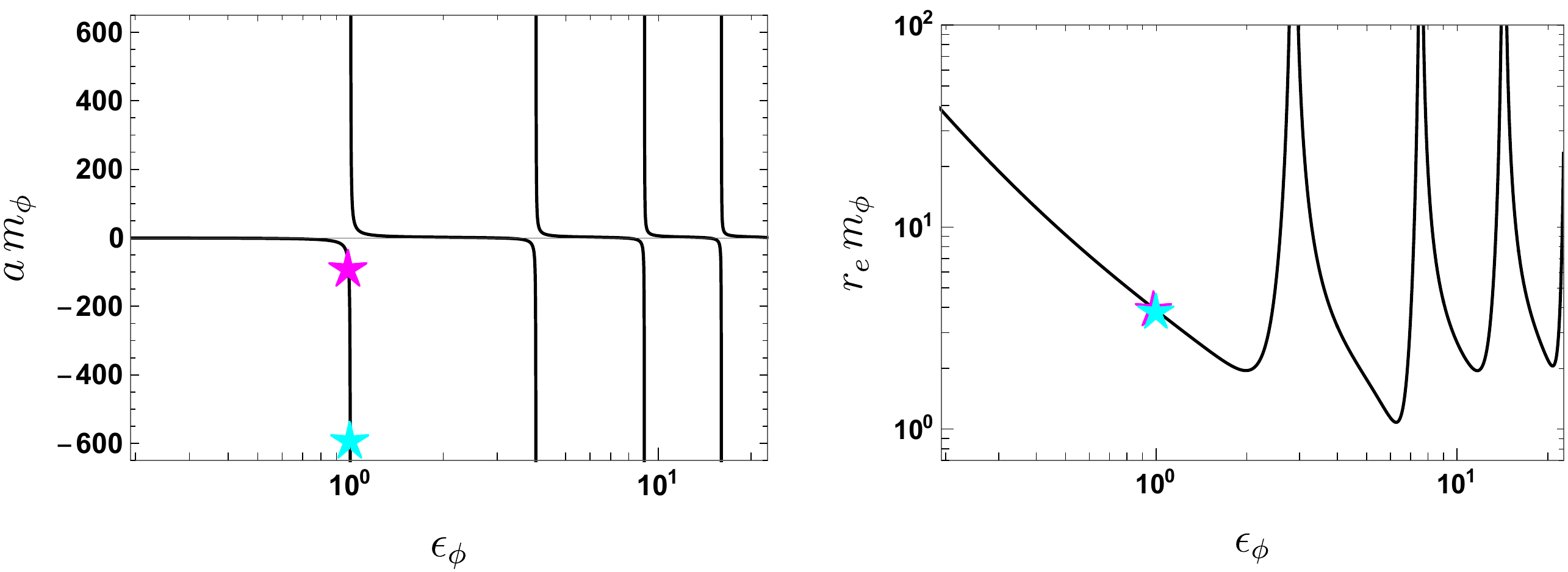}
  \caption{}
  \label{fig:Hulthenparam1}
\end{subfigure}
\caption{
{\bf(a)} 
(\textit{Left}): Velocity dependence of $\sigma/m$ (in black) in the effective-range theory for given DM mass and the effective-range theory parameters $(a,r_e)$.
The dot-dashed curve is the Unitarity bound for the DM self-scattering cross section, $\sigma_{\rm max}/m$.
$\sigma/m$ saturates the Unitarity bound from $k \gtrsim 1/|a|$ (unfilled circle) to $k \lesssim 1/|r_e|$ (filled circle), where the cross section is singly parametrized by the DM mass.
The data points are the inferred values of $\sigma/m$ from the observations on the field dwarf (red)/LSB (blue) galaxies, and galaxy clusters (green)~\cite{Kaplinghat:2015aga}.
The data points in cyan are the values that may explain the anti-correlation between the central DM densities and their orbital pericenter distances~\cite{Correa:2020qam}.
The effective-range theory deviates from the Hulth\'en potential (see the right panel) results for $k\gtrsim 1/|r_e|$ (in brown) as the effective-range expansion fails.
(\textit{Right}): Same as the left panel but for the $s$-wave contribution of $\sigma/m$ in the Hulth\'en potential.
The Hulth\'en-potential parameter sets took are in one-to-one correspondence with the effective-range theory parameter sets in the left panel.
The displayed velocity dependence of $\sigma / m$ is not reliable in the classical regime (in gray), i.e., $k\gtrsim m_\phi$, as the higher partial-wave contributions become important; the filled circle denotes the point $k=m_\phi$.
{\bf (b)} $s$-wave scattering length (left) and effective range (right) in units of $m_{\phi}^{-1}$ as a function of $\epsilon_\phi$ for the Hulth\'en potential. The stars indicate the benchmark parameters took in \cref{fig:ERTvelocity1}, which are near the first quantum resonance, $\epsilon_\phi=1$.
}
\end{figure}

In the left panel of \cref{fig:ERTvelocity1}, we show the velocity dependence of $\sigma / m$ (black) in the effective-range theory.
The data points show the preferred values of $\sigma / m$ as a function of the relative velocity;
they are inferred from the observations on central DM densities of MW's dwarf spheroidal satellite galaxies (cyan), field dwarf spheroidal galaxies (red)/low surface brightness spiral galaxies (blue), and galaxy clusters (green).
There can be seen that the cyan data points disagree with the red data points at $v \sim 30 \, {\rm km/s}$, as discussed in the previous section.
Hereafter we take the former seriously, while being ignorant to the latter;
we give further discussion focusing on the red data points in \cref{sec:sat-vs-field}.
We take the DM mass $m = 20 \, {\rm GeV}$ so that the $s$-wave Unitarity bound $\sigma_{\rm max}/m$ (dot-dashed) crosses the inferred $\sigma / m$'s from the MW's satellites (cyan) and the galaxy clusters (green).
The effective-range theory parameter sets took are $(a, a/r_e) = (-292\,{\rm fm},-152)$ (solid) and $(a, a/r_e) = (-292\,{\rm fm},-22.5)$ (dashed).
Both data sets exhibit $|a/r_e|\gg1$.
This indicates that they are on the quantum (zero-energy) resonance.
The Unitarity bound is saturated from $k \gtrsim 1/|a|$ (unfilled circle) to $k \lesssim 1/|r_e|$ (filled circle); in this regime, the cross section is singly determined by the DM mass.
While the both data sets have the same $a$, the former data set has smaller $|r_e|$ and hence it saturates the Unitarity bound in a wider range of $k$.
Indeed, the solid curve saturates the Unitarity bound all the way up to $v_{\rm rel}\sim 3000\,{\rm km/s}$, while the dashed curve desaturates for $v_{\rm rel}\gtrsim 400\,{\rm km/s}$.
The solid curve exhibits $\sigma / m\propto 1/v_{\rm rel}^2$ at the presented velocity range.
Such strong velocity dependence simultaneously explains the preferred $\sigma/m\sim0.1\,{\rm cm^2/g}$ at $v_{\rm rel}\sim1000\,{\rm km/s}$ and $\sigma/m\sim100\,{\rm cm^2/g}$ at $v_{\rm rel}\sim30\,{\rm km/s}$.
Interestingly, such velocity dependence provides a good fit to the inferred $\sigma/m$'s from the MW's satellites (cyan) at low-velocities.
As we take smaller $|a/r_e|$, $\sigma / m$ desaturate like the dashed curve, but may still explain the cyan data points.
Meanwhile, taking a larger $|a|$ than the data sets would extend the range of $k$ where the Unitarity bound is saturated;
it would retreat the unfilled circles to smaller velocities.
We remark that while an arbitrarily larger $|a|$ would still provide a good fit to the cyan data points, it is not clear how such velocity dependence of $\sigma/m$ affects the structure of DM halos as small as (or smaller than) the MW's satellites.
This is because the SIDM evolution of such small halos may transit into the short mean free path regime~\cite{Balberg:2001qg,Ahn:2004xt} due to the enhanced $\sigma / m$ at low velocities.

It is illustrating to consider the scattering under the Hulth\'en potential:
\eqs{
V(r) = - \frac{\alpha \delta e^{- \delta r}}{1 - e^{- \delta r}} \,.
}
The Hulth\'en potential approximates the Yukawa potential,
\eqs{
V (r) = - \frac{\alpha e^{- m_{\phi} r}}{r} \,,
}
with a proper choice of $\delta$.
Here $m_{\phi}$ is the mass of a mediator and we take $\delta = \sqrt{2 \zeta(3)} m_{\phi}$ with the $\zeta (z)$ being the Riemann zeta function.%
\footnote{
The coefficient of $\sqrt{2 \zeta(3)} = 1.55\dots$ is obtained by equating the Born cross sections with the Hulth\'en and Yukawa
potentials~\cite{Tulin:2013teo}.
The relation of $\delta = 2 \zeta(3) m_{\phi}$ in Ref.~\cite{Chu:2019awd} would be a typo.
A similar prescription for the Sommerfeld-enhancement factor leads to the coefficient of $\zeta(2) = \pi^{2} / 6 = 1.64\dots$~\cite{Cassel:2009wt}.
}
The Hulth\'en potential enjoys an analytic expression of the phase shift:
\eqs{
\delta_{0} = \arg \left( \frac{i \Gamma (\lambda_{+} + \lambda_{-} - 2)}{\Gamma (\lambda_{+}) \Gamma (\lambda_{-})} \right) \,.
\label{eq:Hulthendelta}
}
Here $\Gamma (z)$ is the gamma function and
\eqs{
\lambda_{\pm} = 1 + i \epsilon_{v} \epsilon_{\phi} \pm \sqrt{\epsilon_{\phi} - \epsilon_{v}^{2} \epsilon_{\phi}^{2}} \,, \quad \epsilon_{v} = \frac{v_{\rm rel}}{2 \alpha} \,, \quad \epsilon_{\phi} = \frac{2 \alpha \mu}{\delta} \,.
}
This phase shift results in the scattering length and effective range of
\eqs{
& a = \frac{\psi^{(0)} (1 + \sqrt{\epsilon_{\phi}}) + \psi^{(0)} (1 - \sqrt{\epsilon_{\phi}}) + 2 \gamma_{E}}{\delta} \,, \\
& r_{e} = \frac{2 a}{3} - \frac{1}{3 \delta^{3} \sqrt{\epsilon_{\phi}} a^{2}} \left\{ 3 \left[ \psi^{(1)} (1 + \sqrt{\epsilon_{\phi}}) - \psi^{(1)} (1 - \sqrt{\epsilon_{\phi}}) \right] + \sqrt{\epsilon_{\phi}} \left[ \psi^{(2)} (1 + \sqrt{\epsilon_{\phi}}) + \psi^{(2)} (1 - \sqrt{\epsilon_{\phi}}) + 16 \zeta (3) \right] \right\} \,,
}
respectively.
Here $\psi^{(n)} (z)$ is the polygamma function of order $n$ and $\gamma_{E} = 0.577\dots$ is the Euler–Mascheroni constant.
Above we consider an attractive potential, by taking positive $\alpha$.
One can use the same expressions with an analytic continuation to negative $\alpha$:
$\lambda_{\pm} = 1 + i \epsilon_{v} \epsilon_{\phi} \pm i \sqrt{\epsilon_{\phi} + \epsilon_{v}^{2} \epsilon_{\phi}^{2}}$,
with $\epsilon_{v} = \dfrac{v_{\rm rel}}{2 |\alpha|}$ and $\epsilon_{\phi} = \dfrac{2 |\alpha| \mu}{\delta}$.
Note that for a given DM mass $m$, the effective-range theory parameter set $(a,r_e)$ and the Hulth\'en-potential parameter set $(\alpha,m_\phi)$ has a one-to-one correspondence.

While effective-range theory reproduces well the analytic results of the Hulth\'en potential in the $k\rightarrow0$ limit, it starts to deviate for $k^{-1}\lesssim r_e$;
in other words, when the de Broglie wavelength of the incoming particles is shorter than the effective range.
Such regime is depicted in brown in the left panel of \cref{fig:ERTvelocity1}.
This is expected since the low-momentum effective-range expansion fails at high $k$.
This feature can be seen by comparing the left and the right panel of \cref{fig:ERTvelocity1};
the right panel is the analytic result of the Hulth\'en potential with the parameter sets that give the same effective-range theory parameter sets as in the left panel.

Meanwhile, the velocity dependence of $\sigma/m$ in the Hulth\'en potential is not reliable in the classical regime, $k^{-1}\lesssim m_{\phi}^{-1}$, i.e., where the de Broglie wavelength of the incoming particles is shorter than the range of the Hulth\'en potential.
This is because the higher partial-wave contributions become important in the classical regime.
In the right panel of \cref{fig:ERTvelocity1}, the classical regime is depicted in gray.
Note that the classical regime in the Hulth\'en potential roughly coincides with the regime of $k^{-1}\lesssim r_e$ in the effective-range theory.

\cref{fig:Hulthenparam1} shows $a$ (left) and $r_{e}$ (right) as a function of $\epsilon_{\phi}$ in the units of $m_{\phi}^{-1}$.
The resonances appear at $\epsilon_{\phi} = n^{2}$ ($n = 1, 2, \dots$), where $|a|\rightarrow \infty$ for finite $r_e$.
In the following, we focus on the first resonance, $n = 1$, while discussing higher resonances in \cref{sec:high-res}.
The points depicted by stars corresponds to the parameters took in \cref{fig:ERTvelocity1};
they are both near the first resonance, nearly saturating the Unitarity bound.

Above we consider the elastic scattering and see that we need a light mediator and induced relatively long-range force to reproduce the strong velocity dependence.
The Sommerfeld enhancement for DM annihilation~\cite{Hisano:2002fk,Hisano:2003ec,Hisano:2004ds,Hisano:2005ec,Cirelli:2007xd,Cirelli:2008pk,ArkaniHamed:2008qn,Cassel:2009wt,Feng:2010zp} is controlled by the same potential.
If the Sommerfeld enhancement is also almost on the quantum resonance, this would have multiple implications for the cosmology: strong constraints from the indirect-detection experiments if DM annihilation ends up with the electromagnetic energy injection~\cite{Bringmann:2016din}; and the second stage of the DM freeze-out (i.e., re-annihilation)~\cite{Dent:2009bv,Zavala:2009mi,Feng:2010zp,vandenAarssen:2012ag,Binder:2017lkj}.
The enhancement factor in the Hulth\'en potential is given by~\cite{Duerr:2018mbd}%
\footnote{
Note that our $\epsilon_{\phi}$ is different from Ref.~\cite{Feng:2010zp}.
This expression coincides with Ref.~\cite{Cassel:2009wt}:
\eqs{
S_{\ell} = \left| \frac{\Gamma(\lambda_{+ \ell}) \Gamma(\lambda_{- \ell})}{\Gamma (\lambda_{+ \ell} + \lambda_{- \ell} - 1 + \ell) \Gamma (1 + \ell)} \right|^{2} \,, \quad \lambda_{\pm \ell} = \ell + \lambda_{\pm} \,.
}
}
\eqs{
& S_{s\text{-wave}} = \frac{\pi}{\epsilon_{v}} \frac{\sinh (2 \pi \epsilon_{v} \epsilon_{\phi})}{\cosh (2 \pi \epsilon_{v} \epsilon_{\phi}) - \cos \left(2 \pi \sqrt{\epsilon_{\phi} - \epsilon_{v}^{2} \epsilon_{\phi}^{2}} \right)} \,, \\
& S_{p\text{-wave}} = S_{s\text{-wave}} \frac{(\epsilon_{\phi} - 1)^{2} + 4 \epsilon_{v}^{2} \epsilon_{\phi}^{2}}{1 + 4 \epsilon_{v}^{2} \epsilon_{\phi}^{2}} \,,
}
for $s$-wave and $p$-wave annihilation, respectively.
For annihilation, we take $\delta = \zeta(2) m_{\phi}$.

\begin{figure}
\centering
\begin{subfigure}{1.\textwidth}
\centering
  \includegraphics[width=0.93\linewidth]{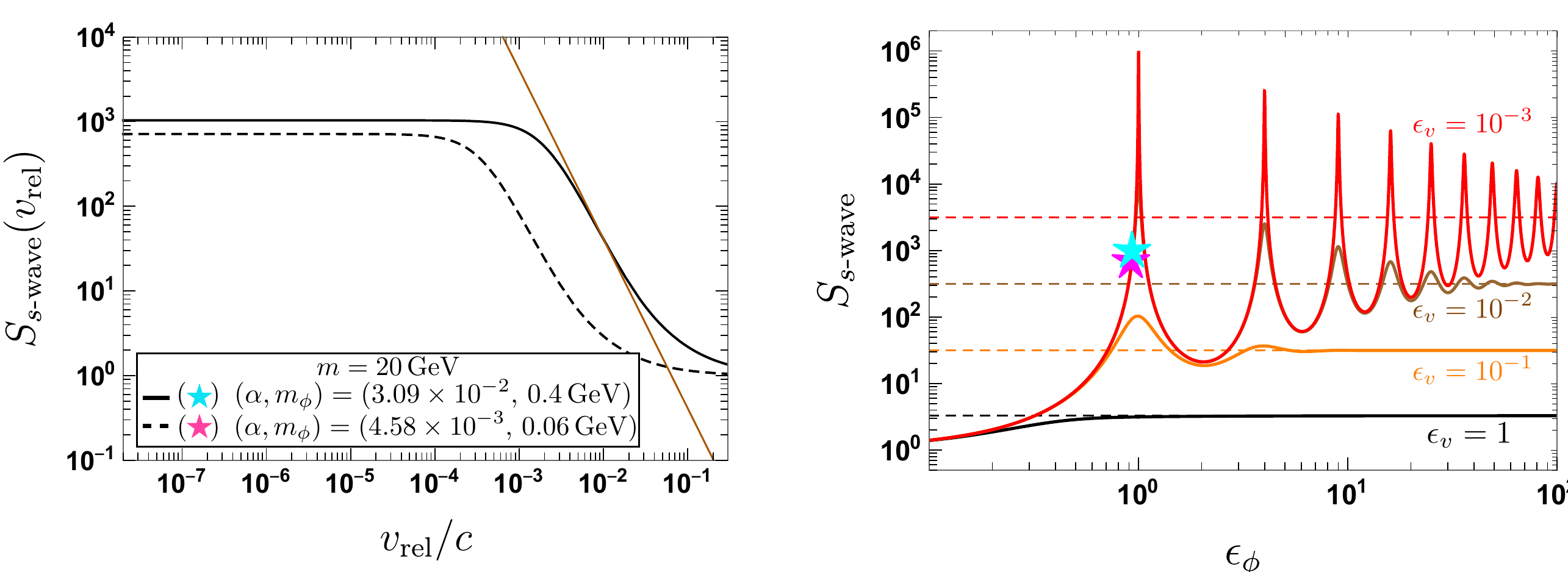}
  \caption{}
  \label{fig:SEFs1}
  \par\bigskip
  \includegraphics[width=0.93\linewidth]{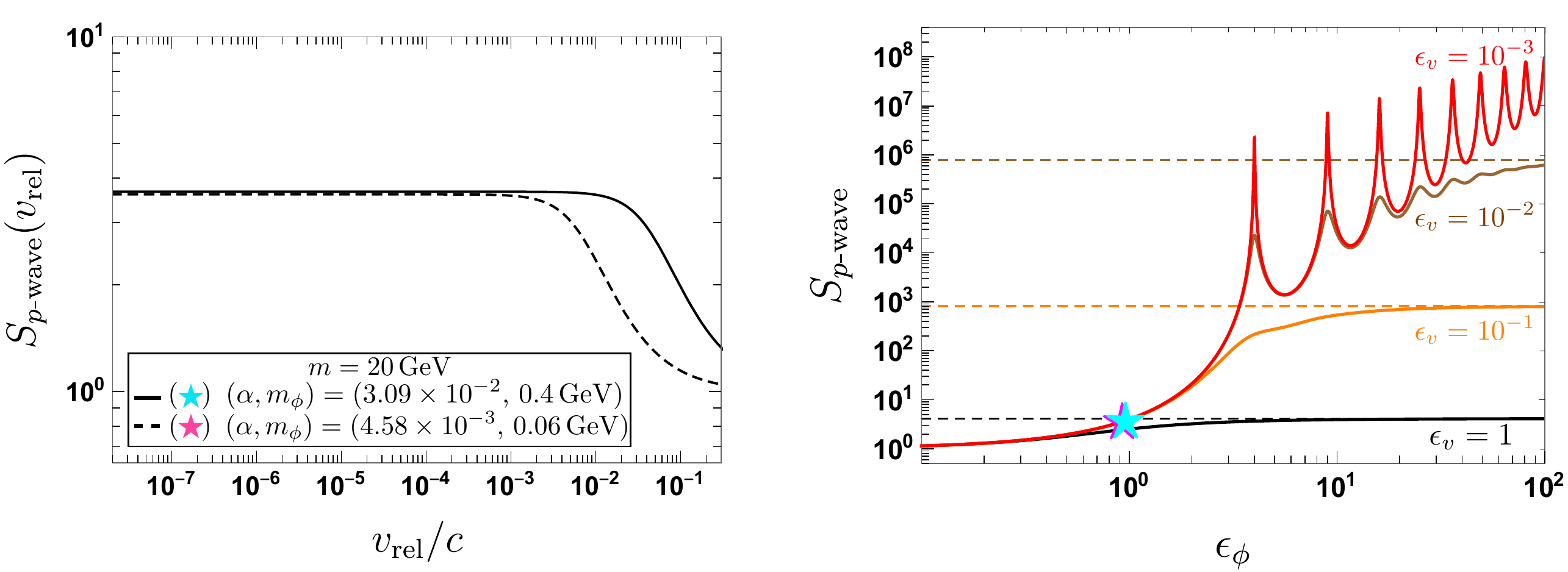}
  \caption{}
  \label{fig:SEFp1}
\end{subfigure}
\caption{
{\bf(a)} ({\it Left}): Velocity dependence of the $s$-wave Sommerfeld-enhancement factor. The Hulth\'en-potential parameters took corresponds to the benchmark parameters in \cref{fig:ERTvelocity1}. The $s$-wave annihilation is near the quantum resonance for the benchmark parameters, as shown in the right panel.
The Sommerfeld-enhancement factor grows as $\propto 1/v_{\rm rel}^2$ (brown) towards low velocity and saturates. ({\it Right}): $\epsilon_\phi$-dependence of the $s$-wave Sommerfeld-enhancement factor.
We take $\epsilon_v = 1, \cdots, 10^{-3}$ from bottom to top.
The dashed lines show the Sommerfeld-enhancement factors in the Coulomb potential that correspond to $\epsilon_\phi \to \infty$.
The points depicted by stars correspond to the benchmark parameters took in \cref{fig:ERTvelocity1} for fixed $\epsilon_v = 10^{-3}$.
{\bf (b)} Same as \cref{fig:SEFs1}, but for the $p$-wave Sommerfeld-enhancement factor.
}
\label{fig:SEF1}
\end{figure}

The left panels of \cref{fig:SEF1} shows the Sommerfeld-enhancement factor as a function of the relative velocity for $s$-wave (top) and $p$-wave (bottom).
The Hulth\'en-potential parameters took corresponds to the parameters depicted as stars in \cref{fig:Hulthenparam1}, which are close to the quantum resonance for elastic scattering.
For these parameters, the $s$-wave Sommerfeld-enhancement factor is also resonantly enhanced towards low-velocity as $\propto1/v^2_{\rm rel}$ and saturates;
as shown in the left panel of \cref{fig:SEFs1}, the parameters are close to the $s$-wave quantum resonance for annihilation.
This may be dangerous in terms of the constraints from observations on the CMB and indirect-detection experiments.
The constraints may be evaded if we consider neutrinos as the final DM annihilation product or consider the $p$-wave annihilation~\cite{Bringmann:2016din,Kamada:2018zxi}.
Indeed, the $p$-wave Sommerfeld-enhancement factors for the parameters are not huge, since they are sufficiently far from the $p$-wave quantum resonance for DM annihilation (see the right panel of \cref{fig:SEFp1}).

In \cref{sec:Lmu-Ltau}, we consider a model with a light mediator based on the gauged $L_{\mu} - L_{\tau}$ model, where DM annihilation is $p$-wave and ends up with neutrinos.
ADM is another good candidate, since it involves a light mediator to deplete the thermal relic abundance of the symmetric component.
ADM may also be expected to be safe from the indirect-detection bounds at the first sight.
In \cref{sec:ADM}, nevertheless, we see that even a (inevitable induced) tiny DM-anti DM oscillation leads to a significant annihilation in the late Universe.

\section{Dark matter with a light mediator: gauged $L_{\mu} - L_{\tau}$ model \label{sec:Lmu-Ltau}}

In this section, we consider DM with a light mediator based on the gauged $L_{\mu} - L_{\tau}$ model~\cite{Kamada:2018zxi}.
In this model, a new gauge boson $Z'$ couples to the muon and tau lepton (and their neutrinos)~\cite{He:1990pn,He:1991qd}.
Interestingly, this new force contributes to the muon $(g-2)_{\mu}$, alleviating a possible tension between the SM prediction and  measurement~\cite{Davier:2010nc,Hagiwara:2011af,Keshavarzi:2019abf} (See also the latest review on $(g-2)_\mu$ in the SM, Ref.~\cite{Aoyama:2020ynm}).%
\footnote{
Two new experiments would shed light on the longstanding tension in $(g-2)_\mu$ in the near future: E989 experiment at Fermilab \cite{Grange:2015fou} and the E34 experiment at J-PARC \cite{Abe:2019thb}.
The lattice studies of hadronic contributions to $(g-2)_\mu$ may also resolve the tension in the future.
The latest study on the hadronic vacuum polarization contribution \cite{Borsanyi:2020mff} claims that the discrepancy between the SM prediction and the experimental result disappears. 
Their result is also in tension with the other result (such as the R-ratio determination), and thus it is expected to scrutinize their result in detail by the other lattice groups.
}
After other constraints are taken into account, the $(g-2)_{\mu}$ discrepancy is ameliorated for the gauge coupling of $g' \simeq 4 \times 10^{-4} \text{-} 10^{-3}$ and the mass of the new gauge boson of $m_{Z'} \simeq 8 \text{-} 200 \, {\rm MeV}$.

Ref.~\cite{Kamada:2018zxi} extends this model with the a vector-like pair of fermions, $N$ and ${\bar N}$.
We assume that the $L_{\mu} - L_{\tau}$ breaking Higgs $\Phi$ carries a unit charge and $N$ (${\bar N}$) carries a (minus) half charge in units of the muon and tau-lepton charges.%
\footnote{
The charge of $\Phi$ determines the neutrino mass matrix by the see-saw mechanism~\cite{Minkowski:1977sc,Yanagida:1979as,Glashow:1979nm,GellMann:1980vs}.
Our choice is a minimal viable one~\cite{Asai:2017ryy} (see also \cref{sec:neu_mass}) and may explain the baryon asymmetry of the Universe~\cite{Asai:2020qax} through the leptogenesis~\cite{Fukugita:1986hr,Giudice:2003jh,Buchmuller:2005eh,Davidson:2008bu}.
}
The resultant ${\mathbb Z}_{2}$ symmetry, which is unbroken by the Higgs vacuum expectation value (VEV), guarantees the stability of the lightest mass eigenstate $N_{1}$ among $N$ and ${\bar N}$, namely, $N_{1}$ being DM.
We consider the pseudo-Dirac dark matter, whose Dirac mass $m_{N}$ is larger than the Majorana mass induced by the VEV of $\Phi$.
The masses are given by the Yukawa coupling $y > 0$:
\eqs{
m_{N_{1(2)}} = m_{N} \pm \frac{\Delta m}{2} \,, \quad \Delta m = \frac{\sqrt{2} y}{g'} m_{Z'} \ll m_{N} \,.
}
Hereafter we assume the $CP$ symmetry to restrict the Yukawa coupling.
For more general discussion, please see Ref.~\cite{Kamada:2018zxi}.
The relevant interactions are
\eqs{
{\cal L}_{\rm int} \supset - \frac{y}{2 \sqrt{2}} \varphi (- {\overline N}_{1} N_{1} + {\overline N}_{2} N_{2}) + i \frac{g'}{2} Z'_{\mu} {\overline N}_{2} \gamma^{\mu} N_{1} \,.
}
Here $N_{1}$ and $N_{2}$ denote Majorana fermions and $\varphi$ is a real scalar resulting from $\Phi$.

The thermal relic abundance of $N_{1}$ is predominantly determined by co-annihilation of $N_{1} N_{1}, N_{2} N_{2} \to Z' Z', \varphi \varphi$ and $N_{1} N_{2} \to Z' \varphi$.
The thermally averaged annihilation cross sections are given by the dimensionless temperature $x = m_{N} / T$:
\eqs{
\langle \sigma v \rangle_{11} = \langle \sigma v \rangle_{22} = \frac{9 y^{4}}{64 \pi m_{N}^{2}} \frac{1}{x} \,, \quad \langle \sigma v \rangle_{12} = \frac{y^{4}}{64 \pi m_{N}^{2}} \,.
}
Here we drop terms proportional to the gauge coupling $g'$, because it is subdominant compared to those with Yukawa coupling $y$.
Note that self-annihilation channels are $p$-wave, while the co-annihilation channel is $s$-wave.
The effective annihilation cross section is given by the relative yield $r$~\cite{Griest:1990kh}:%
\footnote{
As we see below, we consider an ${\cal O}(10) \, {\rm MeV}$ $\varphi$, by taking a ${\cal O}(10^{-6})$ quartic coupling.
This may delay the phase transition of the $L_{\mu} - L_{\tau}$ breaking, possibly inducing a small thermal inflation.
Please see Ref.~\cite{Kamada:2018kmi} for a further discussion.
}
\eqs{
\langle \sigma v \rangle_{\rm eff} = (1 - r)^{2} \langle \sigma v \rangle_{11} + r^{2} \langle \sigma v \rangle_{22} + 2 r (1 - r) \langle \sigma v \rangle_{12} \,, \quad r = \frac{e^{- \Delta m / T}}{1 + e^{- \Delta m / T}} \,.
}
We compare the effective cross section at $x = 20$ with the canonical cross section, $(\sigma v)_{\rm can} = 3 \times 10^{-26} \, {\rm cm^{3} / s}$.
We fix the Yukawa coupling in this way.
The above expression of the effective cross section is derived under the assumption that $N_{1}$ and $N_{2}$ are in chemical equilibrium.
We check that in parameters of interest, the chemical equilibrium is achieved by $N_{2} \leftrightarrow N_{1} + Z'$.

\begin{figure}
\centering
\begin{subfigure}{1.\textwidth}
\centering
  \includegraphics[width=0.85\linewidth]{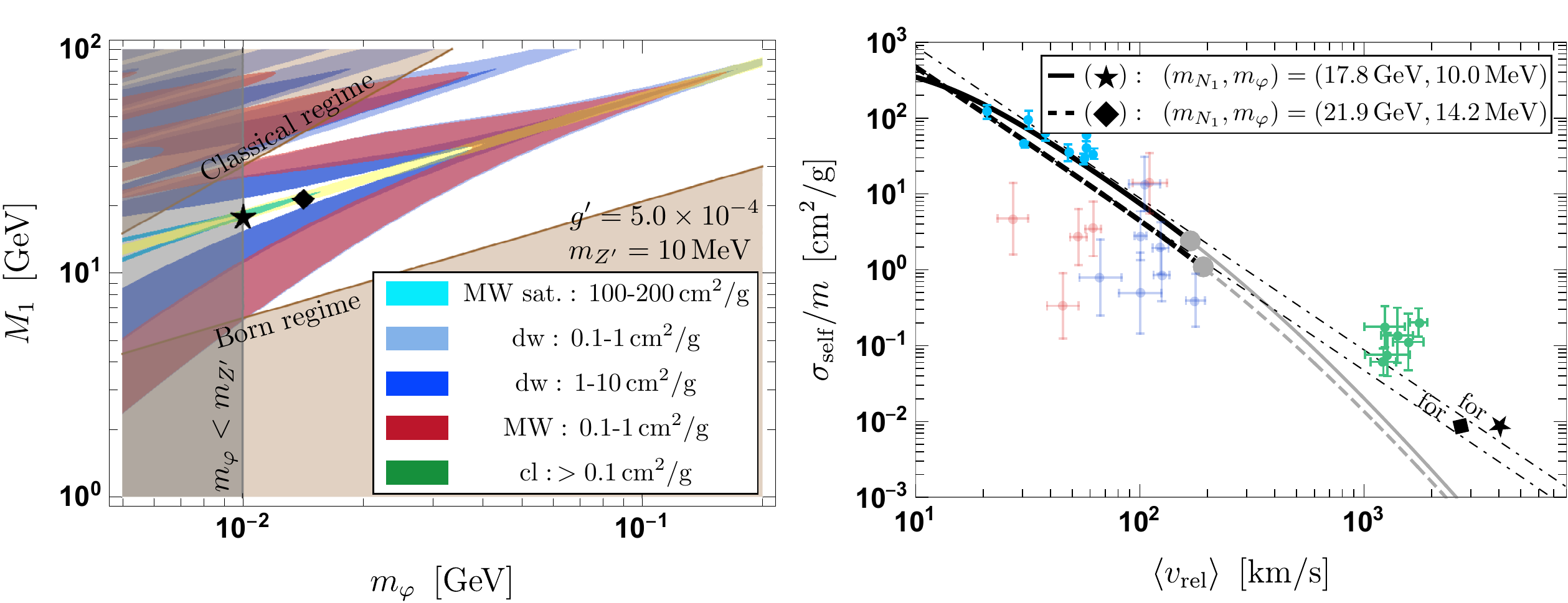}
  \caption{}
  \label{fig:lult_1}
  \par\medskip
  \includegraphics[width=0.85\linewidth]{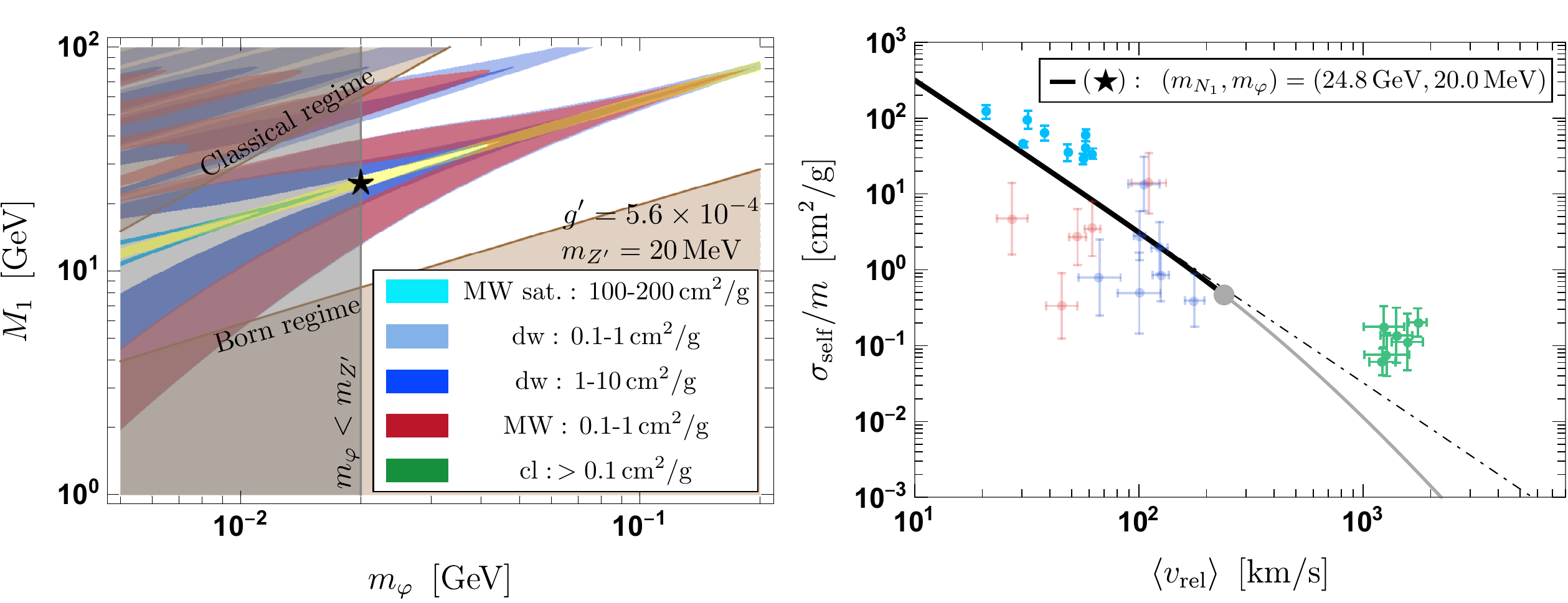}
  \caption{}
  \label{fig:lult_2}
  \par\medskip
  \includegraphics[width=0.85\linewidth]{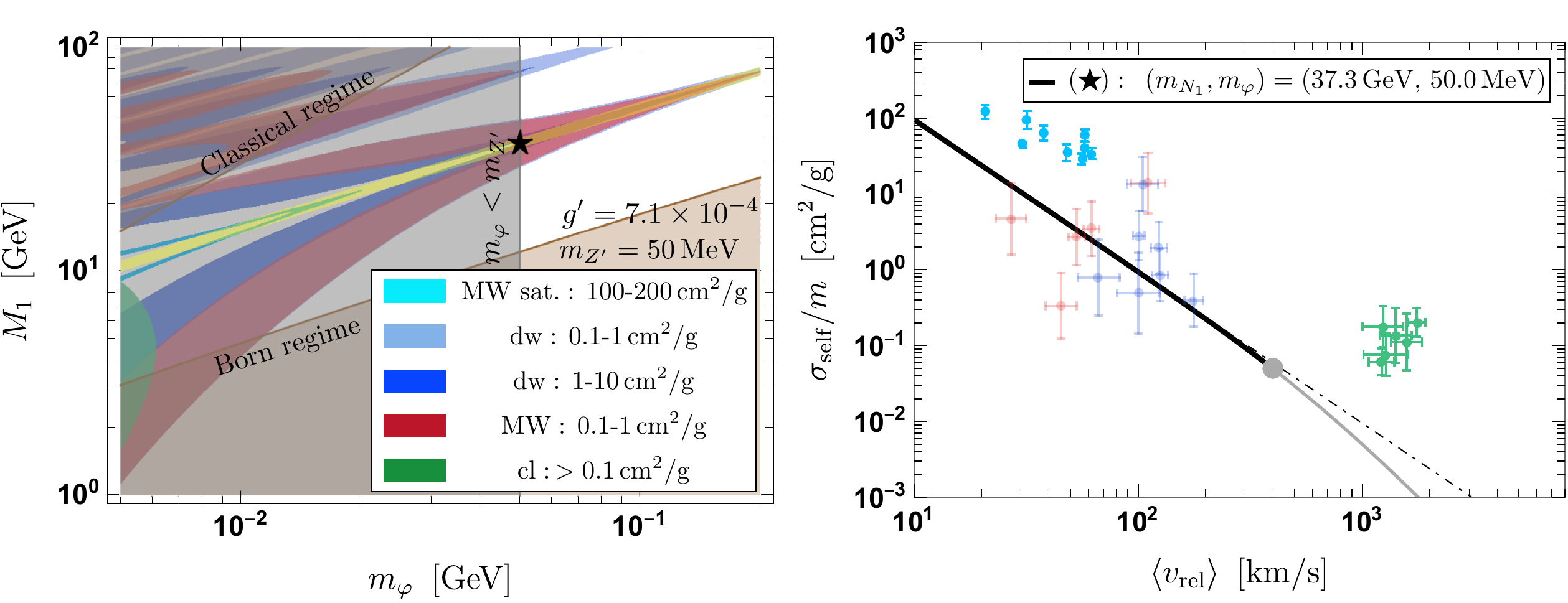}
  \caption{}
  \label{fig:lult_3}
\end{subfigure}
\caption{
(\textit{Left}): Parameter regions for $\sigma/m$ at the velocity scales of MW's dwarf spheroidal satellite galaxies (cyan), field dwarf galaxies (blue), MW-size galaxies (red), and galaxy clusters (green) for given $(m_{Z'},g')$.
The region where $m_{\varphi}<m_{Z'}$ (gray) may be disfavored by the BBN observations~\cite{Kamada:2018zxi}; lifetime of $\varphi$ is longer than $1\,{\rm s}$.
The yellow region depicts the parameters close to the first quantum resonance for elastic scattering, i.e., $0.85<\epsilon_{\phi}<1.15$ where $\epsilon_{\phi}=\alpha m_{N_1}/(\sqrt{2 \zeta(3)}m_{\varphi})$; this range includes the benchmark parameters shown as stars in \cref{fig:Hulthenparam1}.
(\textit{Right}): The velocity dependence of $\sigma/m$ (black) for the benchmark parameter points near the quantum resonance (depicted in the left panel).
The classical regime, i.e., $m_{N_1}v_{\rm rel}/m_\varphi\gtrsim1$, is depicted in gray, where the higher partial-wave contributions become important.
The dot-dashed curves are the Unitarity bound, $\sigma_{\rm max}/m$.
The data points are the same as in \cref{fig:ERTvelocity1}.
}
\label{fig:lult1}
\end{figure}

Again since the gauge coupling $g'$ is tiny, the 2-body potential of DM is also dominated by the Yukawa coupling:
\eqs{
V = - \frac{\alpha e^{- m_{\varphi} r}}{r} \,,
}
where $\alpha = y^{2} / (8 \pi)$.%
\footnote{
$\alpha = y^{2} / (4 \pi)$ in Ref.~\cite{Kamada:2018zxi} is an error, though it does not change the results much.
}
We approximate the Yukawa potential by the Hulth\'en potential.
We assume the Higgs mass $m_{\varphi}$ to be larger than the gauge boson mass $m_{Z'}$, namely, $m_{\varphi} > m_{Z'}$; otherwise the Higgs lifetime exceeds $1 \,{\rm s}$ and cause cosmological problems.
The left panels of \cref{fig:lult1} shows the parameter regions for $\sigma / m = 100$-$200 \, {\rm cm^{2}/g}$ in MW's dwarf spheroidal satellite galaxies (cyan), $0.1$-$1 \, {\rm cm^{2}/g}$ (light blue) and $1$-$10 \, {\rm cm^{2}/g}$ (blue) in field dwarf galaxies, $0.1$-$1 \, {\rm cm^{2}/g}$ in MW-size galaxies (red), and $>0.1 \, {\rm cm^{2}/g}$ (green) in galaxy clusters, where we take $v_{\rm rel}=20$, $30$, $200$, and $1000\,{\rm km/s}$, respectively.
There, we focus on $g^\prime$ and $m_{Z'}$ that ameliorates the $(g-2)_\mu$ discrepancy.
We use the analytic results of the Hulth\'en potential~\cite{Tulin:2012wi}, \cref{eq:Hulthendelta}, to calculate the elastic scattering cross section of $N_1$;
we multiply additional $1/2$ factor to \cref{eq:pwesigma} since we consider the elastic scattering among identical $N_1$ particles.
It should be noted that the analytic results of the Hulth\'en potential is valid only in the resonant regime, i.e., $\alpha m_{N_1}/m_{\varphi}\gtrsim1$ and $m_{N_1}v_{\rm rel}/m_{\varphi}\lesssim1$;
the analytic formula for the cross section in the classical regime, i.e., $\alpha m_{N_1}/m_{\varphi}\gtrsim1$ and $m_{N_1}v_{\rm rel}/m_{\varphi}\gg1$, is given in Ref.~\cite{Feng:2010zp}, and the Born approximation can be used in the Born regime, i.e., $\alpha m_{N_1}/m_{\varphi}\ll1$.
However, we remark that using the corresponding formulas in each regimes do not change our conclusion.
We indicate the classical and Born regimes (brown) in the left panels of \cref{fig:lult1};
for the classical regime, we require the condition for the velocity range that covers the MW's satellites, $v_{\rm rel}\lesssim100\,{\rm km/s}$.

From \cref{fig:lult_1} and \cref{fig:lult_2}, we see that with a light mediator, i.e., $m_{\varphi}\lesssim20\,{\rm MeV}$, $\sigma / m$ can be as large as $\sim100 \, {\rm cm^{2}/g}$ in the MW's satellites, while diminishing towards the galaxy clusters.
As an example, in the right panels of \cref{fig:lult1}, we show the velocity dependence of $\sigma / m$ (black) for the depicted benchmark parameters (as a star or a diamond) in the left panels.
In order to exhibit such large cross section at low-velocities, the parameters should lie very close to the quantum resonance (yellow region in the left panels).
The velocity dependence near the quantum resonance, $\sigma/m\propto1/v_{\rm rel}^2$, fits well the cyan data points. 
The presented velocity dependence of $\sigma / m$'s saturates the Unitarity bound (dot-dashed) at low velocities, $v_{\rm rel}=20\textrm{-}200\,{\rm km/s}$, where $\sigma / m$ is singly parameterized by the DM mass.
The $\sigma / m$'s desaturate from the Unitarity bound from the onset of the classical regime (filled circle) where the higher partial-wave contributions become significant.
Parametrically, we may extend the maximally self-interacting regime by taking a larger $m_\varphi$ (but with a smaller $\alpha$), so that $\sigma / m\propto1/v_{\rm rel}^2$ all the way up to the velocity scales of the galaxy clusters as in \cref{fig:ERTvelocity1}.
However, this is not possible since the Yukawa coupling is fixed by requiring a correct DM relic abundance;
therefore, we have no freedom to vary $m_\varphi$ for a given $\epsilon_\phi$.

With $Z'$ heavier than $m_{Z'}\gtrsim 20\,{\rm MeV}$, it is impossible to achieve such large cross sections ($\sigma/m\sim100\,{\rm cm^2/g}$) at the MW's satellites even at the quantum resonance (if we took the error bars at face values).
This is demonstrated in \cref{fig:lult_3}.
There, we take $m_{Z'}= 50\,{\rm MeV}$ as an example.
The cosmological constraints restricts $m_{\varphi}\gtrsim m_{Z'}$.
Thus, with heavier $Z'$, the constraint also restricts $N_1$ to be heavier near the quantum resonance (yellow).
Eventually, it becomes impossible to achieve $\sim100 \, {\rm cm^{2}/g}$ in the MW's satellites even if $\sigma/m$ saturates the Unitarity bound;
see the right panel of \cref{fig:lult_3}.

DM annihilates into $Z'$ and $\varphi$ predominantly in the $p$-wave, which subsequently decay predominantly into neutrinos.
It seems to follow that this DM is safe from stringent constraints from indirect-detection experiments.
On the other hand, considering the large Sommerfeld-enhancement factor for the $s$-wave, we need to be careful about the subdominant-modes of annihilation.
A leading $s$-wave annihilation channel is $N_{1} N_{1} \to Z' Z'$:
\eqs{
(\sigma v)_{s\text{-wave}} = \frac{g^{\prime 4}}{128 \pi m_{N}^{2}} \simeq  4.6 \times 10^{-36} \, {\rm cm^{3} / s} \left( \frac{g'}{5 \times 10^{-4}} \right)^{4} \left( \frac{20 \, {\rm GeV}}{m_{N}} \right)^{2} \,.
}
Considering the current (future) bound of DM annihilation into neutrinos mainly from Super-Kamiokande (Hyper-Kamiokande)~\cite{Arguelles:2019ouk}, $(\sigma v) \lesssim 10^{-24}$ ($10^{-25}$)\,${\rm cm^{3} / s}$ for $20 \, {\rm GeV}$ DM, the Sommerfeld-enhancement factor of $S \gtrsim 10^{11}$ needs attention.
The electromagnetic decay of $Z' \to e^{+} e^{-}$ via the kinetic mixing between $Z'$ and the SM hypercharge gauge boson needs another attention.
Its natural value is $\epsilon = g' / 70$ from the muon-tau lepton loop.
The electromagentic branching ratio is ${\rm Br} \sim (\epsilon e / g')^{2} \simeq 2 \times 10^{-5}$.
We may compare $(\sigma v) {\rm Br}$ with the constraints on DM electromagentic annihilation, e.g., from the CMB measurements~\cite{Aghanim:2018eyx}, $(\sigma v) \lesssim 10^{-26} \, {\rm cm^{3} / s}$ for $20 \, {\rm GeV}$ DM.

Before concluding this section, we describe the implications and prospects of our results.
We again assume the natural value of $\epsilon = g' / 70$ for the kinetic mixing, unless otherwise noted.
Since the light $L_{\mu} - L_{\tau}$ gauge boson predominantly decays into neutrinos, it heats only neutrinos after the neutrino decoupling, increasing the effective number of neutrino degrees of freedom $N_{\rm eff}$~\cite{Kamada:2015era,Kamada:2018zxi}.
Interestingly, its slight deviation from the SM value, $\Delta N_{\rm eff} \simeq 0.2 \text{-} 0.5$ mitigates the tension in the Hubble expansion rate $H_{0}$ between the local measurements (i.e., local ladder)~\cite{Riess:2016jrr,Riess:2018byc,Wong:2019kwg} and CMB-based measurements (i.e., inverse ladder)~\cite{Aghanim:2018eyx,Aiola:2020azj}.%
\footnote{Here we take $\Delta N_{\rm eff} = 0.2 \text{-} 0.5$ by na\"ively following~\cite{Aghanim:2018eyx}, though it is not a perfect solution to the $H_0$ tension~\cite{Vagnozzi:2019ezj,Arendse:2019hev}. $\Delta N_{\rm eff}>0$ leads to a smaller size of the sound horizon at the time of the decoupling (approximate to the last scattering), preferring a larger $H_0$ to keep the angular scale fixed. It cannot reproduce the ``redshift dependence'' of the inferred $H_0$ in the time-delay of the strongly lensed systems~\cite{Wong:2019kwg} (though not statistically significant). $\Delta N_{\rm eff} > 0$ may also be not preferred in light of the weak Silk damping measured in Ref.~\cite{Aiola:2020azj}.}
The corresponding $Z'$ mass is $m_{Z'} = 10 \text{-} 20 \, {\rm MeV} $~\cite{Escudero:2019gzq}, which coincides with our ``prediction'' surprisingly.
Such a sizable $\Delta N_{\rm eff}$ can be examined by CMB-S4 experiments~\cite{Wu:2014hta,Abazajian:2016yjj}.
The non-standard interactions of solar neutrinos with electrons and nuclei, in neutrino experiments (e.g., COHERENT) and DM experiments (e.g., LZ and Darwin), examine this light $Z'$ region~\cite{Amaral:2020tga, Sadhukhan:2020etu}.
The $Z'$-resonant non-standard interactions of high-energy neutrinos with cosmic neutrino background lead to a ``dip'' in the high-energy neutrino spectrum~\cite{Araki:2014ona,Araki:2015mya,Kamada:2015era}, which may explain (though not statistically significant) the null detection of astrophysical neutrinos with $200 \text{-} 400 \, {\rm TeV}$ in IceCube~\cite{Aartsen:2014gkd,Aartsen:2017mau,Aartsen:2020aqd}.
The light $Z'$ can be probed as missing-energy events in colliders~\cite{Gninenko:2014pea,Gninenko:2018tlp}.

\section{Composite dark matter: asymmetric dark matter \label{sec:ADM}}

In this section, we consider composite ADM based on the dark QCD$\times$QED dynamics with light dark quarks, up $U' (+2/3, +1/3)$ and down $D' (-1/3, +1/3)$ (dark QED charge, $B-L$ charge)~\cite{Ibe:2018juk}.
DM consists of dark nucleons, whose asymmetry has the same origin as the SM baryon asymmetry (i.e., co-genesis).
As a consequence, the mass of the dark nucleons is similar to but slightly heavier than that of SM nucleons as indicated from the coincidence of the mass densities $\Omega_{\rm dm} \sim 5 \Omega_{b}$.
In other words, the dynamical scales of dark QCD and SM QCD are similar to each other: $\Lambda_{\rm QCD'} \sim (10 / N'_{g}) \Lambda_{\rm QCD}$, where $N_{g}'$ is the number of the generations of $U'$ and $D'$.

The simple model of Ref.~\cite{Ibe:2018juk} is featured by the (intermediate-scale) neutrino portal and (low-scale) dark photon portal.
Non-equilibrium decay of the right-handed neutrinos with the masses of $M_{R} > 10^{9} \, {\rm GeV}$ generates the whole baryon
asymmetry (i.e., thermal leptogenesis~\cite{Fukugita:1986hr,Giudice:2003jh,Buchmuller:2005eh,Davidson:2008bu}).
The asymmetry is shared among the dark and SM sectors via the higher-dimensional portal operator.
The portal operator originates from the right-handed neutrinos and dark colored Higgs, $H_C'$, with the mass of $M_{C}$.
The dark photon is assumed to be the lightest particle in the dark sector.
It carries all the dark sector entropy in the late Universe, and transfers it through decay into SM particles via a kinetic mixing with the SM photon.
To this end, the dark photon mass is to be $m_{A'} \gtrsim 10 \, {\rm MeV}$ and the kinetic mixing parameter to be $\epsilon \gtrsim 10^{-10}$~\cite{Ibe:2018juk,Ibe:2019gpv}.
This phenomenological model enjoys compelling ultraviolet physics such as SU(5)$_{\rm SM}\times$SU(4)$_{\rm dark}$~\cite{Ibe:2018tex} and mirror SU(5)$_{\rm SM}\times$SU(5)$_{\rm dark}$~\cite{Ibe:2019ena}.
They involve an intermediate-scale supersymmetry and are discussed in \cref{sec:susy-adm}.

To discuss the phenomenology, we follow the simplified version of the model given by Ref.~\cite{Ibe:2019yra}.
The intermediate-scale portal operators are
\eqs{
  {\cal L} \supset  \frac{1}{M_{*}^{3}} ({\bar U}' {\bar D}' {\bar D'}) (LH) + \frac{1}{M_{*}^{3}} (U^{\prime \dagger} D^{\prime \dagger} {\bar D}^{\prime}) (LH) + {\rm h.c.} \,, \quad \frac{1}{M_{*}^{3}} = \frac{y_{N} Y_{N} Y_{C}}{2 M_{C}^{2} M_{R}} \,.
  \label{eq:int_portal}
}
Hereafter we assume the $CP$ symmetry to restrict the Yukawa coupling.
$Y_{C}$ is the Yukawa coupling of the dark colored Higgs to dark $U' D'$ and ${\bar U}' {\bar D}'$.
$y_{N}$ and $Y_{N}$ are the Yukawa couplings of the right-handed neutrinos to SM $LH$ and dark $H_{C}' {\bar D}'$, respectively.
The former is related with the observed tiny neutrino masses $m_{\nu}$ through the see-saw mechanism~\cite{Minkowski:1977sc,Yanagida:1979as,GellMann:1980vs,Glashow:1979nm}:
\eqs{
y_{N}^{2} \sim 10^{-5} \left( \frac{m_{\nu}}{0.1 \, {\rm eV}} \right)^{1/4} \left( \frac{M_{R}}{10^{9} \, {\rm GeV}} \right) \,.
}
The generated asymmetry is washed out when the other operators that carry different $B-L$ charges are also relevant after the asymmetry is generated.
Such harmful operators are not generated in ultraviolet physics based on a unification model discussed in Refs.~\cite{Ibe:2018tex,Ibe:2019ena}.
These portal operators are to be relevant at least around the time of the thermal leptogenesis: $x_{B-L} = M_{R} / T_{B-L} \simeq 5 \text{--} 10$~\cite{Fukugita:1986hr,Giudice:2003jh,Buchmuller:2005eh,Davidson:2008bu}:
\eqs{
M_{R} < M_{C} \lesssim \frac{10}{x_{B-L}^{5/4}} \left( \frac{m_{\nu}}{0.1 \, {\rm eV}} \right)^{1/4} \sqrt{Y_{N} Y_{C}} M_{R} \,.
}
The first inequality is to prohibit direct decay of the right-handed neutrinos into the dark colored Higgs, which may affect the thermal leptogenesis~\cite{Falkowski:2011xh}.

As shown in Ref.~\cite{Ibe:2011hq,Fukuda:2014xqa,Ibe:2018juk}, if the asymmetry is fully shared among the dark and SM sectors, i.e., if the portal interactions are relevant in the course of the thermal leptogenesis, the DM mass is to be $m_{N'} \simeq 8.5 / N_{g}' \,{\rm GeV}$.
This is because the $B-L$ asymmetries of the SM and dark sectors follow
\eqs{
A_{\rm DM} = \frac{N'_{g} (20 N_{g} + 6 m)}{3 N_{g} (22 N_{g} + 13 m)} A_{\rm SM} \,,
}
where $N_{g}$ is the generation of SM fermions and $m$ is the number of light Higgs doublets.
The relation between the present $B$ and $B-L$ asymmetries within the SM sector depends on the nature of the electroweak symmetry breaking (EWSB)~\cite{Harvey:1990qw,Weinberg:2008zzc}:
\eqs{
&A_{B} = \frac{4 (2 N_{g} + m)}{22 N_{g} + 13 m} A_{\rm SM} \,, \quad \text{sphaleron decoupling after the EWSB} \,, \\
&A_{B} = \frac{8 N_{g} + 4 (m + 2)}{24 N_{g} + 13 (m + 2)} A_{\rm SM} \,, \quad \text{rapid sphaleron after the EWSB} \,, \\
&A_{B} = \frac{2 (2 N_{g} - 1)(2 N_{g} + (m + 2))}{24 N_{g}^{2} + 14 N_{g} - 4 + m (13 N_{g} - 2)} A_{\rm SM} \,, \quad \text{rapid sphaleron + top decoupling after the EWSB}
\,.
}
In the first relation, we assume that the $B-L$ number and hypercharge are conserved, while in the second and third, we assume that the $B-L$ number and electromagnetic charge are conserved.
Hereafter, we use the third relation with $N_{g} = 3$, $m = 1$, and $N_{g}' = 1$, namely, $A_{\rm DM} = \dfrac{44}{237} \dfrac{97}{30} A_{\rm B}$.

\begin{figure}
	\centering
	\includegraphics[width=1\linewidth]{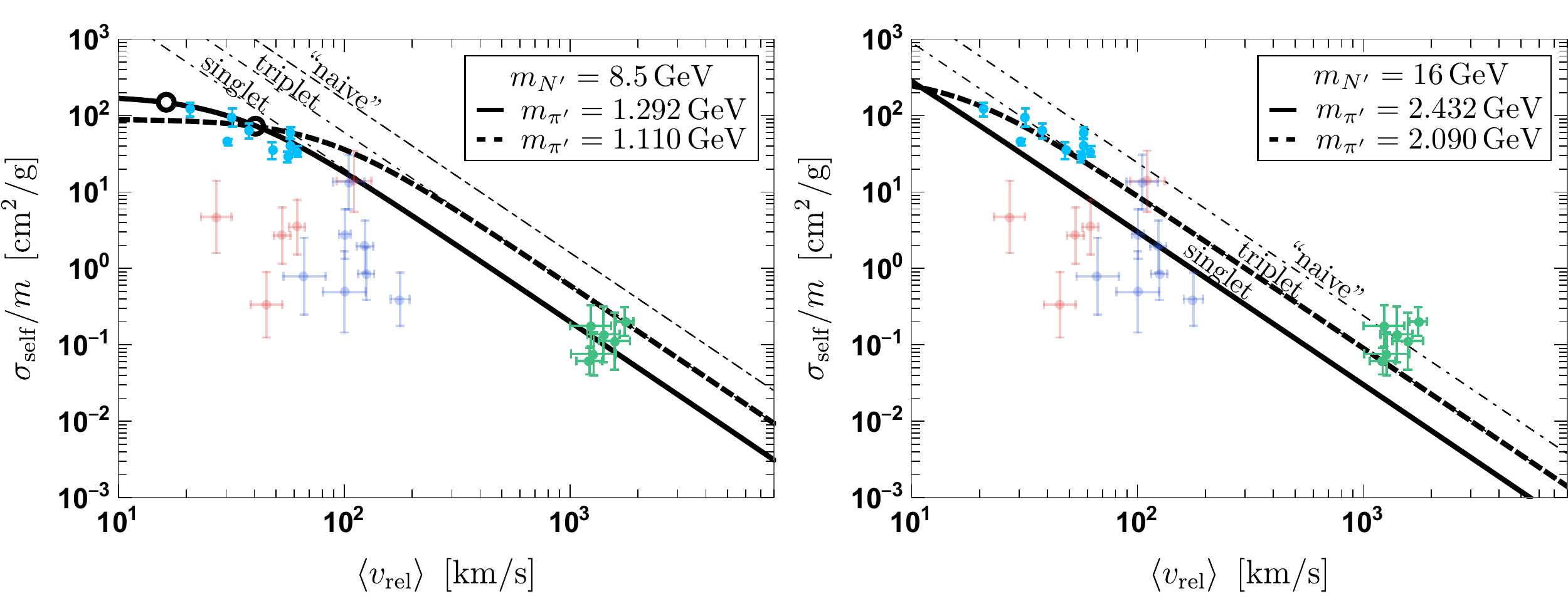}
	\caption{Velocity dependence of the 2-body dark nucleon scattering for given dark nucleon mass $m_{N'}$ and dark pion mass $m_{\pi'}$. The solid (dashed) curves represent the resonant singlet (triplet) scattering case.
	The dot-dashed curves are the Unitarity bounds for the nucleon-nucleon scattering cross section.
	The ``naive" Unitarity bound $\sigma_{\rm max} / m$ would be fully saturated only if both singlet and triplet scatterings were simultaneously on resonance for a given parameter set ($m_{N'}$ and $m_{\pi'}$).
	For the presented $\sigma / m$'s, only one of the channels, i.e., singlet or triplet scattering, is on resonance and saturates its own Unitarity bound, $\sigma_{s,{\rm max}} / m = 1/(2\cdot4)\times\sigma_{\rm max} / m$ or $\sigma_{t,{\rm max}} / m = 3/(2\cdot4)\times\sigma_{\rm max} / m$;
	the unfilled circles depict the point $k=1/|a|$ and $\sigma / m$ saturates the Unitarity bound for higher scattering velocities.
	The effective-range expansion is expected to be a good approximation, i.e., $k < 1/|r_e|$ in the presented velocity range. The data points are the same as in Fig.~\ref{fig:ERTvelocity1}.}
	\label{fig:maxSIDM_ADM}
\end{figure}

Since we consider the QCD-like dark sector, we may employ the QCD-calculation results in the DM phenomenology.
The 2-body dark nucleon scattering is described by the effective-range theory as $\sigma = \dfrac{1}{2} \left(\dfrac{1}{4} \sigma_{s} + \dfrac{3}{4} \sigma_{t} \right)$ with
\eqs{
& a_{s} = \frac{0.58 / \Lambda}{m_{\pi} / \Lambda - 0.57} \,, \quad r_{e s} = \frac{0.63 / \Lambda}{m_{\pi} / \Lambda} + 2.5 / \Lambda \,, \\
& a_{t} = \frac{0.39 / \Lambda}{m_{\pi} / \Lambda - 0.49} \,, \quad r_{e t} = \frac{0.0015 / \Lambda}{(m_{\pi} / \Lambda)^{3}} + 2.2 / \Lambda \,,
}
for the spin singlet and triplet (deuteron) channels, respectively (in SM QCD).%
\footnote{
The binding energy increases with a larger pion mass~\cite{Chen:2010yt,Soto:2011tb}.
Depending on the assumptions, one also finds the opposite~\cite{Epelbaum:2002gb,Yamazaki:2015asa,Bai:2020yml}.
These results are very delicate in light of the perturbative calculations of the effective field theory~\cite{Beane:2001bc}.
}
We only consider the dark neutron-dark proton ($n'$-$p'$) scattering even though we also have the other scattering processes: $n'$-$n'$ and $p'$-$p'$ scattering.%
\footnote{
    The quantum resonance of these scattering processes may require a different dark pion mass from that for $n'$-$p'$ scattering since the dark isospin symmetry, among $n'$ and $p'$, is broken by U(1)$_\mathrm{dark}$ and the current dark quark masses. 
    Therefore, we expect that we can ignore the other dark nucleon scattering on the quantum resonance of $n'$-$p'$ scattering. 
    When the quantum resonance of $p'$-$p'$ and $n'$-$n'$ scattering coincides with that of $n'$-$p'$, we have to take into account each scattering cross section, $\dfrac{1}{2} \dfrac{2}{4} \sigma^{p'p'}_s$ and $\dfrac{1}{2} \dfrac{2}{4} \sigma^{n'n'}_s$.
    Here the additional factor of $2$ originates from the difference in the decomposition of the scattering amplitude into the isospin irreducible representations and from the symmetric factor of the final states.
    It is beyond the scope of this paper to make a further study of the dark nucleon scattering in detail.
}
We assume that the half of the total DM consists of dark protons. 
The DM can interact with both of dark proton and dark neutron, and thus we divide each scattering cross sections by two.
Here we follow Ref.~\cite{Cline:2013zca} and fit the QCD-calculation results given in Ref.~\cite{Chen:2010yt} with $\Lambda = 250 \,{\rm MeV}$.%
This is valid for $m_{\pi} / \Lambda \simeq 0 \text{-} 1.4$.
We approximate the nucleon mass as $m_{N'} \simeq 1.25 N_{c} \Lambda$, where the color number is $N_{c} = 3$ in the present model~\cite{Witten:1979kh}.
We show the velocity dependence of dark nucleon self-scattering cross section in \cref{fig:maxSIDM_ADM} for fixed $m_{N'}$ near the quantum resonance. 

As seen in \cref{fig:maxSIDM_ADM}, $m_{N'} \simeq 8.5 \, {\rm GeV}$ has a constant cross section below $v_{\rm rel} \lesssim 100 {\rm km/s}$, barely reproducing the velocity dependence of $\sigma/m$ inferred by the MW's dwarf spheroidal satellite galaxies.
The binding energies are
\eqs{
&E_{b s} \simeq 0 \,, \quad E_{b t} \simeq 60\,{\rm MeV} \quad \text{for resonant singlet scattering} \,, \\
& E_{b t} \simeq 0 \quad \text{for resonant triplet scattering} \,,
}
where $E_{b}\simeq 0$ means a shallow bound/virtual state.
Meanwhile, $m_{N'} \simeq 16 \, {\rm GeV}$ has a $\sigma / m \propto 1 / v_{\rm rel}^{2}$ in the low-velocity region, providing a better fit.
The binding energies are
\eqs{
&E_{b s} \simeq 0 \,, \quad E_{b t} \simeq 110\,{\rm MeV} \quad \text{for resonant singlet scattering} \,, \\
&E_{b t} \simeq 0 \quad \text{for resonant triplet scattering} \,.
}
These 2-body binding energies are important inputs for the bound-state formation in the Universe (i.e., the dark nucleosynthesis)~\cite{Krnjaic:2014xza, Detmold:2014qqa, Detmold:2014kba, Wise:2014ola, Hardy:2014mqa, Gresham:2017zqi, Gresham:2017cvl, Mahbubani:2019pij}.
In the present model, the dark photon is typically heavier than the binding energy, and thus the 2-body bound-state formation proceeds only through the electron/positron emission and is suppressed by the kinetic mixing parameter.
Though it is beyond the scope of this paper, we may also be able to predict the strength of the dark QCD phase transition.
If it is the strong first order, we have a chance to examine this model in near-future gravitational-wave detectors~\cite{Schwaller:2015tja} and DM may consist of dark-quark nuggets~\cite{Bai:2018dxf}, changing the search strategy (e.g., direct/indirect-detection signals) in the late Universe.

To make $m_{N'} \simeq 16 \, {\rm GeV}$ consistent with $\Omega_{\rm dm} \sim 5 \Omega_{b}$, we need to make portal operators partially irrelevant.
We leave more general analysis, including the direct decay of right-handed neutrinos into the dark colored Higgs, for a future work (see Ref.~\cite{Falkowski:2011xh} for a similar analysis).
Hereafter, instead, we assume that the both cases ($m_{N'} \simeq 8.5 \, {\rm GeV}$, $16 \, {\rm GeV}$) are realized with $M_{C} \simeq \left( 10 / x_{B-L}^{5/4} \right) \left( m_{\nu} / 0.1 \, {\rm eV} \right)^{1/4} \sqrt{Y_{N} Y_{C}} M_{R}$.

\begin{figure}
	\centering
	\includegraphics[width=0.7\linewidth]{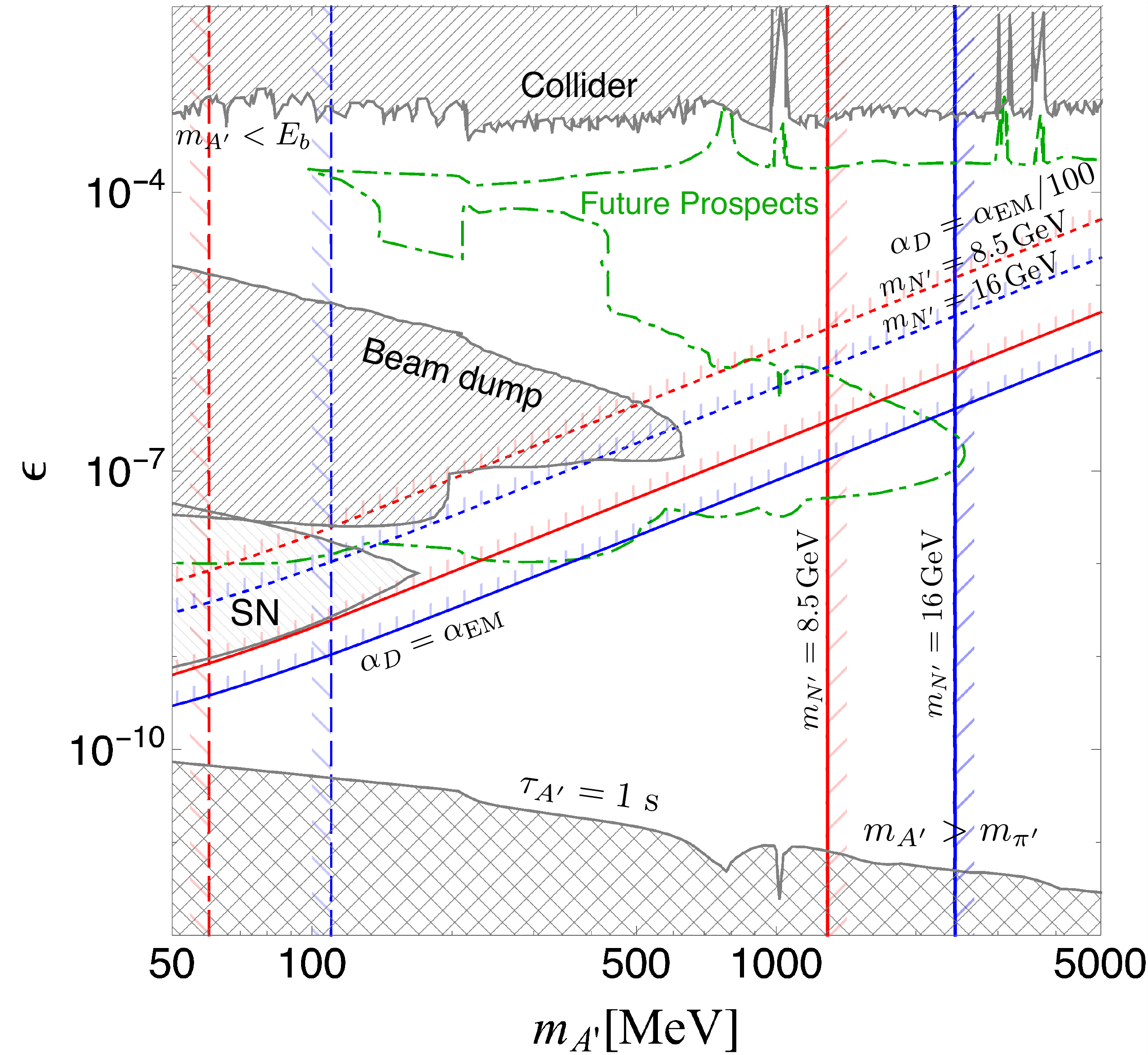}
	\caption{
    Constraints on the dark-photon parameters: dark photon mass $m_{A'}$ and kinetic mixing $\epsilon$.
    The gray meshed region is the current bounds from 
    supernova 1987A constraints (Refs.~\cite{Chang:2016ntp,Chang:2018rso}), beam-dump experiments, and collider experiments (taken from Ref.~\cite{Bauer:2018onh}). 
    The green dot-dashed line shows the future prospects for beam-dump and collider experiments.
    The diagonal blue (red) lines show the upper limits on $\epsilon$ from the PandaX-II experiment~\cite{Ren:2018gyx} for $m_{N'} = 16 \, {\rm GeV}$ ($m_{N'} = 8.5 \, {\rm GeV}$) and fixed U(1)$_{\rm dark}$ coupling, $\alpha_{D} = g_{D}^{2} / (4\pi)$.
    The right side of vertical solid lines indicate the region where the dark photon is heavier than the dark pion for $m_{N'} = 16 \, {\rm GeV}$ (blue) and $m_{N'} = 8.5 \, {\rm GeV}$ (red); the constraints on the dark photon decay would change in the indicated regions.
    The left side of vertical dashed lines indicate the region where dark photon mass is less than the binding energy $E_{bt}$ for $m_{N'} = 16 \, {\rm GeV}$ (blue) and $m_{N'} = 8.5 \, {\rm GeV}$ (red); the dark nucleosynthesis would proceed by emitting the dark photon, and the DM would consist of the dark nuclei.
  }
	\label{fig:DDbound}
\end{figure}

In the present model, the dark photon plays a key role in releasing the dark sector entropy to the visible sector via the kinetic mixing $\epsilon$ with the SM photon.
\cref{fig:DDbound} summarizes the current constraints on the dark-photon parameters $(m_{A'}, \epsilon)$ and the future prospects in beam-dump and collider experiments. 
In collider experiments, dark photons can be produced through the kinetic mixing and subsequently decay into SM particles.
For $m_{A'} < 5~\mathrm{GeV}$, BaBar~\cite{Aubert:2009cp,Lees:2014xha}, LHCb~\cite{Aaij:2017rft}, and KLOE~\cite{Archilli:2011zc,Babusci:2012cr,Anastasi:2015qla,Anastasi:2016ktq} place the upper limit on $\epsilon$.
Belle-II~\cite{Abe:2010gxa,Kou:2018nap} will place the upper limit on $\epsilon$ mainly for $m_{A'} = 10\text{-}5000 \, {\rm MeV}$.
The displaced vertex searches at the LHCb experiment will also explore the region between the collider and beam-dump experiments~\cite{Ilten:2015hya,Ilten:2016tkc}. 
The beam-dump experiments exclude a broad parameter space for $m_{A'} < 1~\mathrm{GeV}$: current constraints are mainly from CHARM~\cite{Bergsma:1985is,Gninenko:2012eq}, LSND~\cite{Athanassopoulos:1997er,Batell:2009di}, and U70/$\nu$Cal~\cite{Blumlein:2011mv,Blumlein:2013cua}.
The dark-photon parameters in this region will be further explored by projected facilities like SHiP~\cite{Alekhin:2015byh,Anelli:2015pba}, FASER~\cite{Feng:2017uoz}, and SeaQuest~\cite{Gardner:2015wea,Berlin:2018pwi}.

The dark photon decay rate is given by
\eqs{
&\Gamma (A' \to e^+ e^-) = \frac{1}{3} \alpha \epsilon^{2} m_{A'} \sqrt{1 - \frac{4 m_{e}^{2}}{m_{A'}^{2}}} \left(1 + \frac{2 m_{e}^{2}}{m_{A'}^{2}} \right) \,, \\
&\Gamma (A' \to \text{hadrons}) = \frac{1}{3} \alpha \epsilon^{2} m_{A'} \sqrt{1 - \frac{4 m_{\mu}^{2}}{m_{A'}^{2}}} \left(1 + \frac{2 m_{\mu}^{2}}{m_{A'}^{2}} \right) R(\sqrt{s} = m_{A'}) \,,
}
where $R (\sqrt{s}) = \dfrac{\sigma (e^+ e^- \to \text{hadrons})}{\sigma (e^+ e^- \to \mu^{+} \mu^{-})}$ takes into account the hadronic resonances (i.e., vector meson dominance).
In the following, we consider the latter for $m_{A'} > 100~\mathrm{MeV}$ taken from Ref.~\cite{Batell:2009yf}.
Since the late-time decay of the dark photon heats only the electron-photon plasma after the neutrino decoupling, $N_\mathrm{eff}$ is lowered and the upper bound is put on the dark photon lifetime.
The dark photon lifetime $\tau_{A'} = 1/\Gamma_{A'}$ exceeds $1 \, \mathrm{s}$ in the gray crosshatched region of \cref{fig:DDbound}.

In the right side of the vertical solid lines of \cref{fig:DDbound}, the dark photon is no longer the lightest particle in the dark sector (the blue-solid vertical line for $m_{N'} = 16 \, {\rm GeV}$, while the red-solid vertical line for $m_{N'} = 8.5 \, {\rm GeV}$). 
In particular, the dark photon is heavier than the dark pion, whose mass is indicated by the strong velocity dependence of the 2-body dark nucleon scattering.
The constraints on the dark photon decay would change in this region.
In the left side of vertical dashed lines, the dark photon mass is less than the binding energy $E_{bt}$ when the resonant singlet scattering is realized (we use the same color scheme for the vertical lines as that in the right region).
In this region, the dark nucleosynthesis would proceed by emitting the dark photon, and then the DM would consists of the dark nuclei.

When the DM consists of the dark proton charged under U(1)$_{\rm dark}$, DM interacts with SM nuclei through the kinetic mixing between the SM photon and dark photon. 
The differential cross section between the dark proton and target nucleus is
\eqs{
\frac{d \sigma_{p'}}{d q^{2}} = \frac{4 \pi \alpha \alpha_{D} \epsilon^{2} Z^{2}}{(q^{2} + m_{A'}^{2})^{2}} \frac{1}{v^{2}} F_{T}^{2} (q^{2}) \,.
}
Here the momentum transfer is $q^{2} = 2 \mu_{T}^{2} v_{\rm rel}^{2} (1 - \cos \theta_{\rm cm})$, where $\mu_{T}$ is the reduced mass of DM and the target nucleus and $\theta_{\rm cm}$ is the center-of-mass scattering angle.
Following Ref.~\cite{DelNobile:2015uua}, we set $q^{2} = 2 \mu_{T}^{2} v_{\odot}^{2}$, where $v_{\odot} = 232 \, {\rm km/s}$ is the typical DM speed at the rest frame of the Earth~\cite{Savage:2008er}.
$F_{T}$ is the nuclear form factor.
We discuss the dark neutron-target nucleus scattering through the magnetic moment in \cref{sec:dd-munp}, which is negligible.
We place the upper limit on the kinetic mixing parameter $\epsilon$ from the direct-detection constraints by the 54 ton-day exposure of the PandaX-II experiment~\cite{Ren:2018gyx} (diagonal red and blue lines in \cref{fig:DDbound}).
We assume that the half of the total DM consists of dark protons.
The DM mass is set to be $16 \, {\rm GeV}$ (blue lines) and $8.5 \, {\rm GeV}$ (red lines), and we take U(1)$_{\rm dark}$ coupling to be $\alpha_{D} = \alpha_{\rm EM} = 1 / 137$ (solid lines) and $\alpha_{D} = \alpha_{\rm EM} / 100$ (dashed lines). 

\begin{figure}
	\centering
	\includegraphics[width=0.6\linewidth]{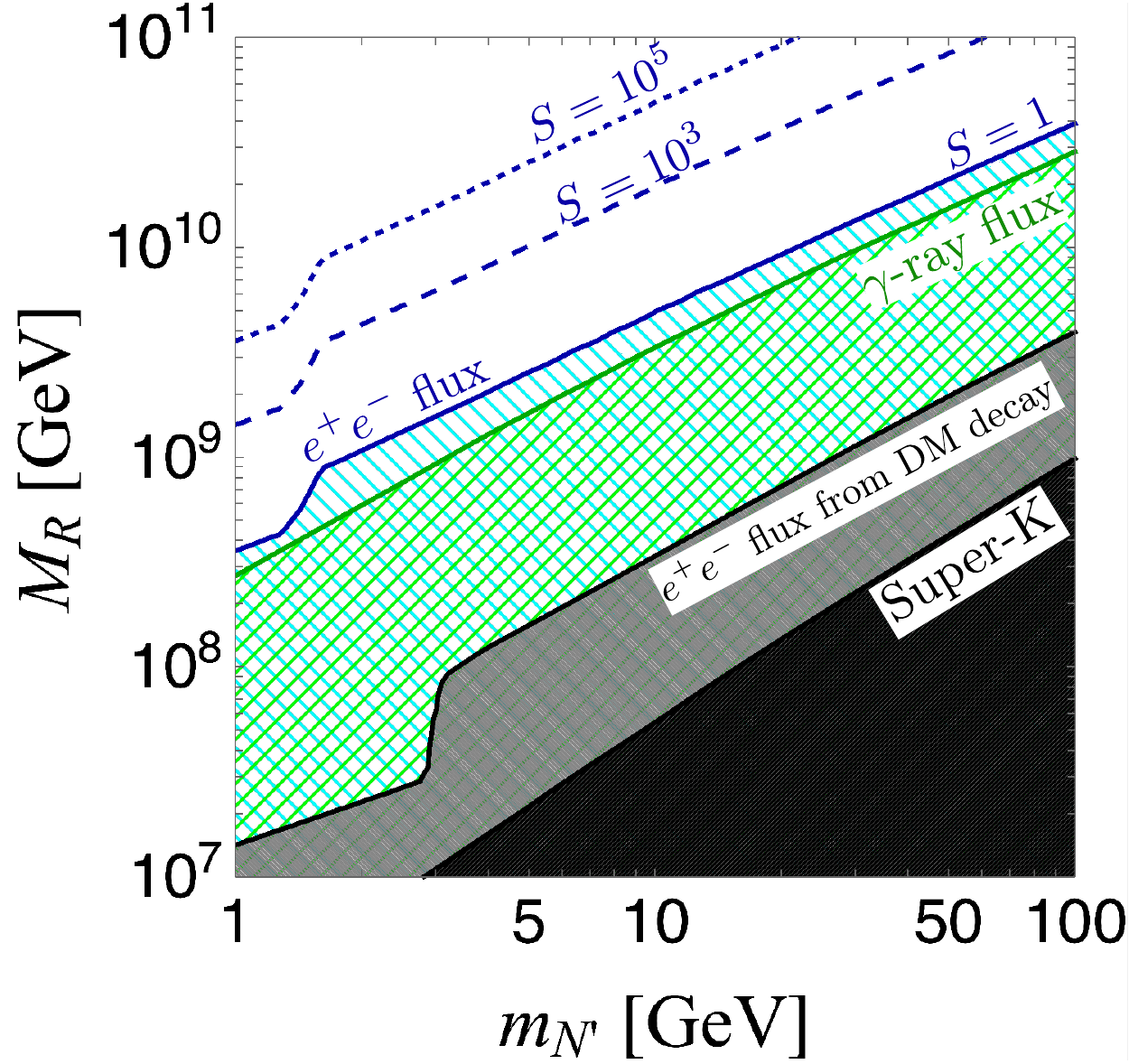}
	\caption{
	Indirect-detection constraints on $(m_{N'}, M_{R})$ plane.
	The black-shaded region is excluded by no excess of a neutrino signal at the Super-Kamiokande experiment.
	The gray-shaded region is constrained by $e^+e^-$ flux from eventual decay of the dark meson from DM decay (constraints by the AMS-02 for $m_{N'} \gtrsim 3~\mathrm{GeV}$ and the Voyager-1 for $m_{N'} \lesssim 3~\mathrm{GeV}$).
  The green-hatched region is excluded by $\gamma$-ray constraints by the Fermi-LAT, and the blue-hatched region is excluded by $e^+ e^-$ flux constraints (by the AMS-02 for $m_{N'} \gtrsim 1.5~\mathrm{GeV}$ and the Voyager-1 for $m_{N'} \lesssim 1.5~\mathrm{GeV}$).
  The blue dashed lines show the $e^+ e^-$ flux constraints with taking a Sommerfeld-enhancement factor of the annihilation cross section to be $S_{\rm eff} = 10^3$ and $10^5$.
  }
	\label{fig:IDbound}
\end{figure}

The intermediate-scale portal operators \cref{eq:int_portal} lead to the DM decay into the dark meson and anti-neutrino~\cite{Fukuda:2014xqa}.
Such decaying DM can be explored in indirect ways: one is a neutrino signal from the DM decay at the Super-Kamiokande, the others are constraints on a flux from eventual decay of the decay product to the SM particles.
No excess of a neutrino signal has been found at the Super-Kamiokande experiment~\cite{Desai:2004pq}, and this sets the lower limit on the DM lifetime~\cite{Covi:2009xn}.
The decay rate is given by
\eqs{
  \Gamma_\mathrm{DM} \simeq 
  \frac{1}{64 \pi} \frac{v_H^2 m_{N'}}{M_\ast^6} |W|^2 \,.
}
Here, $v_H$ is a VEV of the SM Higgs and $W$ is a matrix element for DM to a dark meson. 
We assume the matrix element to be $W \simeq 0.1 \, {\rm GeV}^{2} (m_{N'}/m_n)^2$, to which a lattice calculation result~\cite{Aoki:2013yxa} is na\"ively scaled.
For the decaying DM with a mass of $16 \, {\rm GeV}$, the lower limit is roughly $\tau_{\rm DM} = 1/\Gamma_{\rm DM} \gtrsim 10^{23} \, {\rm s}$.
We na\"ively extrapolate the results of Ref.~\cite{Covi:2009xn} below the DM mass of $10 \, {\rm GeV}$.
The black-shaded region in \cref{fig:IDbound} is excluded by no excess of a neutrino signal in the Super-Kamiokande experiment. 
The decay product, neutral dark mesons, eventually decays into $e^+ e^-$ through kinetic mixing. 
We na\"ively utilize a constraint from the $e^+ e^-$ flux observations (Voyager-1~\cite{Stone150} and AMS-02~\cite{Aguilar:2014mma}) on the DM lifetime given in Ref.~\cite{Boudaud:2016mos}.
The $e^+ e^-$ flux from the DM decay constrains the DM lifetime to be larger than $\tau_{\rm DM} = 1/\Gamma_{\rm DM} \gtrsim 10^{27} \, {\rm s}$ for $m_{N'} \gtrsim 3~\mathrm{GeV}$.
The gray-shaded region in \cref{fig:IDbound} shows the constraint from the $e^+e^-$ flux from the DM decay.
The AMS-02 observation gives a dominant constraint for $m_{N'} \gtrsim 3~\mathrm{GeV}$, while the $e^+ e^-$ flux constraints comes from the Voyager-1 observation for $m_{N'} \lesssim 3~\mathrm{GeV}$.

The same ultraviolet origin of the operator in \cref{eq:int_portal} induces the Majorana mass term for the dark neutrons~\cite{Ibe:2019yra}:
\eqs{
    {\cal L} \supset \frac{1}{2} m_{M} {\bar n}' {\bar n}' + {\rm h.c.} \,, \quad
    m_{M} \simeq \frac{Y_{N}^{2} Y_{C}^{2} \Lambda_{\rm QCD'}^{6}}{4 M_{R} M_{C}^{4}} \,. 
}
Some portion of the dark neutron DM is converted into its anti-particle at the late time $t$ through the dark neutron-anti neutron oscillation: $P_{\bar n} \sim (m_{M} t)^{2}$.
Then dark nucleons $n'$ and $p'$ annihilate with $\bar n'$ into dark pions with the effective cross section,
\eqs{
(\sigma v)_{\rm eff} =\frac{4 \pi}{m_{N'}^{2}} S_{\rm eff} P_{\bar n} \,.
\label{eq:eff_ann}
}
Here $S_{\rm eff}$ is the Sommerfeld-enhancement factor $S_{s\text{-wave}}$ times a fudge factor.
When the dark photon is the lightest particle in the dark sector, dark pions decay into dark photons, and then dark photons eventually decay into $e^+ e^-$ pair.
The final state radiation of this process can emit visible photons.
The late-time annihilation is constrained by the observations of the $\gamma$-ray from the dwarf spheroidal galaxies by the Fermi-LAT~\cite{Ackermann:2015zua} and from the interstellar $e^+ e^-$ flux by the Voyager-1~\cite{Stone150} and the AMS-02~\cite{Aguilar:2014mma}.
We use the $\gamma$-ray limits on the oscillation time scale given by Ref.~\cite{Ibe:2019yra} with $S_{\rm eff} = 1$.
Even though the Fermi-LAT observation constrains the effective annihilation cross section only for $m_{N'} \gtrsim 3~\mathrm{GeV}$, we use a na\"ive extrapolation of the constraint below $3~\mathrm{GeV}$.
The green-hatched region in \cref{fig:IDbound} shows the constraints from the observations of the $\gamma$-ray flux.
We utilize the $e^+ e^-$ flux constraint on the dark matter annihilation cross section in Ref.~\cite{Boudaud:2016mos} as a constraint on $(\sigma v)_{\rm eff}$.

The constraints from the $e^+ e^-$ flux and $\gamma$-ray flux observations depend on the annihilation cross section of dark nucleons. 
The strong velocity dependence of the self-scattering cross section implies a sizable Sommerfeld-enhancement factor as seen in \cref{fig:SEFs1}, and thus we also show the lower limit on $M_R$ from the $e^+ e^-$ flux when we take the factor to be $S_{s\text{-wave}} = 10^3$ (long-dashed line) and $S_{s\text{-wave}} = 10^5$ (short-dashed line).
Though the constraints from $\gamma$-ray flux change when we take the enhancement factor, we do not depict them in \cref{fig:SEFs1}.

The lightest dark $\rho$ meson mass is $m_{\rho'} \simeq 7 \, {\rm GeV}$ ($13.1 \, {\rm GeV}$) for $m_{N'} = 8.5 \, {\rm GeV}$ ($m_{N'} = 16 \, {\rm GeV}$) from a na\"ive scaling of our $\rho$ meson.
From a na\"ive scaling of our $\rho$ meson's mass spectrum, we also expect that higher vector resonances appear as $m_{\rho_{n}}^{2} \sim 76 \, n~\mathrm{GeV}^2 ~(270 \, n~\mathrm{GeV}^2)$ ($n = 1, 2, \dots$)  for $m_{N'} = 8.5 \, {\rm GeV}$ ($m_{N'} = 16 \, {\rm GeV}$).
With a sizable kinetic mixing parameter $\epsilon \gtrsim 10^{-3}$, we have a chance of spectroscopic measurements of such resonances in lepton colliders (e.g., Belle II and BES-III)~\cite{Hochberg:2015vrg,Hochberg:2017khi}.

\section{Conclusion \label{sec:conclusion}}
We have studied the possibility that self-interacting dark matter has a maximal self-scattering cross section from the Unitarity point of view.
When the self-scattering cross section nearly saturates the Unitarity bound, the cross section is singly parameterized by the DM mass.
Interestingly, for the DM mass of $m \sim 10\,{\rm GeV}$, maximally self-interacting dark matter can explain the observed structure of the Universe ranging from $\sigma / m \sim 0.1 \, {\rm cm^{2} / g}$ at galaxy clusters to $\sim 100 \, {\rm cm^{2} / g}$ at the MW's dwarf spheroidal satellite galaxies.
We have demonstrated that a general requirement of the quantum (zero-energy) resonant scattering, by employing the effective-range theory and analytic results with the Hulth\'en potential.
It requires a light force mediator and model parameters to be the special (fine-tuned) values.
We have also demonstrated that DM annihilation tends to be largely boosted by the Sommerfeld enhancement with the same parameters.

It requires DM annihilation to end up with neutrinos or to be the $p$-wave.
As such a model, we have considered the gauged $L_{\mu} - L_{\tau}$ model extended with Dirac DM.
The Dirac DM is slightly split into two Majorana DM by the VEV of the gauged $L_{\mu} - L_{\tau}$ Higgs.
We have taken the Higgs mode to be light so that it mediates the (relatively) long-range force between DM particles.
We have pinned down the parameter points that lead to the DM self-scattering saturating the Unitarity bound.
With them, we can explain the discrepancy in the muon $g - 2$, predict the sizable $\Delta N_{\rm eff}$ mitigating the tension in $H_{0}$, and have rich implications for non-standard neutrino interactions and collider searches.

ADM is another option to evade the indirect-detection bounds from the boosted DM annihilation (at the first sight).
Since ADM requires a lighter state to deplete its symmetric components through efficient annihilation, such a light state is a natural candidate for a force mediator.
We have considered dark nucleon ADM with dark pions being the force mediator, based on the dark QCD$\times$QED dynamics.
The dark pion mass is very predictive to explain the quantum-resonant self-scattering between dark nucleon DM.
For viable cosmology, we have assumed that dark photon is the lightest state in the dark sector, which provide various ways to probe the dark sector: collider and beam-dump experiments and DM direct-detection experiments.
Furthermore, the binding energies of two nucleon states are also predicted, which are important for the dark nucleosynthesis that proceeds with a light dark photon.
The predicted dark vector resonances can be measured by lepton colliders.

It is worthwhile to note that we have compared the velocity-dependent cross sections without the local velocity distribution averaging with the inferred values of $\sigma / m$ from astronomical observations.
This is partially because the inferred values of $\sigma / m$ were interpreted by assuming a constant cross section inside halos of interest.
It would be encouraging to reanalyze the astronomical data with a proper distribution averaging.
The velocity-dependent cross section of maximally self-interacting dark matter may give a benchmark for the reanalyses.

We have shown that the recent astronomical data for structure formation of the Universe are already interesting and have rich implications for DM physics.
Though it is to be clarified which data are robust or suffer from astrophysical uncertainties such as supernova feedback, we are optimistic that the situation will get quickly improved with the fast development of hydrodynamic simulations and more precise observations.
Gravitational probes of DM and related beyond-WIMP DM scenarios enjoy an exciting era.

\subsection*{Acknowledgement}
The work of A. K. and T. K. is supported by IBS under the project code, IBS-R018-D1.
A. K. thanks Keisuke Yanagi for former collaborations, without which this work was not initiated.

\newpage
\appendix

\section{MW's dwarf spheroidal galaxies vs field dwarf galaxies \label{sec:sat-vs-field}}

As seen in \cref{fig:ERTvelocity1}, the inferred values of $\sigma / m$, in the MW's dwarf spheroidal satellite galaxies (cyan)~\cite{Correa:2020qam} and the field dwarf galaxies (red)~\cite{Tulin:2017ara}, seem not compatible with each other, though they have similar collision velocities, $v_{\rm rel} \sim 20 \text{-} 60 \, {\rm km/s}$.
In the main text, we discuss implications for DM phenomenology, taking the former seriously.
In this section, instead, we discuss how our conclusions changed, if the inferred values of $\sigma / m$ from the MW's satellites were lowered.

\begin{figure}
\centering
\begin{subfigure}{1.\textwidth}
\centering
  \includegraphics[width=0.93\linewidth]{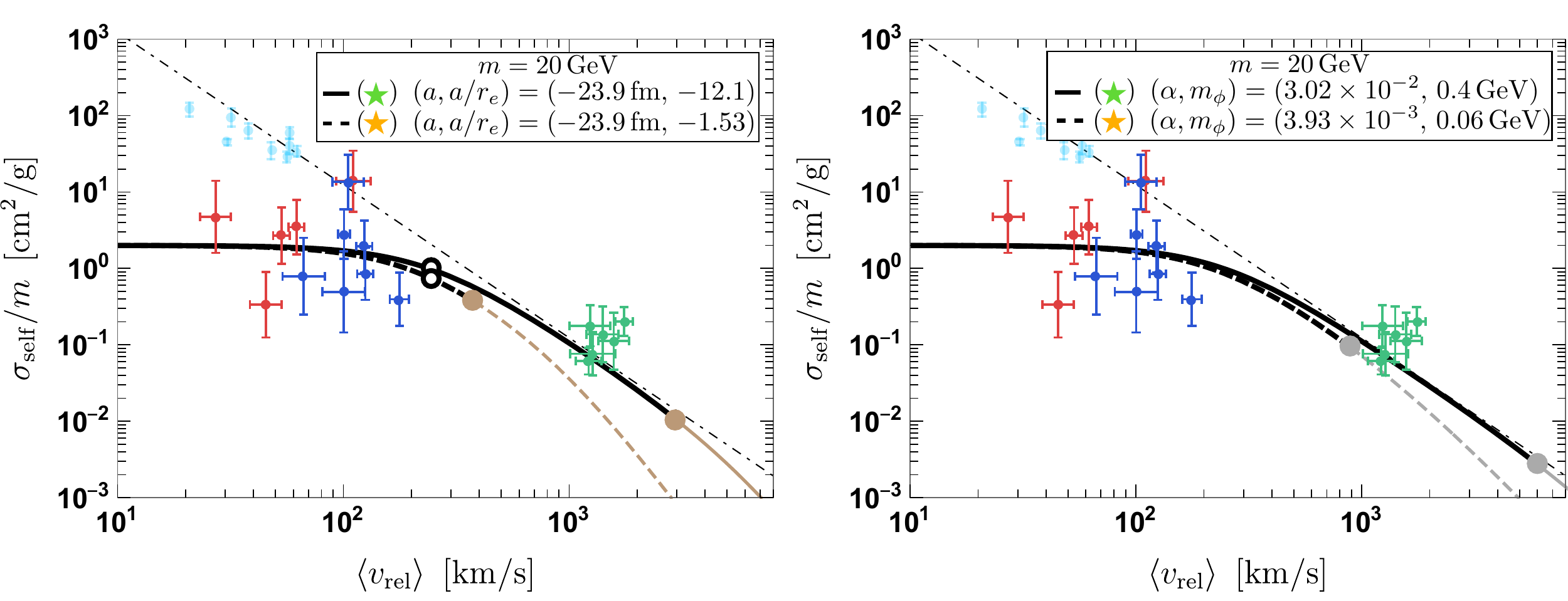}
  \caption{}
  \label{fig:ERTvelocity2}
  \par\medskip
  \includegraphics[width=0.93\linewidth]{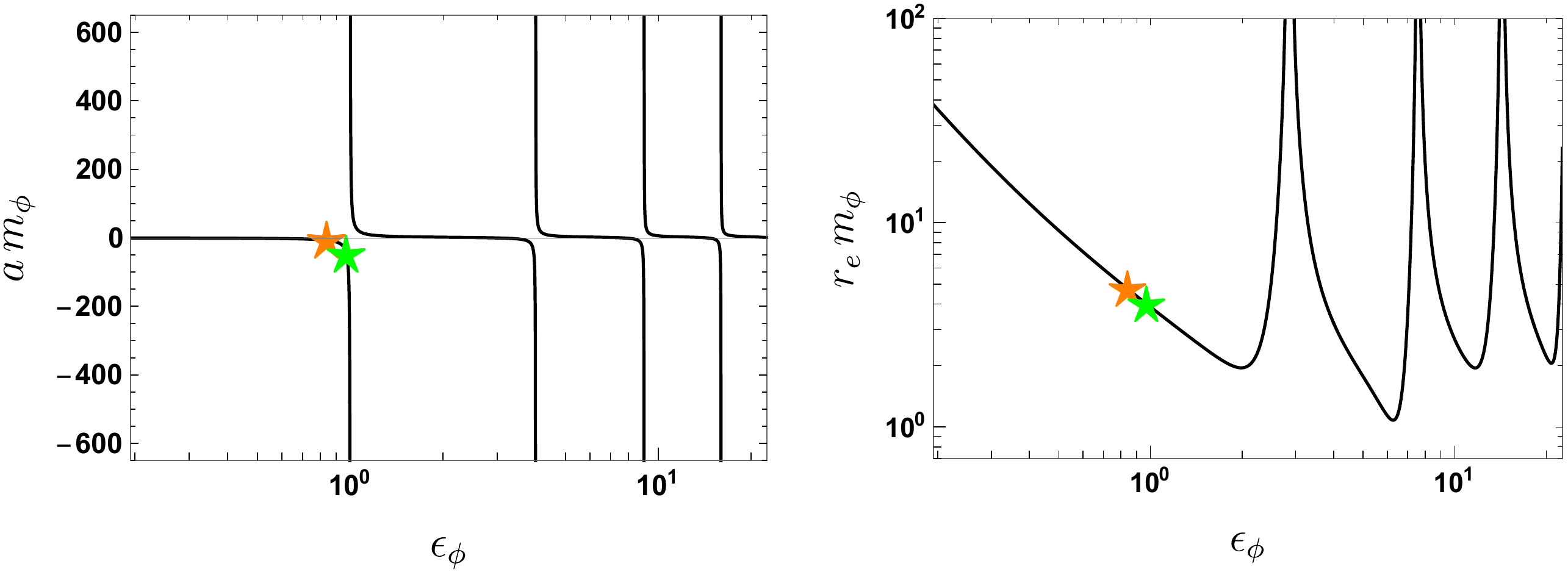}
  \caption{}
  \label{fig:Hulthenparam2}
\end{subfigure}
\caption{
{\bf(a)} Same as \cref{fig:ERTvelocity1}, but for the effective-range theory parameters, $(a,r_e)$, that explains the inferred values of $\sigma/m$ from the observations on the field dwarf galaxies (red).
The Hulth\'en-potential parameter sets took in the right panel exhibits the same effective-range theory parameter sets as the left panel.
{\bf (b)} Same as \cref{fig:Hulthenparam1}, but the stars indicate the benchmark parameters took in \cref{fig:ERTvelocity2}.
Compared to the parameters took in \cref{fig:ERTvelocity1}, the parameters here are relatively away from the quantum resonance at $\epsilon_{\phi}=1$. 
}
\end{figure}

In the left panel of \cref{fig:ERTvelocity2}, we take the effective-range theory parameters so that $\sigma/m$ crosses the inferred values from the field dwarf galaxies (red), rather than the ones from MW's dwarf spheroidal satellite galaxies (cyan) as in \cref{fig:ERTvelocity1}.
In this case, we do not need a strong velocity dependence of $\sigma/m$ as in \cref{fig:ERTvelocity1}.
Notice that the $|a/r_e|$'s took are relatively smaller than the ones in \cref{fig:ERTvelocity1};
this indicates that the parameters are relatively away from the quantum resonance.
This is explicitly shown in \cref{fig:Hulthenparam2}, where the points depicted by stars represent the Hulth\'en-potential parameters that correspond to the effective-range theory parameters in left panel of \cref{fig:ERTvelocity2}.
Notice that the $\epsilon_\phi$'s do not need to be so close to the first quantum resonance, i.e., $\epsilon_{\phi}\simeq1$, to cross the red data points in \cref{fig:ERTvelocity2}.
This is in contrast to the parameters took in \cref{fig:ERTvelocity1}, where the effective-range theory parameters need to be fine-tuned to be near the quantum resonance to cross the cyan data points, as shown in \cref{fig:Hulthenparam1}.

\begin{figure}
\centering
\begin{subfigure}{1.\textwidth}
\centering
  \includegraphics[width=0.93\linewidth]{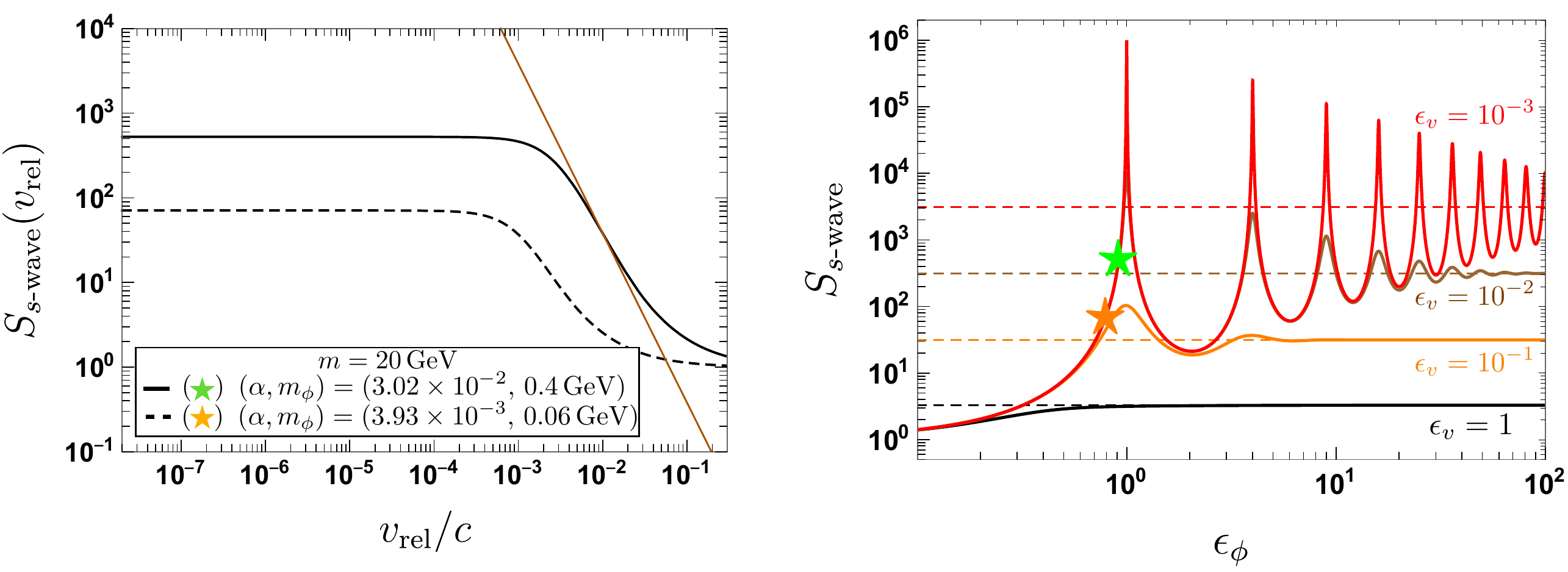}
  \caption{}
  \label{fig:SEFs2}
  \par\medskip
  \includegraphics[width=0.93\linewidth]{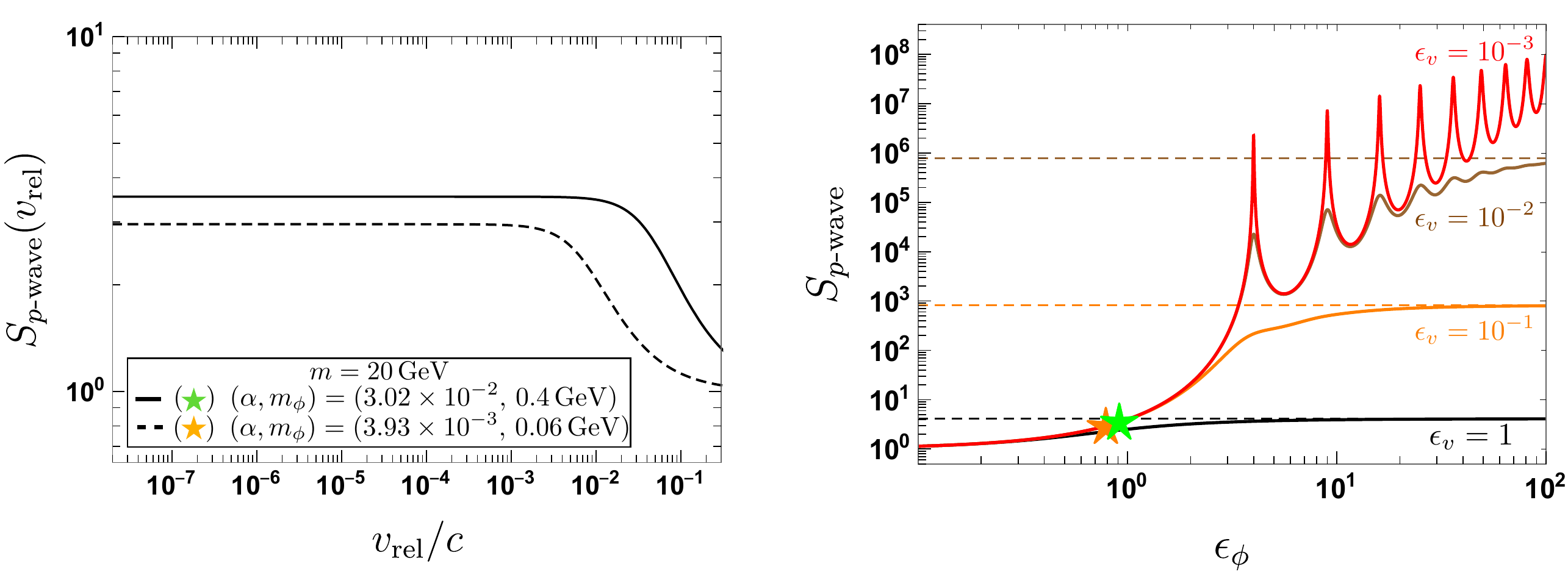}
  \caption{}
  \label{fig:SEFp2}
\end{subfigure}
\caption{
Same as \cref{fig:SEF1}, but with the Hulth\'en-potential parameters took in \cref{fig:ERTvelocity2}. The $s$-wave annihilation is relatively away from the quantum resonance compared the ones in \ref{fig:ERTvelocity1}; the enhancement factors are smaller, and the $s$-wave enhancement factors grow towards low velocity slower than the resonant behavior, i.e., $\propto 1/v_{\rm rel}^2$ (brown).
}
\label{fig:SEF2}
\end{figure}

At the same time, the parameters of \cref{fig:ERTvelocity2} are also relatively away from the quantum resonance for annihilation;
compare the points depicted by stars in \cref{fig:SEF2} with the ones in \cref{fig:SEF1}.
Consequently, they exhibit relatively smaller Sommerfeld-enhancement factors than the ones of \cref{fig:ERTvelocity1}, making them less constrained from the indirect-detection experiments.

\begin{figure}
\centering
\begin{subfigure}{1.\textwidth}
\centering
  \includegraphics[width=0.93\linewidth]{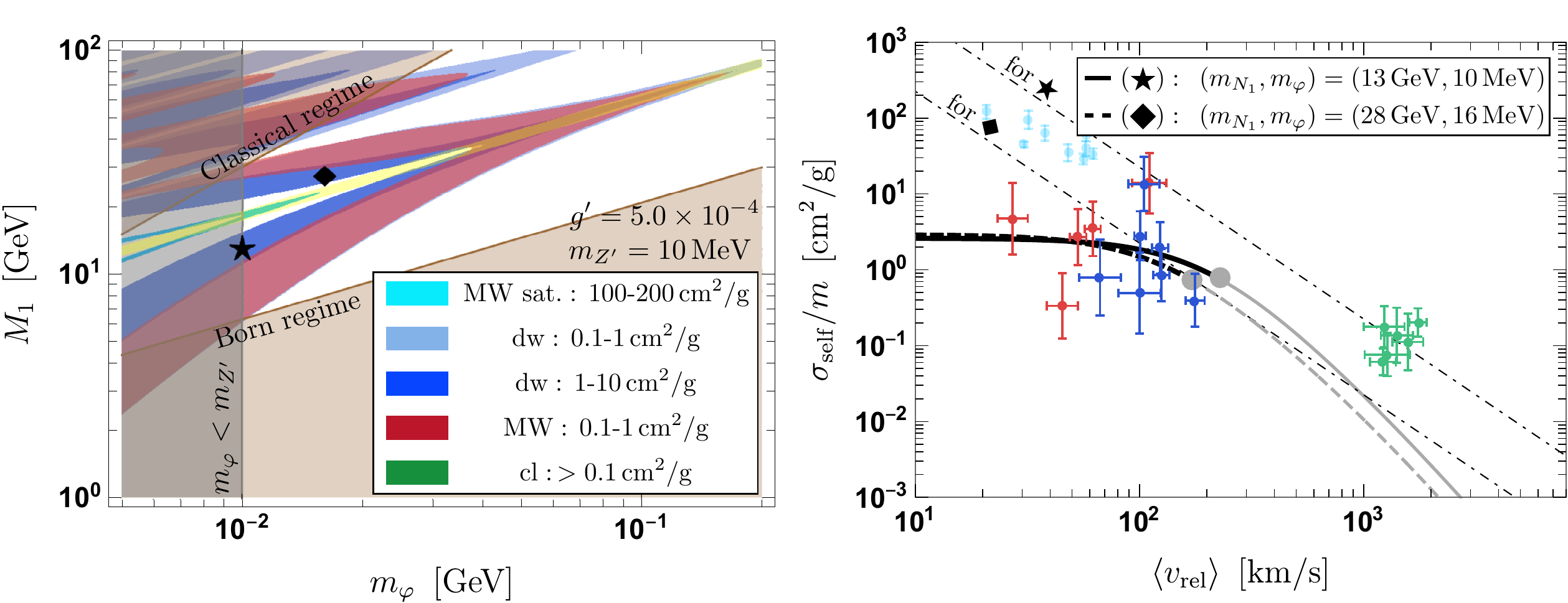}
  \caption{}
  \label{fig:lult_4}
  \par\medskip
  \includegraphics[width=0.93\linewidth]{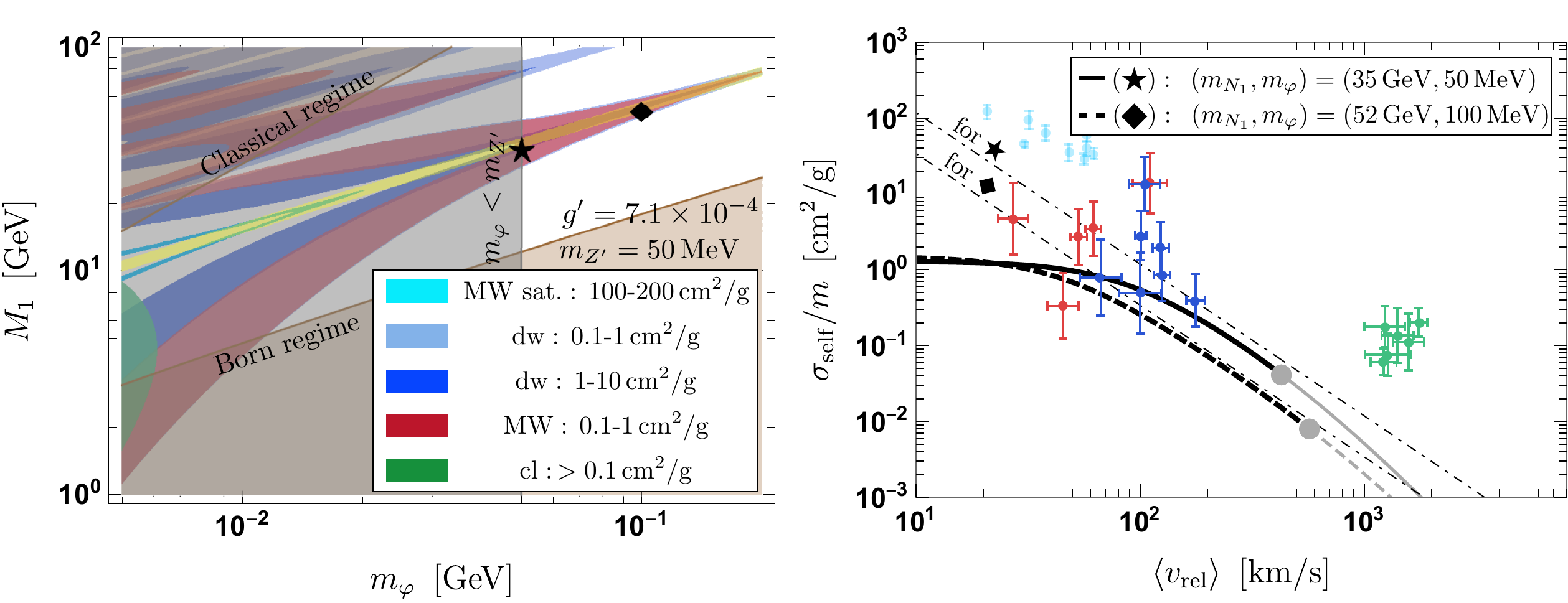}
  \caption{}
  \label{fig:lult_5}
\end{subfigure}
\caption{Same as the left panel of \cref{fig:lult1}, but the with different benchmark parameter points; here, the depicted points (as a star and a diamond) explain the red data points, rather than the cyan ones as in \cref{fig:lult1}.
}
\label{fig:lult2}
\end{figure}

The lack of need for the strong velocity dependence amounts to relaxation of the ``prediction" of the narrow model parameter regions of the gauged $L_{\mu}-L_{\tau}$ model in the main text.
In the left panels of \cref{fig:lult2}, we depict the example parameters of the model that explains the inferred values of $\sigma/m$ from the field dwarf galaxies, i.e., $\sigma/m\sim1\,{\rm cm^2/g}$ at $v_{\rm rel}\sim 20\,{\rm km/s}$.
We show the corresponding velocity dependence of $\sigma/m$ in the right panels.
In \cref{fig:lult_4}, we present the case of $m_{Z'}=10\,{\rm MeV}$, which lies in the $Z'$ mass range preferred to explain the inferred $\sigma/m$ from the MW's satellites, $m_{Z'}\lesssim20\,{\rm MeV}$, as discussed in the main text.
Contrary to the parameter points depicted in \cref{fig:lult_1}, we see that the parameters here do not need to be so fine-tuned to be close to the quantum resonance (yellow) to explain the red data points.
Furthermore, the viable $Z'$ mass range extends towards larger values, compared to the case discussed in the main text. 
For heavier $Z'$, e.g., $m_{Z'}=50\,{\rm MeV}$, it is possible to find parameters to explain the inferred $\sigma/m$ from the field dwarf galaxies, while they need to lie near the quantum resonance.

\begin{figure}
	\centering
	\includegraphics[width=1\linewidth]{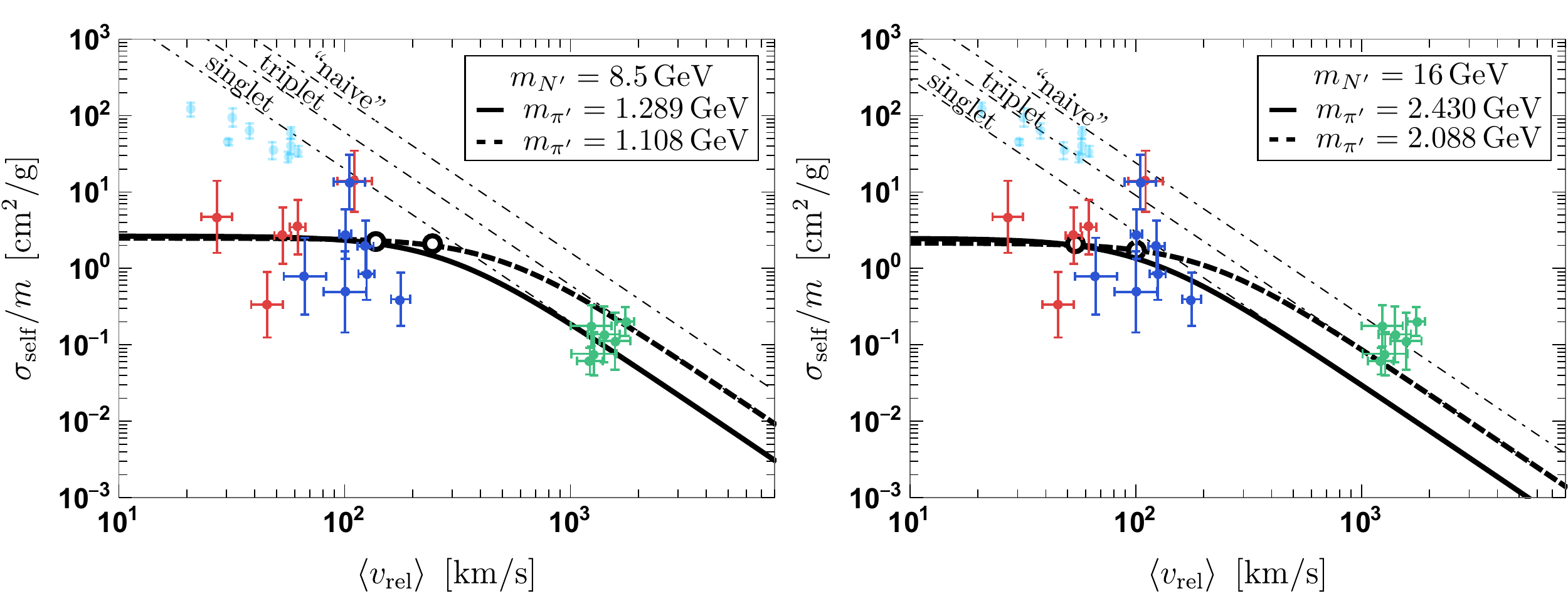}
	\caption{Same as the \cref{fig:maxSIDM_ADM}, but for the dark pion masses that explain the inferred values of $\sigma / m$ from the field dwarf galaxies.
	}
	\label{fig:fdg_ADM}
\end{figure}

For the ADM based on the dark QCD$\times$QED dynamics, the notable change would be that the dark nucleon mass of $m_{N'}\simeq8.5\,{\rm GeV}$ also becomes preferable.
In \cref{fig:maxSIDM_ADM}, we showed that $m_{N'}\simeq8.5\,{\rm GeV}$ poorly fits the inferred velocity dependence of $\sigma/m$ inferred by the MW's dwarf spheroidal satellite galaxies.
But if we focus on $\sigma/m$ inferred by the field dwarf galaxies, $m_{N'}\simeq8.5\,{\rm GeV}$ also provides a good fit to the inferred velocity dependence (see \cref{fig:fdg_ADM}), where we do not need to consider additional entropy production to dilute the DM abundance or consider asymmetric generation of $B-L$ number.
Meanwhile, the preferred dark pion masses in both \cref{fig:maxSIDM_ADM} and \cref{fig:fdg_ADM} do not differ much, i.e., the pion mass (or $m_{\pi'}/\Lambda$) still has to be reasonably tuned near the quantum resonance for elastic scattering.
Similarly, the binding energy of nucleons, $E_{bt}$, decrease by a factor of $\sim 0.9$ as we take pion mass slightly away from the quantum resonance to explain the velocity dependence of $\sigma/m$ inferred by the field dwarf galaxies.
This would slightly retreat the left shaded region in \cref{fig:DDbound} to the further left, in which the dark nucleosynthesis may occur by emitting dark photons.

\section{Velocity averaging \label{sec:v-average}}

In this section, we comment on the velocity averaging of the self-scattering cross section $\sigma / m$.

The horizontal axis of \cref{fig:ERTvelocity1}, $\langle v_{\rm rel}\rangle$, is the average relative velocity between two DM particles.
Consider the collision between particle 1 and 2.
Assuming the Maxwellian velocity distribution for both particles, but with different 1d velocity dispersion: $\sigma_{1}^{2}$ and $\sigma_{2}^{2}$, respectively.
After integrating out the center-of-mass velocity, whose 1d velocity dispersion is $\sigma_{\rm cm}^{2} = \dfrac{\sigma_{1}^{2} \sigma_{2}^{2}}{\sigma_{1}^{2} + \sigma_{2}^{2}}$, we obtain the distribution function for the relative velocity:
\eqs{
d^{3} v_{\rm rel} \frac{1}{(2 \pi \sigma_{\rm rel}^{2})^{3/2}} \exp \left(- \frac{v_{\rm rel}^{2}}{2 \sigma_{\rm rel}^{2}} \right) \,,
}
where the 1d velocity dispersion is related with the average relative velocity as
\eqs{
\sigma_{\rm rel}^{2} = \sigma_{1}^{2} + \sigma_{2}^{2} = \frac{\pi}{8} \langle v_{\rm rel} \rangle^{2} \,.
}
In the gravitationally interacting system, we expect $\sigma_{1} = \sigma_{2} = \sigma$ and thus $\langle v_{\rm rel} \rangle = 4 \sigma / \sqrt{\pi}$ as found in the literature~\cite{Tulin:2017ara}.
Note that this differs from the kinetically equilibrium system, where $m_{1} \sigma_{1}^{2} = m_{2} \sigma_{2}^{2}$.

Meanwhile, the inferred values of $\sigma/m$ at given $\langle v_{\rm rel} \rangle$ in \cref{fig:ERTvelocity1} are achieved by assuming a constant self-scattering cross section inside a halo of interest.
In the case of velocity-dependent $\sigma/m$, a fairer comparison may be done by taking local distribution average for the cross section.
In the main text, we do not take the distribution average for the cross section.
This is partially just for the simplicity.
Another reason is that when astronomical data are interpreted by SIDM, a self-scattering cross section is assumed to be constant.
Thus, to be consistent, we need to reanalyze the data by taking into account the velocity (dispersion) dependence of the (distribution averaged) cross section.

\section{Higher resonances \label{sec:high-res}}

\begin{figure}
\centering
\begin{subfigure}{1.\textwidth}
\centering
  \includegraphics[width=0.93\linewidth]{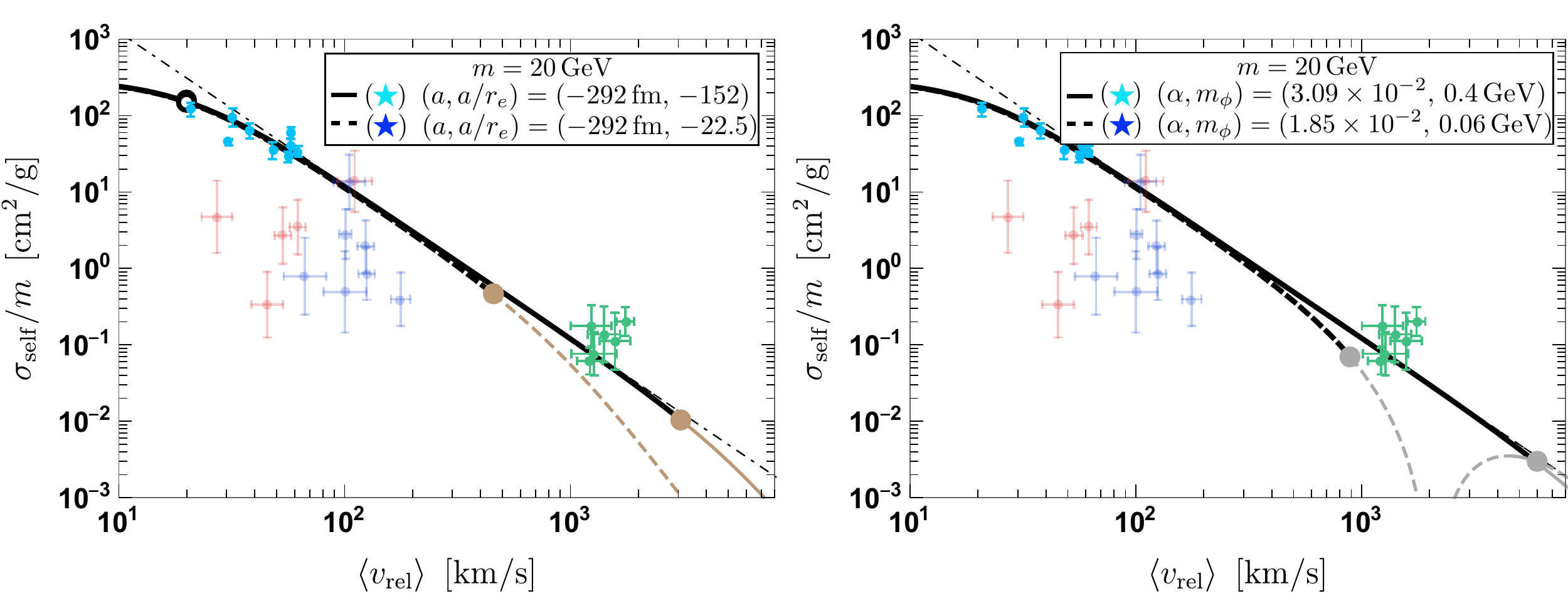}
  \caption{}
  \label{fig:ERTvelocity3}
  \par\medskip
  \includegraphics[width=0.93\linewidth]{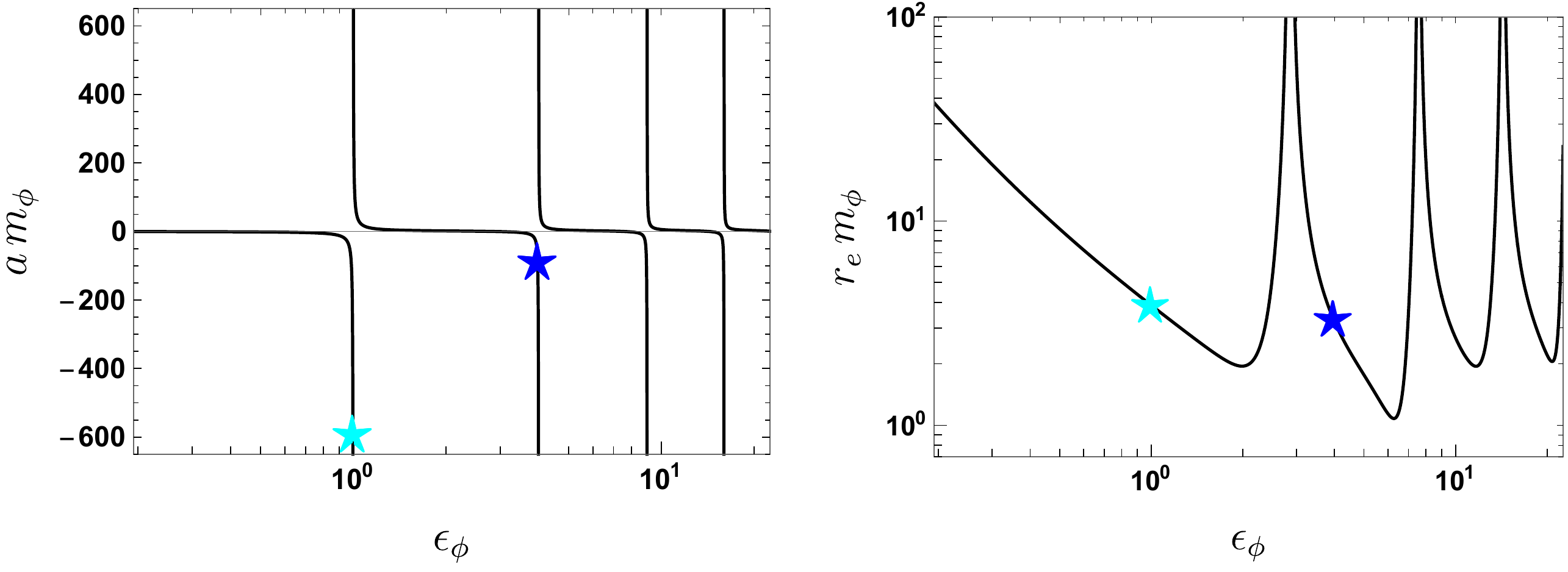}
  \caption{}
  \label{fig:Hulthenparam3}
\end{subfigure}
\caption{Same as \cref{fig:ERTvelocity1}, but comparing the parameter points close to the first (cyan star) and the second (blue star) quantum resonance for elastic scattering, i.e., $n=1$ and $n=2$ for $\epsilon_{\phi} = n^2$, respectively.
In the right panel of \cref{fig:ERTvelocity3}, for the parameter point near the second resonance (dashed), the analytic results for the for the Hulth\'en potential exhibits diminishing $\sigma/m$ at $v_{\rm rel}\sim2000\,{\rm km/s}$.
}
\end{figure}

\begin{figure}
\centering
\begin{subfigure}{1.\textwidth}
\centering
  \includegraphics[width=0.93\linewidth]{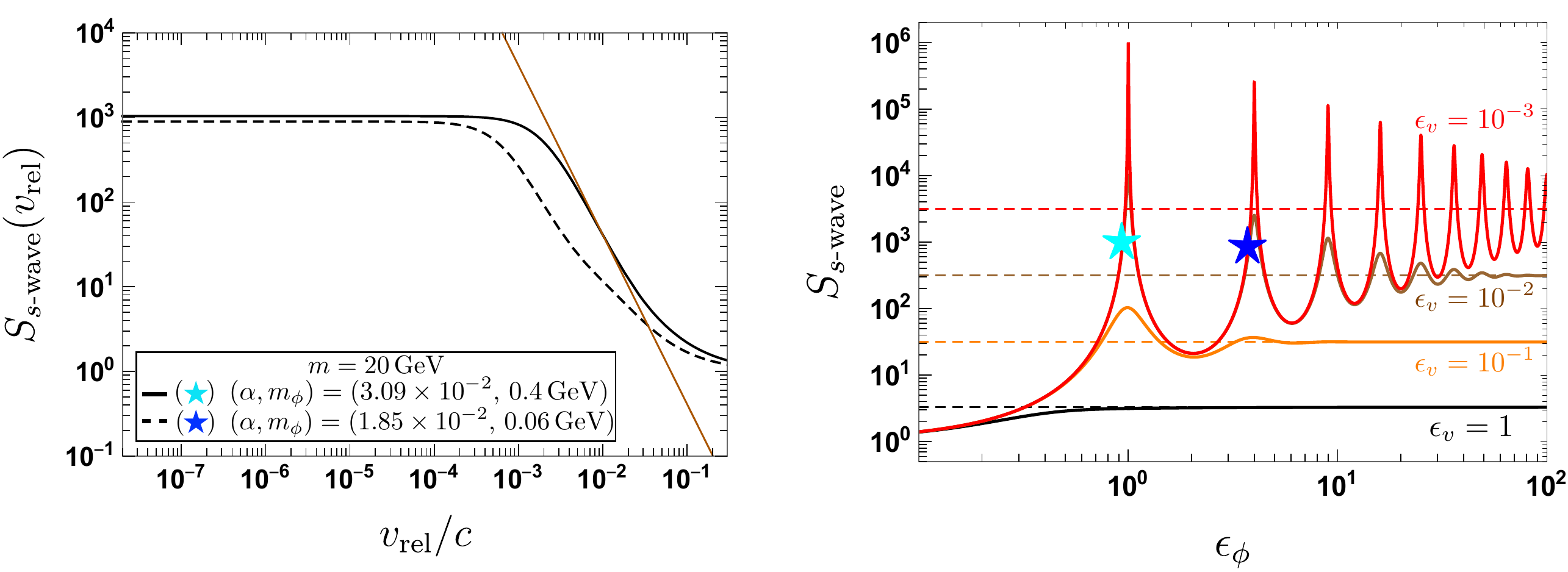}
  \caption{}
  \label{fig:SEFs3}
  \par\medskip
  \includegraphics[width=0.93\linewidth]{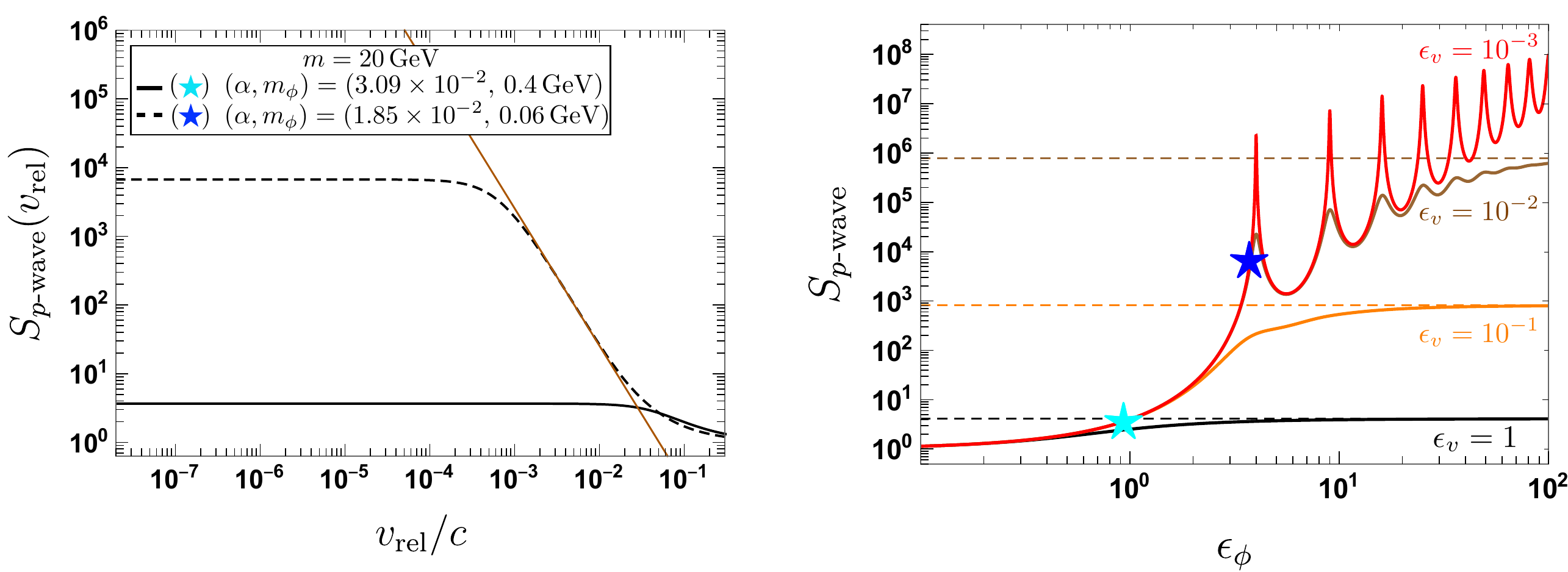}
  \caption{}
  \label{fig:SEFp3}
\end{subfigure}
\caption{
Same as \cref{fig:SEF1}, but for the parameter points depicted in \cref{fig:Hulthenparam3}, which are close to the first (cyan star) and the second quantum resonance (blue star) for elastic scattering.
In the left panels, the brown curve corresponds to $\propto1/v_{\rm rel}^2$.
Contrary to the cyan-star parameter set, which was discussed in \cref{fig:SEF1}, the blue-star parameter set is close to both $s$-wave and $p$-wave resonances for annihilation.
}
\label{fig:SEF3}
\end{figure}

As seen in \cref{fig:Hulthenparam1}, the quantum resonances for elastic scattering appear at $\epsilon_{\phi}=n^2$ ($n=1,2,\dots$).
In the main text, we focused on the parameters near the first one, $n=1$.
While it is certainly encouraging to discuss the higher resonances as well, there are a couple of reasons we try not to.
In this section, we discuss these reasons and remark on the possible differences and difficulties in investigating higher resonances.

The first aspect is that at higher resonances for elastic scattering, $\sigma/m$ could exhibit diminishing values at specific momentum.
This is demonstrated in the right panel of \cref{fig:ERTvelocity3}, where we show a velocity dependence of $\sigma/m$ in the Hulth\'en potential.
There, we see that for the Hulth\'en-potential parameter set near the second resonance, i.e., the blue star of \cref{fig:Hulthenparam3}, $\sigma/m$ is diminishing around $v_{\rm rel}\sim 2000\,{\rm km/s}$.
This is related to the behavior of the phase shift $\delta_0$ at the second resonance.
At the vicinity of the second resonance, $\delta_0$ starts out from $0$ at high-$k$ limit and approaches $\simeq3\pi/2$ in the $k\rightarrow0$ limit;
between the two limits, $\delta_0=\pi$ for some specific value of $k$ and $\sigma/m$ diminishes at such $k$.
One may expect that the existence of this ``dip" generally makes the parameters hard to be compatible with the data points in \cref{fig:ERTvelocity3}.
However, the ``dip" appears in the classical regime (depicted in gray), i.e., $k\gtrsim m_\phi$, where the higher partial-wave contributions are significant.
After taking into account the higher partial-wave contributions, the ``dip" may become relaxed.
We could also consider a larger $m_\phi$ (but with the same $\epsilon_\phi$) to push the onset of the classical regime (filled gray circle) to higher velocities, so that the ``dip" does not overlap with the velocity scales of the astronomical data.

The second aspect is related to the Sommerfeld enhancement of DM annihilation.
As we consider the resonances beyond the first one for elastic scattering, higher partial wave resonances for annihilation may become important.
As an example, in \cref{fig:SEF3}, we show the velocity dependence of the Sommerfeld-enhancement factors for the parameter point near the second resonance (blue star), which corresponds to the parameter took in the right panel of \cref{fig:ERTvelocity3}.
Coincidentally, the parameter lies near the $s$-wave and $p$-wave resonance for annihilation.
For this parameter point, taking $p$-wave annihilating DM to evade the constraints from indirect-detection experiments (as for the cyan parameter point) may not help much.
Likewise, the implications on the constraints may highly depend on the value of the took $\epsilon_{\phi}$ for higher resonances of elastic scattering.
A dedicated investigation on this aspect for higher resonances may be done elsewhere, while we try to focus on the simplest case in this work.

\section{Neutrino masses in the gauged $L_{\mu} - L_{\tau}$ model \label{sec:neu_mass}}

The renormalizable Lagrangian density can be written as
\eqs{
  \label{eq:lag1}
  {\cal L}
   =& {\cal L}_{\rm SM}
    + g^{\prime} Z_{\mu}^{\prime}
    \left( L_{2}^{\dagger} {\bar \sigma}^{\mu} L_{2} - L_{3}^{\dagger} {\bar \sigma}^{\mu} L_{3}
    - {\bar \mu}^{\dagger} \bar{\sigma}^{\mu} {\bar \mu} + {\bar \tau}^{\dagger} \bar{\sigma}^{\mu} {\bar \tau} \right)
    - \frac{1}{4} Z_{\mu \nu}^{\prime} Z^{\prime \mu \nu}
    - \frac{1}{2} \epsilon \, Z_{\mu\nu}^{\prime} B^{\mu \nu}
    \\
   &+ (D_{\mu} \Phi)^{\dagger} D^{\mu} \Phi - V(\Phi, H)
    + i {\bar N}^{\dagger}_{i} \bar{\sigma}^{\mu} D_{\mu} {\bar N}_{i} 
    \\
   & - \lambda_{e} L_{1} H {\bar N}_{e}
    - \lambda_{\mu} L_{2} H {\bar N}_{\mu}
    - \lambda_{\tau} L_{3} H {\bar N}_{\tau}
    - \frac{1}{2} M_{e e} {\bar N}_{e} {\bar N}_{e}
    - M_{\mu \tau} {\bar N}_{\mu} {\bar N}_{\tau}
    - y_{e \mu} \Phi^{*} {\bar N}_{e} {\bar N}_{\mu}
    - y_{e \tau} \Phi {\bar N}_{e} {\bar N}_{\tau} + {\rm h.c.}
}
Here, $L_i ~ (i =2, 3)$ denotes the left-handed leptons in the flavor basis, while $\bar\mu$ and $\bar\tau$ denote the right-handed charged leptons.
The discussion below is applicable to any $Z'$ mass, though we focus on $m_{Z'} \sim 10$\,MeV in the main text.
With a different $Z'$ mass, phenomenology in colliders and cosmology also changes (e.g., see Ref.~\cite{Dror:2020fbh} for very light $Z'$).

After the SM Higgs $H$ and $L_{\mu} - L_{\tau}$ breaking Higgs $\Phi$ develop the VEVs, $v_{H}/ \sqrt{2}$ and $v_{\Phi}/ \sqrt{2}$, the resultant neutrino mass sector takes a form of
\eqs{
  - {\cal L}_{\rm mass}
  &=
  \begin{pmatrix}
   \nu_{e} & \nu_{\mu} & \nu_{\tau}
  \end{pmatrix}
   {\cal M}_{\rm D}
   \begin{pmatrix}
   {\bar N}_{e} \\ {\bar N}_{\mu} \\ {\bar N}_{\tau}
  \end{pmatrix}
  +
  \frac{1}{2}
  \begin{pmatrix}
    {\bar N}_{e} & {\bar N}_{\mu} & {\bar N}_{\tau}
  \end{pmatrix}
  {\cal M}_{\rm N}
  \begin{pmatrix}
    {\bar N}_{e} \\ {\bar N}_{\mu} \\ {\bar N}_{\tau}
  \end{pmatrix}
   + {\rm h.c.} \,, \\
   {\cal M}_{\rm D}
   &=
   \begin{pmatrix}
     Y_{e} & 0 & 0 \\
     0 & Y_{\mu} & 0 \\
     0 & 0 & Y_{\tau}
  \end{pmatrix}
  \,, \quad
  {\cal M}_{\rm N}
   = 
   \begin{pmatrix}
     M_{e e} & M_{e \mu} & M_{e \tau} \\
     M_{e \mu} & 0 & M_{\mu \tau} \\
     M_{e \tau} & M_{\mu \tau} & 0
  \end{pmatrix}
  \,. \label{eq:MDN}
}
Here $(Y_{e}, Y_{\mu}, Y_{\tau}) = \dfrac{v_{H}}{\sqrt{2}} \, (\lambda_{e}, \lambda_{\mu}, \lambda_{\tau})$ and $(M_{e \mu}, M_{e \tau}) = \dfrac{v_{\Phi}}{\sqrt{2}} \, (y_{e \mu}, y_{e \tau})$.
We fix the phases of ${\bar N}_{l}$ so that $M_{e \mu}, M_{e \tau}, M_{\mu \tau} > 0$, while leaving the phases of $\nu_{l}$ for the PMNS parametrization~\cite{Pontecorvo:1957qd, Maki:1962mu}.
The see-saw mechanism~\cite{Minkowski:1977sc,Yanagida:1979as,GellMann:1980vs,Glashow:1979nm} provides the neutrino mass term at low energy as
\eqs{
- {\cal L}_{\rm mass} 
  &\simeq 
  \begin{pmatrix}
   \nu_{e} & \nu_{\mu} & \nu_{\tau}
  \end{pmatrix}
   {\cal M}_{\nu}
   \begin{pmatrix}
   \nu_{e} \\ \nu_{\mu} \\ \nu_{\tau}
  \end{pmatrix}
  =
  \begin{pmatrix}
   \nu_{1} & \nu_{2} & \nu_{3}
  \end{pmatrix}
   \begin{pmatrix}
     m_{1} & 0 & 0 \\
     0 & m_{2} & 0 \\
     0 & 0 & m_{3}
  \end{pmatrix}
  \begin{pmatrix}
   \nu_{1} \\ \nu_{2} \\ \nu_{3}
  \end{pmatrix}
   \,, \\
  {\cal M}_{\nu} 
  &= - {\cal M}_{\rm D} \, {\cal M}_{\rm N}^{-1} \, {\cal M}_{\rm D}^{T} \,, \quad
  \begin{pmatrix}
   \nu_{e} \\ \nu_{\mu} \\ \nu_{\tau}
  \end{pmatrix}
  = U_{\nu}
  \begin{pmatrix}
   \nu_{1} \\ \nu_{2} \\ \nu_{3}
  \end{pmatrix}  \,,
}
where the PMNS matrix is parametrized by
\eqs{
   U_{\nu}   
   = 
   \begin{pmatrix}
     1 & 0 & 0 \\
     0 & \cos \theta_{2 3} & \sin \theta_{2 3} \\
     0 & - \sin \theta_{2 3} & \cos \theta_{2 3}
   \end{pmatrix}
   \begin{pmatrix}
     \cos \theta_{1 3} & 0 & \sin \theta_{1 3} e^{- i \delta} \\
     0 & 1 & 0 \\
     - \sin \theta_{1 3} e^{i \delta} & 0 & \cos \theta_{1 3}
   \end{pmatrix}
   \begin{pmatrix}
     \cos \theta_{1 2} & \sin \theta_{1 2} & 0 \\
     - \sin \theta_{1 2} & \cos \theta_{1 2} & 0 \\
     0 & 0 & 1
   \end{pmatrix}
   \begin{pmatrix}
     1 & 0 & 0 \\
     0 & e^{i \alpha_{2} / 2} & 0 \\
     0 & 0 & e^{i \alpha_{3} / 2}
   \end{pmatrix}  \,,
}
with $\theta_{i j} \in [0, \pi / 2]$ and $\delta, \alpha_{i} \in [0, 2 \pi)$.
and the mass eigenvalues are written as
\eqs{
m_{2}^{2} = m_{1}^{2} + \delta m^{2} \,, \quad m_{3}^{2} = m_{1}^{2} + \Delta m^{2} + \delta m^{2} / 2 \,,
}
for the normal ordering, while $m_{3} < m_{1} < m_{2}$ for the inverted ordering.
The neutrino oscillation parameters are summarized in \cref{tab:neu_osc_param}.
\begin{table}[h]
  \centering
  \begin{tabular}{ccccc}\hline
     Parameter & best fit & 1$\sigma$ range & 2$\sigma$ range \\\hline
     $\delta m^{2} / (10^{-5} \, {\rm eV}^{2})$ & $7.37$ & $7.21$--$7.54$ & $7.07$--$7.73$ \\
     $\Delta m^{2} / (10^{-3} \, {\rm eV}^{2})$ & $2.525$ & $2.495$--$2.567$ & $2.454$--$2.606$ \\    
     $\sin^{2} \theta_{12} / 10^{-1}$ & $2.97$ & $2.81$--$3.14$ & $2.65$--$3.34$ \\
     $\sin^{2} \theta_{23} / 10^{-1}$ & $4.25$ & $4.10$--$4.46$ & $3.95$--$4.70$ \\
     $\sin^{2} \theta_{13} / 10^{-2}$ & $2.15$ & $2.08$--$2.22$ & $1.99$--$2.31$ \\
     $\delta / \pi$ & $1.38$ & $1.18$--$1.61$ & $1.00$--$1.90$ \\ \hline
  \end{tabular}
  \caption{Neutrino oscillation parameters from Ref.~\cite{Capozzi:2017ipn, Asai:2017ryy}.
  }
  \label{tab:neu_osc_param}
\end{table}

It is remarkable that the U(1)$_{L_{\mu} - L_{\tau}}$ symmetry restricts ${\cal M}_{\rm D}$ to be a diagonal matrix.
Thus if $\mathcal{M}_{\rm D}$ has a inverse matrix, the following relation holds:
\eqs{
{\cal M}_{\nu}^{-1} = - {\cal M}_{\rm D}^{-1} \, {\cal M}_{\rm N} \, ({\cal M}_{\rm D}^{-1})^{T} \,.
\label{eq:invnumass}
}
Since ${\cal M}_{\rm D}^{-1}$ is also diagonal, the flavor structure of ${\cal M}_{\nu}^{-1}$ should follow the structure of ${\cal M}_{\rm N}$.
Two zero entries of ${\cal M}_{\rm N}$ (see \cref{eq:MDN}) provides four constraints for the PMNS parameters~\cite{Asai:2017ryy}:
$\delta \simeq \pi / 2$ or $3 \pi /2$, $m \simeq 0.05 \text{-} 0.1 \, {\rm eV}$, $\alpha_{2} / \pi \simeq 0.6$, and $\alpha_{3} / \pi \simeq 1.4$ from $\delta m^{2}$, $\Delta m^{2}$, and $\theta_{ij}$ given in \cref{tab:neu_osc_param}.
The other entries provide 8 constraints, while there are 11 model parameters (6 from $Y_{l}$ and 5 from $M_{l l'}$).
This is because \cref{eq:invnumass} is invariant under the following transformation:
\eqs{
Y_{e} &\to \frac{a b}{c} Y_{e} \,, \quad Y_{\mu} \to \frac{a c}{b} Y_{\mu} \,, \quad Y_{\tau} \to \frac{b c}{a} Y_{\tau} \,, \nonumber \\
M_{e e} &\to \frac{a^{2} b^{2}}{c^{2}} M_{e e} \,, \quad M_{e \mu} \to a^{2} M_{e \mu} \,, \quad M_{e \tau} \to b^{2} M_{e \tau} \,, \quad M_{\mu \tau} \to c^{2} M_{\mu \tau} \,,
\label{eq:numass_trans}
}
with $a, b, c$ being real.
We obtain \cref{tab:model_param} by fitting the other eight model parameters in \cref{eq:invnumass}. 
The mass matrices are rewritten as
\eqs{
{\cal M}_{\rm D}
   =
   \begin{pmatrix}
     Y_{e}' \dfrac{a b}{c} & 0 & 0 \\
     0 & Y_{\mu}' \dfrac{a c}{b} & 0 \\
     0 & 0 & Y_{\tau}' \dfrac{b c}{a}
  \end{pmatrix}
  \,, \quad
  {\cal M}_{\rm N}
   = 
   \begin{pmatrix}
     M_{e e}' \dfrac{a^{2} b^{2}}{c^{2}} & M_{e \mu}' a^{2} & M_{e \tau}' b^{2} \\
     M_{e \mu}' a^{2} & 0 & M_{\mu \tau}' c^{2} \\
     M_{e \tau}' b^{2} & M_{\mu \tau}' c^{2} & 0
  \end{pmatrix} \,.
  \label{eq:MDN_para}
}
\begin{table}[h]
  \centering
  \begin{tabular}{ccccc}\hline
     Parameter & best fit in \cref{tab:neu_osc_param} \\ \hline
     $Y_{e}' / (1 \, {\rm eV})$ & $0.360 - 0.0819 \, i$ \\
     $Y_{\mu}' / (1 \, {\rm eV})$ & $0.153 + 0.189 \, i$ \\    
     $Y_{\tau}' / (1 \, {\rm eV})$ & $- 0.180 - 0.219 \, i$ \\
     $M_{e e}' / (1 \, {\rm eV})$ & $- 1.98 + 0.0522 \, i$ \\ \hline
     $M_{e \mu}' / (1 \, {\rm eV})$ & $1$ \\
     $M_{e \tau}' / (1 \, {\rm eV})$ & $1$ \\
     $M_{\mu \tau}' / (1 \, {\rm eV})$ & $1$ \\ \hline
  \end{tabular}
  \caption{Model parameters from \cref{eq:invnumass}.
  We fix $M'_{e \mu} = M'_{e \tau} = M'_{\mu \tau} = 1 \, {\rm eV}$ (prime denotes this fixing) by using \cref{eq:numass_trans}.
  }
  \label{tab:model_param}
\end{table}

\section{Supersymmetric realization of composite ADM \label{sec:susy-adm}}

A supersymmetric realization of composite ADM scenario~\cite{Ibe:2018tex,Ibe:2019ena} may cause the fast DM decay as in nucleon decay in supersymmetric grand unified theories~\cite{Goto:1998qg,Murayama:2001ur}.
Heavy particles in ultraviolet physics induce the intermediate-scale portal operators with the mass dimension-six rather than the mass dimension-seven, 
\eqs{
  \mathcal{L} \supset \int d^2 \theta \frac{y_{N} Y_{N} Y_{\bar C}}{M_{C} M_{R}} ({\bar U}' {\bar D}' {\bar D'}) (LH) + \mathrm{h.c.}
  \label{eq:decay_susy}
}
The equilibrium condition now reads
\eqs{
M_{R} < M_{C} \lesssim \frac{100}{x_{B-L}^{3/2}} \left( \frac{m_{\nu}}{0.1 \, {\rm eV}} \right)^{1/2} Y_{N} Y_{C} M_{R} \,.
}
The decay rate is given by
\eqs{
  \Gamma_\mathrm{DM} \simeq 
  \frac{L^{2}}{32 \pi} \frac{m_{N'}}{M_{*}^{4}} \frac{v_H^{2} M_{1/2}^{2}}{M_{S}^{4}} |W|^2 \,, \quad \frac{1}{M_{*}^2} = \frac{y_{N} Y_{N} Y_{C}}{M_{C} M_{R}} \,.
}
$M_{S}$ and $M_{1/2}$ denote typical mass scales of supersymmetric scalar particles and supersymmetric fermionic particles, respectively.
$L$ denotes a typical one-loop factor.
The intermediate-scale portal operator \cref{eq:decay_susy} vanishes when the generation $N'_{g} = 1$ due to anti-symmetrization over color indices. 
Hereafter, we assume $N'_{g} = 2$.

The supersymmetric particles change the relation between the $B-L$ asymmetries in the SM and dark sectors:
\eqs{
A_{\rm DM} = \frac{2 N'_{g} (20 N_{g} + 3 m)}{3 N_{g} (13 m + 44 N_{g})} A_{\rm SM} \,,
}
with the full supersymmetric particles are available.
On the other hand, they do not change the relation between the present relation between the $B$ and $B-L$ asymmetries in the SM sector, 
$A_{B} = \dfrac{30}{97} A_{\rm SM}$,
with the decoupling limit of supersymmetric particles and heavy Higgses.
As a result, $m_{N'} = 4.3 \, {\rm GeV}$ for $m = 2$ (MSSM), $N_{g} = 3$, and $N'_{g} = 2$. 

As with non-supersymmetric realization, we also have the Majorana mass term for the dark neutrons in the supersymmetric realization.
The corresponding superpotential has the mass dimension seven, and is given by 
\eqs{
    \mathcal{L} \supset - \frac{Y_{N}^{2} Y_{C}^{2}}{2 M_{R} M_{C}^{2}} \int d^{2} \theta ({\bar U}' {\bar D}' {\bar D'})^{2} + \mathrm{h.c.}
}
This operator induces the dimension nine operator at the mass scale of supersymmetric particles with two-loop diagrams~\cite{Costa:1982uv}, and gives the Majorana mass term for the dark neutrons.
A typical size of the Majorana mass is
\eqs{
    m_{N} \simeq \frac{Y_{N}^{2} Y_{C}^{2}}{2 M_{R} M_{C}^{2}} \frac{L^{2} M_{1/2}^{2}}{M_{S}^{4}} \Lambda_{\rm QCD'}^{6} \,.
}

\begin{figure}
	\centering
	\includegraphics[width=0.6\linewidth]{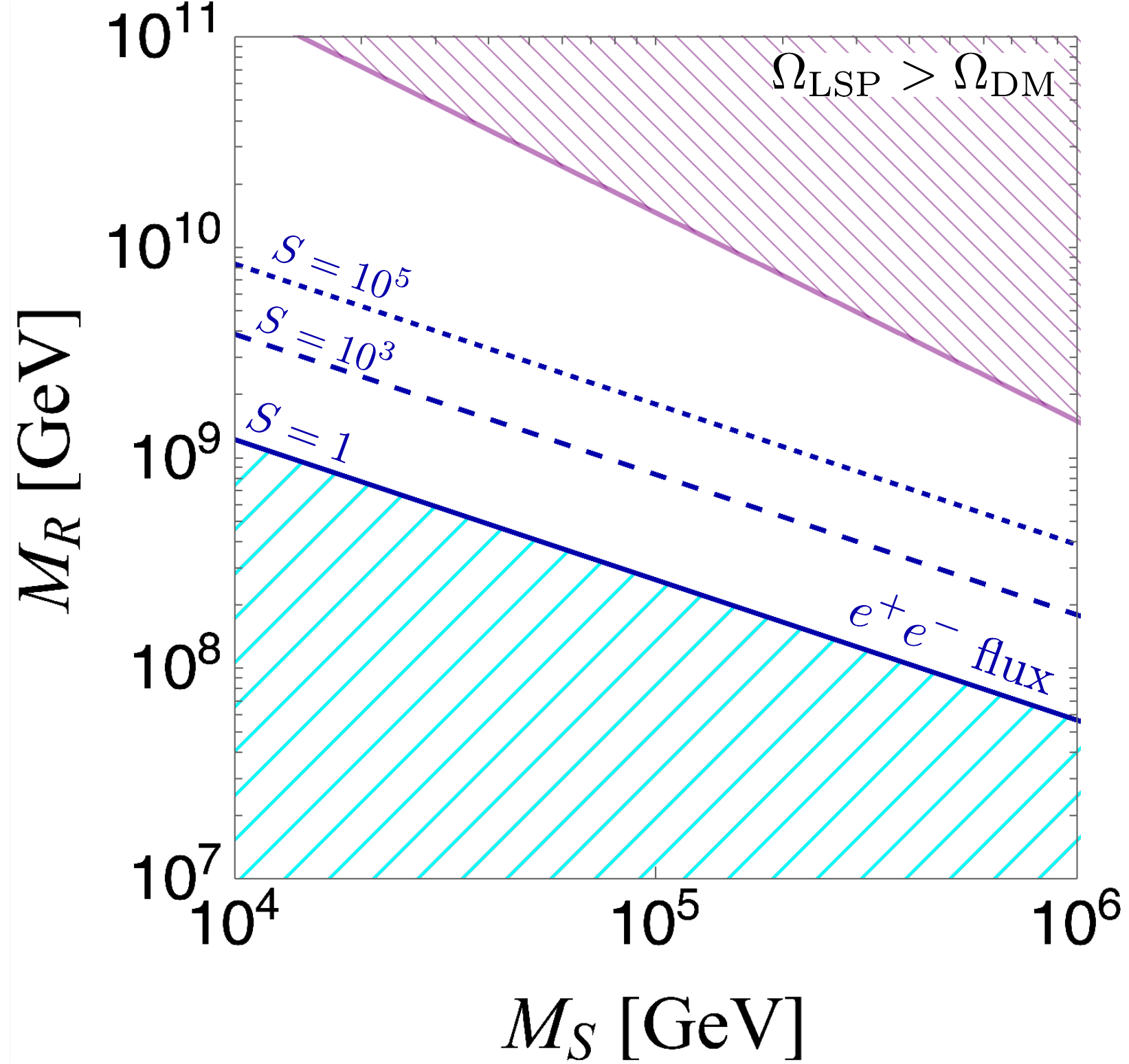}
	\caption{
    Constraints on $M_{R}$-$M_{S}$ plane in a supersymmetric realization of composite ADM.
    The blue-hatched region and the blue-dashed lines are the same as in \cref{fig:IDbound}. 
    The LSP abundance from the gravitino decay exceeds the observed DM abundance in the meshed-purple region in thermal leptogenesis scenario.
  }
	\label{fig:Decay}
\end{figure}

\cref{fig:Decay} shows the indirect-detection constraints on the $M_{R}$-$M_{S}$ plane from cosmic ray observations. 
In this figure, we assume that the loop factor is $L = 0.01$, and $M_{1/2} = 10^{-4} M_{S}$ that is a spectrum motivated by split supersymmetry scenario \cite{ArkaniHamed:2004fb,Giudice:2004tc,Wells:2004di}.
We show only the constraint from the $e^+ e^-$ flux observations in this figure (the blue-hatched region), and we ignore the other bounds from the neutrino signal from the DM decay, the $e^+e^-$ flux from the DM decay, and the $\gamma$-ray flux from the ADM annihilation.
As we see in \cref{fig:IDbound}, the $\gamma$-ray constraint on the annihilation cross section is comparable with the $e^+ e^-$ flux constraint. 

In supersymmetric realization, the lightest supersymmetric particle (LSP) can also be stable due to the $R$-parity. 
We have the LSPs in each of visible and dark sectors, and then the heavier state can decay into the lighter one through portal interactions at the late time.
The supersymmetric counterpart of the kinetic mixing between the dark photon and SM hypercharge gauge boson leads the prompt decay of the heavier state to the lighter one when the higgsinos are the lightest particles in each of sectors~\cite{Ibe:2018tex}.
Here, the higgsino in the visible sector, called visible higgsino, is the supersymmetric partner of the SM Higgs, while the one in the dark sector, called dark higgsino, is that of the U(1)$_{\rm dark}$ breaking Higgs.
Even though the dark higgsino is less constrained than the visible higgsino, the LSP abundance from the gravitino decay would cause the overclosure of the Universe~\cite{Kawasaki:2008qe,Kawasaki:2017bqm}. 
The purple-meshed region in \cref{fig:Decay} shows the LSP abundance exceeds the observed DM abundance when $T_{R} \simeq M_{R}$ and the LSP with a mass of $M_{1/2}$.

The constraints from the LSP abundance may be naturally relaxed in the models with an intermediate-scale dark grand unification: SU(5)$_{\rm SM}\times$SU(4)$_{\rm dark}$~\cite{Ibe:2018tex} and mirror SU(5)$_{\rm SM}\times$SU(5)$_{\rm dark}$~\cite{Ibe:2019ena}. 
They associate an intermediate-scale dark monopole in the SU(4)$_{\rm dark} \to$SU(3)$_{\rm dark} \times$ U(1)$_{\rm dark}$ phase transition.
Their abundance is determined by the pair-annihilation as~\cite{Preskill:1979zi,Khoze:2014woa}%
\footnote{
Note that Ref.~\cite{Khoze:2014woa} considers SO(3)$\to$U(1) and thus, the monopole charge is $4 \pi / g_{D}$, while in the present case, it is $2 \pi / g_{D}$.
}
\eqs{
\frac{n_{m}}{s} \sim \frac{1}{B} \frac{2 g_{D}^{4}}{\pi}\sqrt{\frac{45}{4 \pi^{3} g_{*}}} \frac{M_{m}}{M_{\rm Pl}} = 6.9 \times 10^{-14} \left( \frac{\alpha_{D}}{\alpha_{\rm EM}} \right)^{3} \left( \frac{M_{m}}{10^{10} \, {\rm GeV}} \right) \,.
}
Here we use $g_{*} = g_{*{\rm MSSM}} + g_{*{\rm dark}} + g_{* N}$ and $g_{*{\rm MSSM}} = 228.75$ (full MSSM multiplets),
$g_{*{\rm dark}} = 131.25$ (2 generations of $U'/{\bar U}'$ and $D'/{\bar D}'$, $g'$, $A'$, and $\phi_{D}/{\bar \phi}_{D}$ multiplets), and $g_{* N} = 11.25$ (3 generations of $N$ multiplets).
$M_m$ denotes the mass of the dark monopole, and $M_\mathrm{Pl}$ the reduced Planck mass.
$B$ is a dimensionless quantity defined by $B = q^{2}_{*} \dfrac{\zeta(3)}{4 \pi^{2} g_D^2}$, where $q^{2}_{*} = \dfrac{91}{3}$ is the summation of the dark charges squared with a weight of $1$ $(3/4)$ for bosons (fermions).

The entropy production factor is
\eqs{
\frac{S_{\rm after}}{S_{\rm before}} = \frac{T_{\rm eq}}{T_{\rm ann}} = \left( \frac{g_{*} (T_{\rm ann}) T_{\rm eq}^{3}}{g_{*} (T_{c}) T_{c}^{3}} \right)^{1/4} 
\simeq 0.002 \left( \frac{\alpha_{D}}{\alpha_{\rm EM}} \right)^{21/8} \left( \frac{1 \, {\rm GeV}}{m_{A'}}\right)^{3/4} \left( \frac{M_{m}}{10^{10} \, {\rm GeV}} \right)^{3/2} \,.
}
Note that this expression is valid only for $S_{\rm after} / S_{\rm before} > 1$; otherwise, $S_{\rm after} / S_{\rm before} = 1$ (i.e., no entropy production).
Here $T_{c} \sim m_{A'} / g_{D}$ is the critical temperature of the U(1)$_{\rm dark}$ phase transition.
$T_{\rm ann}$ is the temperature at which monopole annihilates.
In the second equality, we take $g_{*} (T_{c}) = g_{*} (T_{\rm ann})$.
$T_{\rm eq}$ is the temperature at which monopole domination begins:
\eqs{
T_{\rm eq} = \frac{4}{3} M_{m} \frac{n_{m}}{s} \,.
}

In the above discussion, we assume that the SM and dark sectors are in equilibrium in the course of the monopole domination, while it may not be valid for a tiny kinetic mixing.
Monopole annihilation may also produce the LSP~\cite{Harigaya:2014waa}, while we leave further analysis for a future work.

\section{Direct detection through the magnetic moment of the dark neutron \label{sec:dd-munp}}

Similarly to the dark proton scattering with SM nuclei, the dark neutron carries the magnetic moment under U(1)$_{\rm dark}$:
\eqs{
{\cal L} \supset \frac{\mu_{n'}}{2} \bar{n}' \sigma^{\mu \nu} n' F^{A'}_{\mu \nu} \,.
}
Here we take the magnetic moment to be $\mu_{n'} = (g_{n'} / 2) (e_{D} / 2m_{N'})$ with the $g$-factors being $g_{n'} = - 3.83$ ($g_{p'} = 5.59$ for the dark proton) in analogy to the SM nucleons.
The matrix element for the dark neutron-SM proton scattering is
\eqs{
& {\cal M}_{n'} = \frac{\mu_{n'} / e_{D}}{2 m_{N'}} \left[ q^{2} + 4 m_{N'} {\cal O}^{\rm NR}_{5} + 2 g_{p} \frac{m_{N'}}{m_{N}} \left( {\cal O}^{\rm NR}_{4} q^{2} - {\cal O}^{\rm NR}_{6} \right) \right] {\cal M}_{p'} \,,
}
where ${\cal M}_{p'}$ is the matrix element for the dark proton-SM proton scattering, and
\eqs{
{\cal O}^{\rm NR}_{4} = \vec{s}_{n'} \cdot \vec{s}_{p} \,, \quad {\cal O}^{\rm NR}_{5} = i \vec{s}_{n'} \cdot (\vec{q} \times \vec{v}^{\perp}) \,, \quad {\cal O}^{\rm NR}_{6} = (\vec{s}_{n'} \cdot \vec{q}) (\vec{s}_{p} \cdot \vec{q}) \,. \\
}
We refer readers to Ref.~\cite{DelNobile:2013sia} for exchange of SM massless photon, rather than massive dark photon.
$\vec{s}_{p (n')}$ is the spin vector of the SM proton (dark neutron, i.e., DM) and $\vec{v}^{\perp} = \vec{v} - \vec{q} / (2 \mu_{N})$.
We can further evaluate the matrix element for the dark neutron-target nucleus scattering with the form factors of $F^{N_{1} N_{2}}_{i, j}$~\cite{Fitzpatrick:2012ib,DelNobile:2013sia}, while we leave it for a future work.
In orders of magnitude, the direct-detection bounds on $\epsilon$ from dark-neutron scattering is weakened by a factor of $1 / v^{2} \sim 10^{6}$ compared to the dark proton-target nuclei scattering.

\bibliographystyle{utphys}
\bibliography{ref}

\providecommand{\href}[2]{#2}\begingroup\raggedright\begin{thebibliography}{100}

\bibitem{Bullock:2017xww}
J.~S. Bullock and M.~Boylan-Kolchin, ``{Small-Scale Challenges to the
  $\Lambda$CDM Paradigm},''
  \href{http://dx.doi.org/10.1146/annurev-astro-091916-055313}{{\em Ann. Rev.
  Astron. Astrophys.} {\bfseries 55} (2017) 343--387},
\href{http://arxiv.org/abs/1707.04256}{{\ttfamily arXiv:1707.04256
  [astro-ph.CO]}}.

\bibitem{Spergel:1999mh}
D.~N. Spergel and P.~J. Steinhardt, ``{Observational evidence for
  selfinteracting cold dark matter},''
  \href{http://dx.doi.org/10.1103/PhysRevLett.84.3760}{{\em Phys. Rev. Lett.}
  {\bfseries 84} (2000) 3760--3763},
\href{http://arxiv.org/abs/astro-ph/9909386}{{\ttfamily arXiv:astro-ph/9909386
  [astro-ph]}}.

\bibitem{Tulin:2017ara}
S.~Tulin and H.-B. Yu, ``{Dark Matter Self-interactions and Small Scale
  Structure},'' \href{http://dx.doi.org/10.1016/j.physrep.2017.11.004}{{\em
  Phys. Rept.} {\bfseries 730} (2018) 1--57},
\href{http://arxiv.org/abs/1705.02358}{{\ttfamily arXiv:1705.02358 [hep-ph]}}.

\bibitem{Buckley:2017ijx}
M.~R. Buckley and A.~H.~G. Peter, ``{Gravitational probes of dark matter
  physics},'' \href{http://dx.doi.org/10.1016/j.physrep.2018.07.003}{{\em Phys.
  Rept.} {\bfseries 761} (2018) 1--60},
\href{http://arxiv.org/abs/1712.06615}{{\ttfamily arXiv:1712.06615
  [astro-ph.CO]}}.

\bibitem{Arcadi:2017kky}
G.~Arcadi, M.~Dutra, P.~Ghosh, M.~Lindner, Y.~Mambrini, M.~Pierre, S.~Profumo,
  and F.~S. Queiroz, ``{The waning of the WIMP? A review of models, searches,
  and constraints},''
  \href{http://dx.doi.org/10.1140/epjc/s10052-018-5662-y}{{\em Eur. Phys. J.}
  {\bfseries C78} no.~3, (2018) 203},
\href{http://arxiv.org/abs/1703.07364}{{\ttfamily arXiv:1703.07364 [hep-ph]}}.

\bibitem{Roszkowski:2017nbc}
L.~Roszkowski, E.~M. Sessolo, and S.~Trojanowski, ``{WIMP dark matter
  candidates and searches?current status and future prospects},''
  \href{http://dx.doi.org/10.1088/1361-6633/aab913}{{\em Rept. Prog. Phys.}
  {\bfseries 81} no.~6, (2018) 066201},
\href{http://arxiv.org/abs/1707.06277}{{\ttfamily arXiv:1707.06277 [hep-ph]}}.

\bibitem{Kaplinghat:2015aga}
M.~Kaplinghat, S.~Tulin, and H.-B. Yu, ``{Dark Matter Halos as Particle
  Colliders: Unified Solution to Small-Scale Structure Puzzles from Dwarfs to
  Clusters},'' \href{http://dx.doi.org/10.1103/PhysRevLett.116.041302}{{\em
  Phys. Rev. Lett.} {\bfseries 116} no.~4, (2016) 041302},
\href{http://arxiv.org/abs/1508.03339}{{\ttfamily arXiv:1508.03339
  [astro-ph.CO]}}.

\bibitem{Newman:2012nw}
A.~B. Newman, T.~Treu, R.~S. Ellis, and D.~J. Sand, ``{The Density Profiles of
  Massive, Relaxed Galaxy Clusters: II. Separating Luminous and Dark Matter in
  Cluster Cores},'' \href{http://dx.doi.org/10.1088/0004-637X/765/1/25}{{\em
  Astrophys. J.} {\bfseries 765} (2013) 25},
\href{http://arxiv.org/abs/1209.1392}{{\ttfamily arXiv:1209.1392
  [astro-ph.CO]}}.

\bibitem{Randall:2007ph}
S.~W. Randall, M.~Markevitch, D.~Clowe, A.~H. Gonzalez, and M.~Brada\v{c},
  ``{Constraints on the Self-Interaction Cross-Section of Dark Matter from
  Numerical Simulations of the Merging Galaxy Cluster 1E 0657-56},''
  \href{http://dx.doi.org/10.1086/587859}{{\em Astrophys. J.} {\bfseries 679}
  (2008) 1173--1180}, \href{http://arxiv.org/abs/0704.0261}{{\ttfamily
  arXiv:0704.0261 [astro-ph]}}.

\bibitem{Harvey:2018uwf}
D.~Harvey, A.~Robertson, R.~Massey, and I.~G. McCarthy, ``{Observable tests of
  self-interacting dark matter in galaxy clusters: BCG wobbles in a constant
  density core},'' \href{http://dx.doi.org/10.1093/mnras/stz1816}{{\em Mon.
  Not. Roy. Astron. Soc.} {\bfseries 488} no.~2, (2019) 1572--1579},
  \href{http://arxiv.org/abs/1812.06981}{{\ttfamily arXiv:1812.06981
  [astro-ph.CO]}}.

\bibitem{Sagunski:2020spe}
L.~Sagunski, S.~Gad-Nasr, B.~Colquhoun, A.~Robertson, and S.~Tulin,
  ``{Velocity-dependent Self-interacting Dark Matter from Groups and Clusters
  of Galaxies},'' \href{http://arxiv.org/abs/2006.12515}{{\ttfamily
  arXiv:2006.12515 [astro-ph.CO]}}.

\bibitem{Newman:2015kzv}
A.~B. Newman, R.~S. Ellis, and T.~Treu, ``{Luminous and Dark Matter Profiles
  From Galaxies to Clusters: Bridging the gap With Group-scale Lenses},''
  \href{http://dx.doi.org/10.1088/0004-637X/814/1/26}{{\em Astrophys. J.}
  {\bfseries 814} no.~1, (2015) 26}.

\bibitem{Oman:2015xda}
K.~A. Oman {\em et~al.}, ``{The unexpected diversity of dwarf galaxy rotation
  curves},'' \href{http://dx.doi.org/10.1093/mnras/stv1504}{{\em Mon. Not. Roy.
  Astron. Soc.} {\bfseries 452} no.~4, (2015) 3650--3665},
\href{http://arxiv.org/abs/1504.01437}{{\ttfamily arXiv:1504.01437
  [astro-ph.GA]}}.

\bibitem{Kamada:2016euw}
A.~Kamada, M.~Kaplinghat, A.~B. Pace, and H.-B. Yu, ``{Self-Interacting dark
  matter can explain diverse galactic rotation curves},''
  \href{http://dx.doi.org/10.1103/PhysRevLett.119.111102}{{\em Phys. Rev.
  Lett.} {\bfseries 119} no.~11, (2017) 111102},
  \href{http://arxiv.org/abs/1611.02716}{{\ttfamily arXiv:1611.02716
  [astro-ph.GA]}}.

\bibitem{Ren:2018jpt}
T.~Ren, A.~Kwa, M.~Kaplinghat, and H.-B. Yu, ``{Reconciling the Diversity and
  Uniformity of Galactic Rotation Curves with Self-Interacting Dark Matter},''
  \href{http://dx.doi.org/10.1103/PhysRevX.9.031020}{{\em Phys. Rev. X}
  {\bfseries 9} no.~3, (2019) 031020},
  \href{http://arxiv.org/abs/1808.05695}{{\ttfamily arXiv:1808.05695
  [astro-ph.GA]}}.

\bibitem{Tulin:2012wi}
S.~Tulin, H.-B. Yu, and K.~M. Zurek, ``{Resonant Dark Forces and Small Scale
  Structure},'' \href{http://dx.doi.org/10.1103/PhysRevLett.110.111301}{{\em
  Phys. Rev. Lett.} {\bfseries 110} no.~11, (2013) 111301},
  \href{http://arxiv.org/abs/1210.0900}{{\ttfamily arXiv:1210.0900 [hep-ph]}}.

\bibitem{Tulin:2013teo}
S.~Tulin, H.-B. Yu, and K.~M. Zurek, ``{Beyond Collisionless Dark Matter:
  Particle Physics Dynamics for Dark Matter Halo Structure},''
  \href{http://dx.doi.org/10.1103/PhysRevD.87.115007}{{\em Phys. Rev. D}
  {\bfseries 87} no.~11, (2013) 115007},
  \href{http://arxiv.org/abs/1302.3898}{{\ttfamily arXiv:1302.3898 [hep-ph]}}.

\bibitem{Kahlhoefer:2017umn}
F.~Kahlhoefer, K.~Schmidt-Hoberg, and S.~Wild, ``{Dark matter self-interactions
  from a general spin-0 mediator},''
  \href{http://dx.doi.org/10.1088/1475-7516/2017/08/003}{{\em JCAP} {\bfseries
  08} (2017) 003}, \href{http://arxiv.org/abs/1704.02149}{{\ttfamily
  arXiv:1704.02149 [hep-ph]}}.

\bibitem{Ma:2017ucp}
E.~Ma, ``{Inception of Self-Interacting Dark Matter with Dark Charge
  Conjugation Symmetry},''
  \href{http://dx.doi.org/10.1016/j.physletb.2017.06.067}{{\em Phys. Lett. B}
  {\bfseries 772} (2017) 442--445},
  \href{http://arxiv.org/abs/1704.04666}{{\ttfamily arXiv:1704.04666
  [hep-ph]}}.

\bibitem{Duch:2017khv}
M.~Duch, B.~Grzadkowski, and D.~Huang, ``{Strongly self-interacting vector dark
  matter via freeze-in},''
  \href{http://dx.doi.org/10.1007/JHEP01(2018)020}{{\em JHEP} {\bfseries 01}
  (2018) 020}, \href{http://arxiv.org/abs/1710.00320}{{\ttfamily
  arXiv:1710.00320 [hep-ph]}}.

\bibitem{Duerr:2018mbd}
M.~Duerr, K.~Schmidt-Hoberg, and S.~Wild, ``{Self-interacting dark matter with
  a stable vector mediator},''
  \href{http://dx.doi.org/10.1088/1475-7516/2018/09/033}{{\em JCAP} {\bfseries
  09} (2018) 033}, \href{http://arxiv.org/abs/1804.10385}{{\ttfamily
  arXiv:1804.10385 [hep-ph]}}.

\bibitem{Kamada:2018zxi}
A.~Kamada, K.~Kaneta, K.~Yanagi, and H.-B. Yu, ``{Self-interacting dark matter
  and muon $(g-2)$ in a gauged U$(1)_{L_{\mu} - L_{\tau}}$ model},''
  \href{http://dx.doi.org/10.1007/JHEP06(2018)117}{{\em JHEP} {\bfseries 06}
  (2018) 117}, \href{http://arxiv.org/abs/1805.00651}{{\ttfamily
  arXiv:1805.00651 [hep-ph]}}.

\bibitem{Kamada:2018kmi}
A.~Kamada, M.~Yamada, and T.~T. Yanagida, ``{Self-interacting dark matter with
  a vector mediator: kinetic mixing with the $
  \mathrm{U}{(1)}_{{\left(B-L\right)}_3} $ gauge boson},''
  \href{http://dx.doi.org/10.1007/JHEP03(2019)021}{{\em JHEP} {\bfseries 03}
  (2019) 021}, \href{http://arxiv.org/abs/1811.02567}{{\ttfamily
  arXiv:1811.02567 [hep-ph]}}.

\bibitem{Duch:2019vjg}
M.~Duch, B.~Grzadkowski, and D.~Huang, ``{Strong Dark Matter Self-Interaction
  from a Stable Scalar Mediator},''
  \href{http://dx.doi.org/10.1007/JHEP03(2020)096}{{\em JHEP} {\bfseries 03}
  (2020) 096}, \href{http://arxiv.org/abs/1910.01238}{{\ttfamily
  arXiv:1910.01238 [hep-ph]}}.

\bibitem{Duch:2017nbe}
M.~Duch and B.~Grzadkowski, ``{Resonance enhancement of dark matter
  interactions: the case for early kinetic decoupling and velocity dependent
  resonance width},'' \href{http://dx.doi.org/10.1007/JHEP09(2017)159}{{\em
  JHEP} {\bfseries 09} (2017) 159},
  \href{http://arxiv.org/abs/1705.10777}{{\ttfamily arXiv:1705.10777
  [hep-ph]}}.

\bibitem{Chu:2018fzy}
X.~Chu, C.~Garcia-Cely, and H.~Murayama, ``{Velocity Dependence from Resonant
  Self-Interacting Dark Matter},''
  \href{http://dx.doi.org/10.1103/PhysRevLett.122.071103}{{\em Phys. Rev.
  Lett.} {\bfseries 122} no.~7, (2019) 071103},
  \href{http://arxiv.org/abs/1810.04709}{{\ttfamily arXiv:1810.04709
  [hep-ph]}}.

\bibitem{CyrRacine:2012fz}
F.-Y. Cyr-Racine and K.~Sigurdson, ``{Cosmology of atomic dark matter},''
  \href{http://dx.doi.org/10.1103/PhysRevD.87.103515}{{\em Phys. Rev. D}
  {\bfseries 87} no.~10, (2013) 103515},
  \href{http://arxiv.org/abs/1209.5752}{{\ttfamily arXiv:1209.5752
  [astro-ph.CO]}}.

\bibitem{Cline:2013pca}
J.~M. Cline, Z.~Liu, G.~Moore, and W.~Xue, ``{Scattering properties of dark
  atoms and molecules},''
  \href{http://dx.doi.org/10.1103/PhysRevD.89.043514}{{\em Phys. Rev. D}
  {\bfseries 89} no.~4, (2014) 043514},
  \href{http://arxiv.org/abs/1311.6468}{{\ttfamily arXiv:1311.6468 [hep-ph]}}.

\bibitem{Cline:2013zca}
J.~M. Cline, Z.~Liu, G.~Moore, and W.~Xue, ``{Composite strongly interacting
  dark matter},'' \href{http://dx.doi.org/10.1103/PhysRevD.90.015023}{{\em
  Phys. Rev. D} {\bfseries 90} no.~1, (2014) 015023},
  \href{http://arxiv.org/abs/1312.3325}{{\ttfamily arXiv:1312.3325 [hep-ph]}}.

\bibitem{Boddy:2014yra}
K.~K. Boddy, J.~L. Feng, M.~Kaplinghat, and T.~M.~P. Tait, ``{Self-Interacting
  Dark Matter from a Non-Abelian Hidden Sector},''
  \href{http://dx.doi.org/10.1103/PhysRevD.89.115017}{{\em Phys. Rev. D}
  {\bfseries 89} no.~11, (2014) 115017},
  \href{http://arxiv.org/abs/1402.3629}{{\ttfamily arXiv:1402.3629 [hep-ph]}}.

\bibitem{Boddy:2014qxa}
K.~K. Boddy, J.~L. Feng, M.~Kaplinghat, Y.~Shadmi, and T.~M.~P. Tait,
  ``{Strongly interacting dark matter: Self-interactions and keV lines},''
  \href{http://dx.doi.org/10.1103/PhysRevD.90.095016}{{\em Phys. Rev. D}
  {\bfseries 90} no.~9, (2014) 095016},
  \href{http://arxiv.org/abs/1408.6532}{{\ttfamily arXiv:1408.6532 [hep-ph]}}.

\bibitem{Boddy:2016bbu}
K.~K. Boddy, M.~Kaplinghat, A.~Kwa, and A.~H.~G. Peter, ``{Hidden Sector
  Hydrogen as Dark Matter: Small-scale Structure Formation Predictions and the
  Importance of Hyperfine Interactions},''
  \href{http://dx.doi.org/10.1103/PhysRevD.94.123017}{{\em Phys. Rev. D}
  {\bfseries 94} no.~12, (2016) 123017},
  \href{http://arxiv.org/abs/1609.03592}{{\ttfamily arXiv:1609.03592
  [hep-ph]}}.

\bibitem{Ibe:2018juk}
M.~Ibe, A.~Kamada, S.~Kobayashi, and W.~Nakano, ``{Composite Asymmetric Dark
  Matter with a Dark Photon Portal},''
  \href{http://dx.doi.org/10.1007/JHEP11(2018)203}{{\em JHEP} {\bfseries 11}
  (2018) 203}, \href{http://arxiv.org/abs/1805.06876}{{\ttfamily
  arXiv:1805.06876 [hep-ph]}}.

\bibitem{Chu:2018faw}
X.~Chu, C.~Garcia-Cely, and H.~Murayama, ``{Finite-size dark matter and its
  effect on small-scale structure},''
  \href{http://dx.doi.org/10.1103/PhysRevLett.124.041101}{{\em Phys. Rev.
  Lett.} {\bfseries 124} no.~4, (2020) 041101},
  \href{http://arxiv.org/abs/1901.00075}{{\ttfamily arXiv:1901.00075
  [hep-ph]}}.

\bibitem{Loeb:2010gj}
A.~Loeb and N.~Weiner, ``{Cores in Dwarf Galaxies from Dark Matter with a
  Yukawa Potential},''
  \href{http://dx.doi.org/10.1103/PhysRevLett.106.171302}{{\em Phys. Rev.
  Lett.} {\bfseries 106} (2011) 171302},
\href{http://arxiv.org/abs/1011.6374}{{\ttfamily arXiv:1011.6374
  [astro-ph.CO]}}.

\bibitem{Schutz:2014nka}
K.~Schutz and T.~R. Slatyer, ``{Self-Scattering for Dark Matter with an Excited
  State},'' \href{http://dx.doi.org/10.1088/1475-7516/2015/01/021}{{\em JCAP}
  {\bfseries 1501} no.~01, (2015) 021},
\href{http://arxiv.org/abs/1409.2867}{{\ttfamily arXiv:1409.2867 [hep-ph]}}.

\bibitem{McDermott:2017vyk}
S.~D. McDermott, ``{Is Self-Interacting Dark Matter Undergoing Dark Fusion?},''
  \href{http://dx.doi.org/10.1103/PhysRevLett.120.221806}{{\em Phys. Rev.
  Lett.} {\bfseries 120} no.~22, (2018) 221806},
\href{http://arxiv.org/abs/1711.00857}{{\ttfamily arXiv:1711.00857 [hep-ph]}}.

\bibitem{Vogelsberger:2018bok}
M.~Vogelsberger, J.~Zavala, K.~Schutz, and T.~R. Slatyer, ``{Evaporating the
  Milky Way halo and its satellites with inelastic self-interacting dark
  matter},''
\href{http://arxiv.org/abs/1805.03203}{{\ttfamily arXiv:1805.03203
  [astro-ph.GA]}}.

\bibitem{Kamada:2019wjo}
A.~Kamada and H.~J. Kim, ``{Escalating core formation with dark matter
  self-heating},'' \href{http://dx.doi.org/10.1103/PhysRevD.102.043009}{{\em
  Phys. Rev. D} {\bfseries 102} no.~4, (2020) 043009},
  \href{http://arxiv.org/abs/1911.09717}{{\ttfamily arXiv:1911.09717
  [hep-ph]}}.

\bibitem{Gilmore:2007fy}
G.~Gilmore, M.~I. Wilkinson, R.~F.~G. Wyse, J.~T. Kleyna, A.~Koch, N.~W. Evans,
  and E.~K. Grebel, ``{The Observed properties of Dark Matter on small spatial
  scales},'' \href{http://dx.doi.org/10.1086/518025}{{\em Astrophys. J.}
  {\bfseries 663} (2007) 948--959},
\href{http://arxiv.org/abs/astro-ph/0703308}{{\ttfamily arXiv:astro-ph/0703308
  [ASTRO-PH]}}.

\bibitem{Valli:2017ktb}
M.~Valli and H.-B. Yu, ``{Dark matter self-interactions from the internal
  dynamics of dwarf spheroidals},''
  \href{http://dx.doi.org/10.1038/s41550-018-0560-7}{{\em Nature Astron.}
  {\bfseries 2} (2018) 907--912},
  \href{http://arxiv.org/abs/1711.03502}{{\ttfamily arXiv:1711.03502
  [astro-ph.GA]}}.

\bibitem{Read:2018pft}
J.~Read, M.~Walker, and P.~Steger, ``{The case for a cold dark matter cusp in
  Draco},'' \href{http://dx.doi.org/10.1093/mnras/sty2286}{{\em Mon. Not. Roy.
  Astron. Soc.} {\bfseries 481} no.~1, (2018) 860--877},
  \href{http://arxiv.org/abs/1805.06934}{{\ttfamily arXiv:1805.06934
  [astro-ph.GA]}}.

\bibitem{Hayashi:2020syu}
K.~Hayashi, M.~Ibe, S.~Kobayashi, Y.~Nakayama, and S.~Shirai, ``{Probing Dark
  Matter Self-interaction with Ultra-faint Dwarf Galaxies},''
  \href{http://arxiv.org/abs/2008.02529}{{\ttfamily arXiv:2008.02529
  [astro-ph.CO]}}.

\bibitem{Balberg:2001qg}
S.~Balberg and S.~L. Shapiro, ``{Gravothermal collapse of selfinteracting dark
  matter halos and the origin of massive black holes},''
  \href{http://dx.doi.org/10.1103/PhysRevLett.88.101301}{{\em Phys. Rev. Lett.}
  {\bfseries 88} (2002) 101301},
\href{http://arxiv.org/abs/astro-ph/0111176}{{\ttfamily arXiv:astro-ph/0111176
  [astro-ph]}}.

\bibitem{Nishikawa:2019lsc}
H.~Nishikawa, K.~K. Boddy, and M.~Kaplinghat, ``{Accelerated core collapse in
  tidally stripped self-interacting dark matter halos},''
  \href{http://dx.doi.org/10.1103/PhysRevD.101.063009}{{\em Phys. Rev. D}
  {\bfseries 101} no.~6, (2020) 063009},
  \href{http://arxiv.org/abs/1901.00499}{{\ttfamily arXiv:1901.00499
  [astro-ph.GA]}}.

\bibitem{Kaplinghat:2019svz}
M.~Kaplinghat, M.~Valli, and H.-B. Yu, ``{Too Big To Fail in Light of Gaia},''
  \href{http://dx.doi.org/10.1093/mnras/stz2511}{{\em Mon. Not. Roy. Astron.
  Soc.} {\bfseries 490} no.~1, (2019) 231--242},
  \href{http://arxiv.org/abs/1904.04939}{{\ttfamily arXiv:1904.04939
  [astro-ph.GA]}}.

\bibitem{Correa:2020qam}
C.~A. Correa, ``{Constraining Velocity-dependent Self-Interacting Dark Matter
  with the Milky Way's dwarf spheroidal galaxies},''
  \href{http://arxiv.org/abs/2007.02958}{{\ttfamily arXiv:2007.02958
  [astro-ph.GA]}}.

\bibitem{Sameie:2019zfo}
O.~Sameie, H.-B. Yu, L.~V. Sales, M.~Vogelsberger, and J.~Zavala,
  ``{Self-Interacting Dark Matter Subhalos in the Milky Way\textquoteright{}s
  Tides},'' \href{http://dx.doi.org/10.1103/PhysRevLett.124.141102}{{\em Phys.
  Rev. Lett.} {\bfseries 124} no.~14, (2020) 141102},
  \href{http://arxiv.org/abs/1904.07872}{{\ttfamily arXiv:1904.07872
  [astro-ph.GA]}}.

\bibitem{Blatt:1949zz}
J.~M. Blatt and J.~Jackson, ``{On the Interpretation of Neutron-Proton
  Scattering Data by the Schwinger Variational Method},''
  \href{http://dx.doi.org/10.1103/PhysRev.76.18}{{\em Phys. Rev.} {\bfseries
  76} (1949) 18--37}.

\bibitem{Bethe:1949yr}
H.~Bethe, ``{Theory of the Effective Range in Nuclear Scattering},''
  \href{http://dx.doi.org/10.1103/PhysRev.76.38}{{\em Phys. Rev.} {\bfseries
  76} (1949) 38--50}.

\bibitem{Kaplan:2005es}
D.~B. Kaplan, ``{Five lectures on effective field theory},''
\newblock 10, 2005.
\newblock \href{http://arxiv.org/abs/nucl-th/0510023}{{\ttfamily
  arXiv:nucl-th/0510023}}.

\bibitem{Chu:2019awd}
X.~Chu, C.~Garcia-Cely, and H.~Murayama, ``{A Practical and Consistent
  Parametrization of Dark Matter Self-Interactions},''
  \href{http://dx.doi.org/10.1088/1475-7516/2020/06/043}{{\em JCAP} {\bfseries
  06} (2020) 043}, \href{http://arxiv.org/abs/1908.06067}{{\ttfamily
  arXiv:1908.06067 [hep-ph]}}.

\bibitem{Kahlhoefer:2013dca}
F.~Kahlhoefer, K.~Schmidt-Hoberg, M.~T. Frandsen, and S.~Sarkar, ``{Colliding
  clusters and dark matter self-interactions},''
  \href{http://dx.doi.org/10.1093/mnras/stt2097}{{\em Mon. Not. Roy. Astron.
  Soc.} {\bfseries 437} no.~3, (2014) 2865--2881},
  \href{http://arxiv.org/abs/1308.3419}{{\ttfamily arXiv:1308.3419
  [astro-ph.CO]}}.

\bibitem{Ahn:2004xt}
K.-J. Ahn and P.~R. Shapiro, ``{Formation and evolution of self-interacting
  dark matter haloes},''
  \href{http://dx.doi.org/10.1111/j.1365-2966.2005.09492.x}{{\em Mon. Not. Roy.
  Astron. Soc.} {\bfseries 363} (2005) 1092--1124},
  \href{http://arxiv.org/abs/astro-ph/0412169}{{\ttfamily
  arXiv:astro-ph/0412169}}.

\bibitem{Cassel:2009wt}
S.~Cassel, ``{Sommerfeld factor for arbitrary partial wave processes},''
  \href{http://dx.doi.org/10.1088/0954-3899/37/10/105009}{{\em J. Phys. G}
  {\bfseries 37} (2010) 105009},
  \href{http://arxiv.org/abs/0903.5307}{{\ttfamily arXiv:0903.5307 [hep-ph]}}.

\bibitem{Hisano:2002fk}
J.~Hisano, S.~Matsumoto, and M.~M. Nojiri, ``{Unitarity and higher order
  corrections in neutralino dark matter annihilation into two photons},''
  \href{http://dx.doi.org/10.1103/PhysRevD.67.075014}{{\em Phys. Rev.}
  {\bfseries D67} (2003) 075014},
\href{http://arxiv.org/abs/hep-ph/0212022}{{\ttfamily arXiv:hep-ph/0212022
  [hep-ph]}}.

\bibitem{Hisano:2003ec}
J.~Hisano, S.~Matsumoto, and M.~M. Nojiri, ``{Explosive dark matter
  annihilation},'' \href{http://dx.doi.org/10.1103/PhysRevLett.92.031303}{{\em
  Phys. Rev. Lett.} {\bfseries 92} (2004) 031303},
\href{http://arxiv.org/abs/hep-ph/0307216}{{\ttfamily arXiv:hep-ph/0307216
  [hep-ph]}}.

\bibitem{Hisano:2004ds}
J.~Hisano, S.~Matsumoto, M.~M. Nojiri, and O.~Saito, ``{Non-perturbative effect
  on dark matter annihilation and gamma ray signature from the galactic
  center},'' \href{http://dx.doi.org/10.1103/PhysRevD.71.063528}{{\em Phys.
  Rev.} {\bfseries D71} (2005) 063528},
\href{http://arxiv.org/abs/hep-ph/0412403}{{\ttfamily arXiv:hep-ph/0412403
  [hep-ph]}}.

\bibitem{Hisano:2005ec}
J.~Hisano, S.~Matsumoto, O.~Saito, and M.~Senami, ``{Heavy wino-like neutralino
  dark matter annihilation into antiparticles},''
  \href{http://dx.doi.org/10.1103/PhysRevD.73.055004}{{\em Phys. Rev.}
  {\bfseries D73} (2006) 055004},
\href{http://arxiv.org/abs/hep-ph/0511118}{{\ttfamily arXiv:hep-ph/0511118
  [hep-ph]}}.

\bibitem{Cirelli:2007xd}
M.~Cirelli, A.~Strumia, and M.~Tamburini, ``{Cosmology and Astrophysics of
  Minimal Dark Matter},''
  \href{http://dx.doi.org/10.1016/j.nuclphysb.2007.07.023}{{\em Nucl. Phys. B}
  {\bfseries 787} (2007) 152--175},
  \href{http://arxiv.org/abs/0706.4071}{{\ttfamily arXiv:0706.4071 [hep-ph]}}.

\bibitem{Cirelli:2008pk}
M.~Cirelli, M.~Kadastik, M.~Raidal, and A.~Strumia, ``{Model-independent
  implications of the e+-, ${\bar p}$ cosmic ray spectra on properties of Dark
  Matter},'' \href{http://dx.doi.org/10.1016/j.nuclphysb.2008.11.031}{{\em
  Nucl. Phys. B} {\bfseries 813} (2009) 1--21},
  \href{http://arxiv.org/abs/0809.2409}{{\ttfamily arXiv:0809.2409 [hep-ph]}}.
  [Addendum: Nucl.Phys.B 873, 530--533 (2013)].

\bibitem{ArkaniHamed:2008qn}
N.~Arkani-Hamed, D.~P. Finkbeiner, T.~R. Slatyer, and N.~Weiner, ``{A Theory of
  Dark Matter},'' \href{http://dx.doi.org/10.1103/PhysRevD.79.015014}{{\em
  Phys. Rev. D} {\bfseries 79} (2009) 015014},
  \href{http://arxiv.org/abs/0810.0713}{{\ttfamily arXiv:0810.0713 [hep-ph]}}.

\bibitem{Feng:2010zp}
J.~L. Feng, M.~Kaplinghat, and H.-B. Yu, ``{Sommerfeld Enhancements for Thermal
  Relic Dark Matter},''
  \href{http://dx.doi.org/10.1103/PhysRevD.82.083525}{{\em Phys. Rev. D}
  {\bfseries 82} (2010) 083525},
  \href{http://arxiv.org/abs/1005.4678}{{\ttfamily arXiv:1005.4678 [hep-ph]}}.

\bibitem{Bringmann:2016din}
T.~Bringmann, F.~Kahlhoefer, K.~Schmidt-Hoberg, and P.~Walia, ``{Strong
  constraints on self-interacting dark matter with light mediators},''
  \href{http://dx.doi.org/10.1103/PhysRevLett.118.141802}{{\em Phys. Rev.
  Lett.} {\bfseries 118} no.~14, (2017) 141802},
  \href{http://arxiv.org/abs/1612.00845}{{\ttfamily arXiv:1612.00845
  [hep-ph]}}.

\bibitem{Dent:2009bv}
J.~B. Dent, S.~Dutta, and R.~J. Scherrer, ``{Thermal Relic Abundances of
  Particles with Velocity-Dependent Interactions},''
  \href{http://dx.doi.org/10.1016/j.physletb.2010.03.018}{{\em Phys. Lett. B}
  {\bfseries 687} (2010) 275--279},
  \href{http://arxiv.org/abs/0909.4128}{{\ttfamily arXiv:0909.4128
  [astro-ph.CO]}}.

\bibitem{Zavala:2009mi}
J.~Zavala, M.~Vogelsberger, and S.~D. White, ``{Relic density and CMB
  constraints on dark matter annihilation with Sommerfeld enhancement},''
  \href{http://dx.doi.org/10.1103/PhysRevD.81.083502}{{\em Phys. Rev. D}
  {\bfseries 81} (2010) 083502},
  \href{http://arxiv.org/abs/0910.5221}{{\ttfamily arXiv:0910.5221
  [astro-ph.CO]}}.

\bibitem{vandenAarssen:2012ag}
L.~G. van~den Aarssen, T.~Bringmann, and Y.~C. Goedecke, ``{Thermal decoupling
  and the smallest subhalo mass in dark matter models with Sommerfeld-enhanced
  annihilation rates},''
  \href{http://dx.doi.org/10.1103/PhysRevD.85.123512}{{\em Phys. Rev. D}
  {\bfseries 85} (2012) 123512},
  \href{http://arxiv.org/abs/1202.5456}{{\ttfamily arXiv:1202.5456 [hep-ph]}}.

\bibitem{Binder:2017lkj}
T.~Binder, M.~Gustafsson, A.~Kamada, S.~M.~R. Sandner, and M.~Wiesner,
  ``{Reannihilation of self-interacting dark matter},''
  \href{http://dx.doi.org/10.1103/PhysRevD.97.123004}{{\em Phys. Rev. D}
  {\bfseries 97} no.~12, (2018) 123004},
  \href{http://arxiv.org/abs/1712.01246}{{\ttfamily arXiv:1712.01246
  [astro-ph.CO]}}.

\bibitem{He:1990pn}
X.~He, G.~C. Joshi, H.~Lew, and R.~Volkas, ``{NEW Z-prime PHENOMENOLOGY},''
  \href{http://dx.doi.org/10.1103/PhysRevD.43.R22}{{\em Phys. Rev. D}
  {\bfseries 43} (1991) 22--24}.

\bibitem{He:1991qd}
X.-G. He, G.~C. Joshi, H.~Lew, and R.~Volkas, ``{Simplest Z-prime model},''
  \href{http://dx.doi.org/10.1103/PhysRevD.44.2118}{{\em Phys. Rev. D}
  {\bfseries 44} (1991) 2118--2132}.

\bibitem{Davier:2010nc}
M.~Davier, A.~Hoecker, B.~Malaescu, and Z.~Zhang, ``{Reevaluation of the
  Hadronic Contributions to the Muon g-2 and to alpha(MZ)},''
  \href{http://dx.doi.org/10.1140/epjc/s10052-012-1874-8}{{\em Eur. Phys. J. C}
  {\bfseries 71} (2011) 1515}, \href{http://arxiv.org/abs/1010.4180}{{\ttfamily
  arXiv:1010.4180 [hep-ph]}}. [Erratum: Eur.Phys.J.C 72, 1874 (2012)].

\bibitem{Hagiwara:2011af}
K.~Hagiwara, R.~Liao, A.~D. Martin, D.~Nomura, and T.~Teubner, ``{(g-2)$\_mu$
  and alpha(M$\_Z^2$) re-evaluated using new precise data},''
  \href{http://dx.doi.org/10.1088/0954-3899/38/8/085003}{{\em J. Phys. G}
  {\bfseries 38} (2011) 085003},
  \href{http://arxiv.org/abs/1105.3149}{{\ttfamily arXiv:1105.3149 [hep-ph]}}.

\bibitem{Keshavarzi:2019abf}
A.~Keshavarzi, D.~Nomura, and T.~Teubner, ``{$g-2$ of charged leptons, $\alpha
  (M^2_Z)$ , and the hyperfine splitting of muonium},''
  \href{http://dx.doi.org/10.1103/PhysRevD.101.014029}{{\em Phys. Rev. D}
  {\bfseries 101} no.~1, (2020) 014029},
  \href{http://arxiv.org/abs/1911.00367}{{\ttfamily arXiv:1911.00367
  [hep-ph]}}.

\bibitem{Aoyama:2020ynm}
T.~Aoyama {\em et~al.}, ``{The anomalous magnetic moment of the muon in the
  Standard Model},'' \href{http://arxiv.org/abs/2006.04822}{{\ttfamily
  arXiv:2006.04822 [hep-ph]}}.

\bibitem{Grange:2015fou}
{\bfseries E989} Collaboration, J.~Grange {\em et~al.}, ``{Muon (g-2) Technical
  Design Report},'' \href{http://arxiv.org/abs/1501.06858}{{\ttfamily
  arXiv:1501.06858 [physics.ins-det]}}.

\bibitem{Abe:2019thb}
M.~Abe {\em et~al.}, ``{A New Approach for Measuring the Muon Anomalous
  Magnetic Moment and Electric Dipole Moment},''
  \href{http://dx.doi.org/10.1093/ptep/ptz030}{{\em PTEP} {\bfseries 2019}
  no.~5, (2019) 053C02}, \href{http://arxiv.org/abs/1901.03047}{{\ttfamily
  arXiv:1901.03047 [physics.ins-det]}}.

\bibitem{Borsanyi:2020mff}
S.~Borsanyi {\em et~al.}, ``{Leading hadronic contribution to the muon magnetic
  moment from lattice QCD},'' \href{http://arxiv.org/abs/2002.12347}{{\ttfamily
  arXiv:2002.12347 [hep-lat]}}.

\bibitem{Minkowski:1977sc}
P.~Minkowski, ``{$\mu \to e\gamma$ at a Rate of One Out of $10^{9}$ Muon
  Decays?},''
\href{http://dx.doi.org/10.1016/0370-2693(77)90435-X}{{\em Phys. Lett.}
  {\bfseries 67B} (1977) 421--428}.

\bibitem{Yanagida:1979as}
T.~Yanagida, ``{HORIZONTAL SYMMETRY AND MASSES OF NEUTRINOS},''
{\em Conf. Proc.} {\bfseries C7902131} (1979) 95--99.

\bibitem{Glashow:1979nm}
S.~Glashow, ``{The Future of Elementary Particle Physics},''
  \href{http://dx.doi.org/10.1007/978-1-4684-7197-7\_15}{{\em NATO Sci. Ser. B}
  {\bfseries 61} (1980) 687}.

\bibitem{GellMann:1980vs}
M.~Gell-Mann, P.~Ramond, and R.~Slansky, ``{Complex Spinors and Unified
  Theories},'' {\em Conf. Proc.} {\bfseries C790927} (1979) 315--321,
\href{http://arxiv.org/abs/1306.4669}{{\ttfamily arXiv:1306.4669 [hep-th]}}.

\bibitem{Asai:2017ryy}
K.~Asai, K.~Hamaguchi, and N.~Nagata, ``{Predictions for the neutrino
  parameters in the minimal gauged U(1)$_{L_\mu-L_\tau}$ model},''
  \href{http://dx.doi.org/10.1140/epjc/s10052-017-5348-x}{{\em Eur. Phys. J.}
  {\bfseries C77} no.~11, (2017) 763},
\href{http://arxiv.org/abs/1705.00419}{{\ttfamily arXiv:1705.00419 [hep-ph]}}.

\bibitem{Asai:2020qax}
K.~Asai, K.~Hamaguchi, N.~Nagata, and S.-Y. Tseng, ``{Leptogenesis in the
  minimal gauged U(1)$_{L_\mu-L_\tau}$ model and the sign of the cosmological
  baryon asymmetry},'' \href{http://arxiv.org/abs/2005.01039}{{\ttfamily
  arXiv:2005.01039 [hep-ph]}}.

\bibitem{Fukugita:1986hr}
M.~Fukugita and T.~Yanagida, ``{Baryogenesis Without Grand Unification},''
\href{http://dx.doi.org/10.1016/0370-2693(86)91126-3}{{\em Phys. Lett.}
  {\bfseries B174} (1986) 45--47}.

\bibitem{Giudice:2003jh}
G.~Giudice, A.~Notari, M.~Raidal, A.~Riotto, and A.~Strumia, ``{Towards a
  complete theory of thermal leptogenesis in the SM and MSSM},''
  \href{http://dx.doi.org/10.1016/j.nuclphysb.2004.02.019}{{\em Nucl. Phys. B}
  {\bfseries 685} (2004) 89--149},
  \href{http://arxiv.org/abs/hep-ph/0310123}{{\ttfamily arXiv:hep-ph/0310123}}.

\bibitem{Buchmuller:2005eh}
W.~Buchmuller, R.~D. Peccei, and T.~Yanagida, ``{Leptogenesis as the origin of
  matter},''
  \href{http://dx.doi.org/10.1146/annurev.nucl.55.090704.151558}{{\em Ann. Rev.
  Nucl. Part. Sci.} {\bfseries 55} (2005) 311--355},
\href{http://arxiv.org/abs/hep-ph/0502169}{{\ttfamily arXiv:hep-ph/0502169
  [hep-ph]}}.

\bibitem{Davidson:2008bu}
S.~Davidson, E.~Nardi, and Y.~Nir, ``{Leptogenesis},''
  \href{http://dx.doi.org/10.1016/j.physrep.2008.06.002}{{\em Phys. Rept.}
  {\bfseries 466} (2008) 105--177},
  \href{http://arxiv.org/abs/0802.2962}{{\ttfamily arXiv:0802.2962 [hep-ph]}}.

\bibitem{Griest:1990kh}
K.~Griest and D.~Seckel, ``{Three exceptions in the calculation of relic
  abundances},'' \href{http://dx.doi.org/10.1103/PhysRevD.43.3191}{{\em Phys.
  Rev. D} {\bfseries 43} (1991) 3191--3203}.

\bibitem{Arguelles:2019ouk}
C.~A. Argüelles, A.~Diaz, A.~Kheirandish, A.~Olivares-Del-Campo, I.~Safa, and
  A.~C. Vincent, ``{Dark Matter Annihilation to Neutrinos: An Updated,
  Consistent \& Compelling Compendium of Constraints},''
  \href{http://arxiv.org/abs/1912.09486}{{\ttfamily arXiv:1912.09486
  [hep-ph]}}.

\bibitem{Aghanim:2018eyx}
{\bfseries Planck} Collaboration, N.~Aghanim {\em et~al.}, ``{Planck 2018
  results. VI. Cosmological parameters},''
  \href{http://arxiv.org/abs/1807.06209}{{\ttfamily arXiv:1807.06209
  [astro-ph.CO]}}.

\bibitem{Kamada:2015era}
A.~Kamada and H.-B. Yu, ``{Coherent Propagation of PeV Neutrinos and the Dip in
  the Neutrino Spectrum at IceCube},''
  \href{http://dx.doi.org/10.1103/PhysRevD.92.113004}{{\em Phys. Rev. D}
  {\bfseries 92} no.~11, (2015) 113004},
  \href{http://arxiv.org/abs/1504.00711}{{\ttfamily arXiv:1504.00711
  [hep-ph]}}.

\bibitem{Riess:2016jrr}
A.~G. Riess {\em et~al.}, ``{A 2.4 per cent Determination of the Local Value of
  the Hubble Constant},''
  \href{http://dx.doi.org/10.3847/0004-637X/826/1/56}{{\em Astrophys. J.}
  {\bfseries 826} no.~1, (2016) 56},
  \href{http://arxiv.org/abs/1604.01424}{{\ttfamily arXiv:1604.01424
  [astro-ph.CO]}}.

\bibitem{Riess:2018byc}
A.~G. Riess {\em et~al.}, ``{Milky Way Cepheid Standards for Measuring Cosmic
  Distances and Application to Gaia DR2: Implications for the Hubble
  Constant},'' \href{http://dx.doi.org/10.3847/1538-4357/aac82e}{{\em
  Astrophys. J.} {\bfseries 861} no.~2, (2018) 126},
  \href{http://arxiv.org/abs/1804.10655}{{\ttfamily arXiv:1804.10655
  [astro-ph.CO]}}.

\bibitem{Wong:2019kwg}
K.~C. Wong {\em et~al.}, ``{H0LiCOW XIII. A 2.4\% measurement of $H_{0}$ from
  lensed quasars: $5.3\sigma$ tension between early and late-Universe
  probes},'' \href{http://arxiv.org/abs/1907.04869}{{\ttfamily arXiv:1907.04869
  [astro-ph.CO]}}.

\bibitem{Aiola:2020azj}
S.~Aiola {\em et~al.}, ``{The Atacama Cosmology Telescope: DR4 Maps and
  Cosmological Parameters},'' \href{http://arxiv.org/abs/2007.07288}{{\ttfamily
  arXiv:2007.07288 [astro-ph.CO]}}.

\bibitem{Vagnozzi:2019ezj}
S.~Vagnozzi, ``{New physics in light of the $H_0$ tension: An alternative
  view},'' \href{http://dx.doi.org/10.1103/PhysRevD.102.023518}{{\em Phys. Rev.
  D} {\bfseries 102} no.~2, (2020) 023518},
  \href{http://arxiv.org/abs/1907.07569}{{\ttfamily arXiv:1907.07569
  [astro-ph.CO]}}.

\bibitem{Arendse:2019hev}
N.~Arendse {\em et~al.}, ``{Cosmic dissonance: new physics or systematics
  behind a short sound horizon?},''
  \href{http://dx.doi.org/10.1051/0004-6361/201936720}{{\em Astron. Astrophys.}
  {\bfseries 639} (2020) A57},
  \href{http://arxiv.org/abs/1909.07986}{{\ttfamily arXiv:1909.07986
  [astro-ph.CO]}}.

\bibitem{Escudero:2019gzq}
M.~Escudero, D.~Hooper, G.~Krnjaic, and M.~Pierre, ``{Cosmology with A Very
  Light L$_{\mu}$ $-$ L$_{\tau}$ Gauge Boson},''
  \href{http://dx.doi.org/10.1007/JHEP03(2019)071}{{\em JHEP} {\bfseries 03}
  (2019) 071}, \href{http://arxiv.org/abs/1901.02010}{{\ttfamily
  arXiv:1901.02010 [hep-ph]}}.

\bibitem{Wu:2014hta}
W.~Wu, J.~Errard, C.~Dvorkin, C.~Kuo, A.~Lee, P.~McDonald, A.~Slosar, and
  O.~Zahn, ``{A Guide to Designing Future Ground-based Cosmic Microwave
  Background Experiments},''
  \href{http://dx.doi.org/10.1088/0004-637X/788/2/138}{{\em Astrophys. J.}
  {\bfseries 788} (2014) 138}, \href{http://arxiv.org/abs/1402.4108}{{\ttfamily
  arXiv:1402.4108 [astro-ph.CO]}}.

\bibitem{Abazajian:2016yjj}
{\bfseries CMB-S4} Collaboration, K.~N. Abazajian {\em et~al.}, ``{CMB-S4
  Science Book, First Edition},''
  \href{http://arxiv.org/abs/1610.02743}{{\ttfamily arXiv:1610.02743
  [astro-ph.CO]}}.

\bibitem{Amaral:2020tga}
d.~Amaral, Dorian Warren~Praia, D.~G. Cerdeno, P.~Foldenauer, and E.~Reid,
  ``{Solar neutrino probes of the muon anomalous magnetic moment in the gauged
  $U(1)_{L_\mu-L_\tau}$},'' \href{http://arxiv.org/abs/2006.11225}{{\ttfamily
  arXiv:2006.11225 [hep-ph]}}.

\bibitem{Sadhukhan:2020etu}
S.~Sadhukhan and M.~P. Singh, ``{Neutrino Floor in Leptophilic $U(1)$ Models:
  Modification in $U(1)_{L_{\mu}-L_{\tau}}$},''
  \href{http://arxiv.org/abs/2006.05981}{{\ttfamily arXiv:2006.05981
  [hep-ph]}}.

\bibitem{Araki:2014ona}
T.~Araki, F.~Kaneko, Y.~Konishi, T.~Ota, J.~Sato, and T.~Shimomura, ``{Cosmic
  neutrino spectrum and the muon anomalous magnetic moment in the gauged
  $L_{\mu}-L_{\tau}$ model},''
  \href{http://dx.doi.org/10.1103/PhysRevD.91.037301}{{\em Phys. Rev. D}
  {\bfseries 91} no.~3, (2015) 037301},
  \href{http://arxiv.org/abs/1409.4180}{{\ttfamily arXiv:1409.4180 [hep-ph]}}.

\bibitem{Araki:2015mya}
T.~Araki, F.~Kaneko, T.~Ota, J.~Sato, and T.~Shimomura, ``{MeV scale leptonic
  force for cosmic neutrino spectrum and muon anomalous magnetic moment},''
  \href{http://dx.doi.org/10.1103/PhysRevD.93.013014}{{\em Phys. Rev. D}
  {\bfseries 93} no.~1, (2016) 013014},
  \href{http://arxiv.org/abs/1508.07471}{{\ttfamily arXiv:1508.07471
  [hep-ph]}}.

\bibitem{Aartsen:2014gkd}
{\bfseries IceCube} Collaboration, M.~Aartsen {\em et~al.}, ``{Observation of
  High-Energy Astrophysical Neutrinos in Three Years of IceCube Data},''
  \href{http://dx.doi.org/10.1103/PhysRevLett.113.101101}{{\em Phys. Rev.
  Lett.} {\bfseries 113} (2014) 101101},
  \href{http://arxiv.org/abs/1405.5303}{{\ttfamily arXiv:1405.5303
  [astro-ph.HE]}}.

\bibitem{Aartsen:2017mau}
{\bfseries IceCube} Collaboration, M.~Aartsen {\em et~al.}, ``{The IceCube
  Neutrino Observatory - Contributions to ICRC 2017 Part II: Properties of the
  Atmospheric and Astrophysical Neutrino Flux},''
  \href{http://arxiv.org/abs/1710.01191}{{\ttfamily arXiv:1710.01191
  [astro-ph.HE]}}.

\bibitem{Aartsen:2020aqd}
{\bfseries IceCube} Collaboration, M.~Aartsen {\em et~al.}, ``{Characteristics
  of the diffuse astrophysical electron and tau neutrino flux with six years of
  IceCube high energy cascade data},''
  \href{http://dx.doi.org/10.1103/PhysRevLett.125.121104}{{\em Phys. Rev.
  Lett.} {\bfseries 125} no.~12, (2020) 121104},
  \href{http://arxiv.org/abs/2001.09520}{{\ttfamily arXiv:2001.09520
  [astro-ph.HE]}}.

\bibitem{Gninenko:2014pea}
S.~Gninenko, N.~Krasnikov, and V.~Matveev, ``{Muon g-2 and searches for a new
  leptophobic sub-GeV dark boson in a missing-energy experiment at CERN},''
  \href{http://dx.doi.org/10.1103/PhysRevD.91.095015}{{\em Phys. Rev. D}
  {\bfseries 91} (2015) 095015},
  \href{http://arxiv.org/abs/1412.1400}{{\ttfamily arXiv:1412.1400 [hep-ph]}}.

\bibitem{Gninenko:2018tlp}
S.~Gninenko and N.~Krasnikov, ``{Probing the muon $g\_\mu-2$ anomaly, $L\_{\mu}
  - L\_{\tau}$ gauge boson and Dark Matter in dark photon experiments},''
  \href{http://dx.doi.org/10.1016/j.physletb.2018.06.043}{{\em Phys. Lett. B}
  {\bfseries 783} (2018) 24--28},
  \href{http://arxiv.org/abs/1801.10448}{{\ttfamily arXiv:1801.10448
  [hep-ph]}}.

\bibitem{Ibe:2019gpv}
M.~Ibe, S.~Kobayashi, Y.~Nakayama, and S.~Shirai, ``{Cosmological constraint on
  dark photon from N$_{eff}$},''
  \href{http://dx.doi.org/10.1007/JHEP04(2020)009}{{\em JHEP} {\bfseries 04}
  (2020) 009}, \href{http://arxiv.org/abs/1912.12152}{{\ttfamily
  arXiv:1912.12152 [hep-ph]}}.

\bibitem{Ibe:2018tex}
M.~Ibe, A.~Kamada, S.~Kobayashi, T.~Kuwahara, and W.~Nakano, ``{Ultraviolet
  Completion of a Composite Asymmetric Dark Matter Model with a Dark Photon
  Portal},'' \href{http://dx.doi.org/10.1007/JHEP03(2019)173}{{\em JHEP}
  {\bfseries 03} (2019) 173}, \href{http://arxiv.org/abs/1811.10232}{{\ttfamily
  arXiv:1811.10232 [hep-ph]}}.

\bibitem{Ibe:2019ena}
M.~Ibe, A.~Kamada, S.~Kobayashi, T.~Kuwahara, and W.~Nakano, ``{Baryon-Dark
  Matter Coincidence in Mirrored Unification},''
  \href{http://dx.doi.org/10.1103/PhysRevD.100.075022}{{\em Phys. Rev. D}
  {\bfseries 100} no.~7, (2019) 075022},
  \href{http://arxiv.org/abs/1907.03404}{{\ttfamily arXiv:1907.03404
  [hep-ph]}}.

\bibitem{Ibe:2019yra}
M.~Ibe, S.~Kobayashi, R.~Nagai, and W.~Nakano, ``{Oscillating Composite
  Asymmetric Dark Matter},''
  \href{http://dx.doi.org/10.1007/JHEP01(2020)027}{{\em JHEP} {\bfseries 01}
  (2020) 027}, \href{http://arxiv.org/abs/1907.11464}{{\ttfamily
  arXiv:1907.11464 [hep-ph]}}.

\bibitem{Falkowski:2011xh}
A.~Falkowski, J.~T. Ruderman, and T.~Volansky, ``{Asymmetric Dark Matter from
  Leptogenesis},'' \href{http://dx.doi.org/10.1007/JHEP05(2011)106}{{\em JHEP}
  {\bfseries 05} (2011) 106}, \href{http://arxiv.org/abs/1101.4936}{{\ttfamily
  arXiv:1101.4936 [hep-ph]}}.

\bibitem{Ibe:2011hq}
M.~Ibe, S.~Matsumoto, and T.~T. Yanagida, ``{The GeV-scale dark matter with
  B--L asymmetry},''
  \href{http://dx.doi.org/10.1016/j.physletb.2012.01.032}{{\em Phys. Lett. B}
  {\bfseries 708} (2012) 112--118},
  \href{http://arxiv.org/abs/1110.5452}{{\ttfamily arXiv:1110.5452 [hep-ph]}}.

\bibitem{Fukuda:2014xqa}
H.~Fukuda, S.~Matsumoto, and S.~Mukhopadhyay, ``{Asymmetric dark matter in
  early Universe chemical equilibrium always leads to an antineutrino
  signal},'' \href{http://dx.doi.org/10.1103/PhysRevD.92.013008}{{\em Phys.
  Rev. D} {\bfseries 92} no.~1, (2015) 013008},
  \href{http://arxiv.org/abs/1411.4014}{{\ttfamily arXiv:1411.4014 [hep-ph]}}.

\bibitem{Harvey:1990qw}
J.~A. Harvey and M.~S. Turner, ``{Cosmological baryon and lepton number in the
  presence of electroweak fermion number violation},''
  \href{http://dx.doi.org/10.1103/PhysRevD.42.3344}{{\em Phys. Rev. D}
  {\bfseries 42} (1990) 3344--3349}.

\bibitem{Weinberg:2008zzc}
S.~Weinberg, {\em {Cosmology}}.
\newblock 9, 2008.

\bibitem{Chen:2010yt}
J.-W. Chen, T.-K. Lee, C.-P. Liu, and Y.-S. Liu, ``{Quark Mass Dependence of
  Two Nucleon Observables},''
  \href{http://dx.doi.org/10.1103/PhysRevC.86.054001}{{\em Phys. Rev. C}
  {\bfseries 86} (2012) 054001},
  \href{http://arxiv.org/abs/1012.0453}{{\ttfamily arXiv:1012.0453 [nucl-th]}}.

\bibitem{Soto:2011tb}
J.~Soto and J.~Tarr\'us, ``{On the quark mass dependence of nucleon-nucleon
  S-wave scattering lengths},''
  \href{http://dx.doi.org/10.1103/PhysRevC.85.044001}{{\em Phys. Rev. C}
  {\bfseries 85} (2012) 044001},
  \href{http://arxiv.org/abs/1112.4426}{{\ttfamily arXiv:1112.4426 [nucl-th]}}.

\bibitem{Epelbaum:2002gb}
E.~Epelbaum, U.-G. Meissner, and W.~Glockle, ``{Nuclear forces in the chiral
  limit},'' \href{http://dx.doi.org/10.1016/S0375-9474(02)01393-3}{{\em Nucl.
  Phys. A} {\bfseries 714} (2003) 535--574},
  \href{http://arxiv.org/abs/nucl-th/0207089}{{\ttfamily
  arXiv:nucl-th/0207089}}.

\bibitem{Yamazaki:2015asa}
T.~Yamazaki, K.-i. Ishikawa, Y.~Kuramashi, and A.~Ukawa, ``{Study of quark mass
  dependence of binding energy for light nuclei in 2+1 flavor lattice QCD},''
  \href{http://dx.doi.org/10.1103/PhysRevD.92.014501}{{\em Phys. Rev. D}
  {\bfseries 92} no.~1, (2015) 014501},
  \href{http://arxiv.org/abs/1502.04182}{{\ttfamily arXiv:1502.04182
  [hep-lat]}}.

\bibitem{Bai:2020yml}
Q.-Q. Bai, C.-X. Wang, Y.~Xiao, and L.-S. Geng, ``{Pion-mass dependence of the
  nucleon-nucleon interaction},''
  \href{http://dx.doi.org/10.1016/j.physletb.2020.135745}{{\em Phys. Lett.}
  {\bfseries B} (2020) 135745},
  \href{http://arxiv.org/abs/2007.01638}{{\ttfamily arXiv:2007.01638
  [nucl-th]}}.

\bibitem{Beane:2001bc}
S.~Beane, P.~F. Bedaque, M.~Savage, and U.~van Kolck, ``{Towards a perturbative
  theory of nuclear forces},''
  \href{http://dx.doi.org/10.1016/S0375-9474(01)01324-0}{{\em Nucl. Phys. A}
  {\bfseries 700} (2002) 377--402},
  \href{http://arxiv.org/abs/nucl-th/0104030}{{\ttfamily
  arXiv:nucl-th/0104030}}.

\bibitem{Witten:1979kh}
E.~Witten, ``{Baryons in the 1/n Expansion},''
  \href{http://dx.doi.org/10.1016/0550-3213(79)90232-3}{{\em Nucl. Phys. B}
  {\bfseries 160} (1979) 57--115}.

\bibitem{Krnjaic:2014xza}
G.~Krnjaic and K.~Sigurdson, ``{Big Bang Darkleosynthesis},''
  \href{http://dx.doi.org/10.1016/j.physletb.2015.11.001}{{\em Phys. Lett.}
  {\bfseries B751} (2015) 464--468},
\href{http://arxiv.org/abs/1406.1171}{{\ttfamily arXiv:1406.1171 [hep-ph]}}.

\bibitem{Detmold:2014qqa}
W.~Detmold, M.~McCullough, and A.~Pochinsky, ``{Dark Nuclei I: Cosmology and
  Indirect Detection},''
  \href{http://dx.doi.org/10.1103/PhysRevD.90.115013}{{\em Phys. Rev.}
  {\bfseries D90} no.~11, (2014) 115013},
\href{http://arxiv.org/abs/1406.2276}{{\ttfamily arXiv:1406.2276 [hep-ph]}}.

\bibitem{Detmold:2014kba}
W.~Detmold, M.~McCullough, and A.~Pochinsky, ``{Dark nuclei. II. Nuclear
  spectroscopy in two-color QCD},''
  \href{http://dx.doi.org/10.1103/PhysRevD.90.114506}{{\em Phys. Rev.}
  {\bfseries D90} no.~11, (2014) 114506},
\href{http://arxiv.org/abs/1406.4116}{{\ttfamily arXiv:1406.4116 [hep-lat]}}.

\bibitem{Wise:2014ola}
M.~B. Wise and Y.~Zhang, ``{Yukawa Bound States of a Large Number of
  Fermions},'' \href{http://dx.doi.org/10.1007/JHEP10(2015)165,
  10.1007/JHEP02(2015)023}{{\em JHEP} {\bfseries 02} (2015) 023},
  \href{http://arxiv.org/abs/1411.1772}{{\ttfamily arXiv:1411.1772 [hep-ph]}}.
[Erratum: JHEP10,165(2015)].

\bibitem{Hardy:2014mqa}
E.~Hardy, R.~Lasenby, J.~March-Russell, and S.~M. West, ``{Big Bang Synthesis
  of Nuclear Dark Matter},''
  \href{http://dx.doi.org/10.1007/JHEP06(2015)011}{{\em JHEP} {\bfseries 06}
  (2015) 011},
\href{http://arxiv.org/abs/1411.3739}{{\ttfamily arXiv:1411.3739 [hep-ph]}}.

\bibitem{Gresham:2017zqi}
M.~I. Gresham, H.~K. Lou, and K.~M. Zurek, ``{Nuclear Structure of Bound States
  of Asymmetric Dark Matter},''
  \href{http://dx.doi.org/10.1103/PhysRevD.96.096012}{{\em Phys. Rev.}
  {\bfseries D96} no.~9, (2017) 096012},
\href{http://arxiv.org/abs/1707.02313}{{\ttfamily arXiv:1707.02313 [hep-ph]}}.

\bibitem{Gresham:2017cvl}
M.~I. Gresham, H.~K. Lou, and K.~M. Zurek, ``{Early Universe synthesis of
  asymmetric dark matter nuggets},''
  \href{http://dx.doi.org/10.1103/PhysRevD.97.036003}{{\em Phys. Rev.}
  {\bfseries D97} no.~3, (2018) 036003},
\href{http://arxiv.org/abs/1707.02316}{{\ttfamily arXiv:1707.02316 [hep-ph]}}.

\bibitem{Mahbubani:2019pij}
R.~Mahbubani, M.~Redi, and A.~Tesi, ``{Indirect detection of composite
  asymmetric dark matter},''
  \href{http://dx.doi.org/10.1103/PhysRevD.101.103037}{{\em Phys. Rev. D}
  {\bfseries 101} no.~10, (2020) 103037},
  \href{http://arxiv.org/abs/1908.00538}{{\ttfamily arXiv:1908.00538
  [hep-ph]}}.

\bibitem{Schwaller:2015tja}
P.~Schwaller, ``{Gravitational Waves from a Dark Phase Transition},''
  \href{http://dx.doi.org/10.1103/PhysRevLett.115.181101}{{\em Phys. Rev.
  Lett.} {\bfseries 115} no.~18, (2015) 181101},
  \href{http://arxiv.org/abs/1504.07263}{{\ttfamily arXiv:1504.07263
  [hep-ph]}}.

\bibitem{Bai:2018dxf}
Y.~Bai, A.~J. Long, and S.~Lu, ``{Dark Quark Nuggets},''
  \href{http://dx.doi.org/10.1103/PhysRevD.99.055047}{{\em Phys. Rev. D}
  {\bfseries 99} no.~5, (2019) 055047},
  \href{http://arxiv.org/abs/1810.04360}{{\ttfamily arXiv:1810.04360
  [hep-ph]}}.

\bibitem{Chang:2016ntp}
J.~H. Chang, R.~Essig, and S.~D. McDermott, ``{Revisiting Supernova 1987A
  Constraints on Dark Photons},''
  \href{http://dx.doi.org/10.1007/JHEP01(2017)107}{{\em JHEP} {\bfseries 01}
  (2017) 107},
\href{http://arxiv.org/abs/1611.03864}{{\ttfamily arXiv:1611.03864 [hep-ph]}}.

\bibitem{Chang:2018rso}
J.~H. Chang, R.~Essig, and S.~D. McDermott, ``{Supernova 1987A Constraints on
  Sub-GeV Dark Sectors, Millicharged Particles, the QCD Axion, and an
  Axion-like Particle},'' \href{http://dx.doi.org/10.1007/JHEP09(2018)051}{{\em
  JHEP} {\bfseries 09} (2018) 051},
\href{http://arxiv.org/abs/1803.00993}{{\ttfamily arXiv:1803.00993 [hep-ph]}}.

\bibitem{Bauer:2018onh}
M.~Bauer, P.~Foldenauer, and J.~Jaeckel, ``{Hunting All the Hidden Photons},''
  \href{http://dx.doi.org/10.1007/JHEP07(2018)094}{{\em JHEP} {\bfseries 07}
  (2018) 094},
\href{http://arxiv.org/abs/1803.05466}{{\ttfamily arXiv:1803.05466 [hep-ph]}}.

\bibitem{Ren:2018gyx}
{\bfseries PandaX-II} Collaboration, X.~Ren {\em et~al.}, ``{Constraining Dark
  Matter Models with a Light Mediator at the PandaX-II Experiment},''
  \href{http://dx.doi.org/10.1103/PhysRevLett.121.021304}{{\em Phys. Rev.
  Lett.} {\bfseries 121} no.~2, (2018) 021304},
  \href{http://arxiv.org/abs/1802.06912}{{\ttfamily arXiv:1802.06912
  [hep-ph]}}.

\bibitem{Aubert:2009cp}
{\bfseries BaBar} Collaboration, B.~Aubert {\em et~al.}, ``{Search for Dimuon
  Decays of a Light Scalar Boson in Radiative Transitions Upsilon ---> gamma
  A0},'' \href{http://dx.doi.org/10.1103/PhysRevLett.103.081803}{{\em Phys.
  Rev. Lett.} {\bfseries 103} (2009) 081803},
  \href{http://arxiv.org/abs/0905.4539}{{\ttfamily arXiv:0905.4539 [hep-ex]}}.

\bibitem{Lees:2014xha}
{\bfseries BaBar} Collaboration, J.~Lees {\em et~al.}, ``{Search for a Dark
  Photon in $e^+e^-$ Collisions at BaBar},''
  \href{http://dx.doi.org/10.1103/PhysRevLett.113.201801}{{\em Phys. Rev.
  Lett.} {\bfseries 113} no.~20, (2014) 201801},
  \href{http://arxiv.org/abs/1406.2980}{{\ttfamily arXiv:1406.2980 [hep-ex]}}.

\bibitem{Aaij:2017rft}
{\bfseries LHCb} Collaboration, R.~Aaij {\em et~al.}, ``{Search for Dark
  Photons Produced in 13 TeV $pp$ Collisions},''
  \href{http://dx.doi.org/10.1103/PhysRevLett.120.061801}{{\em Phys. Rev.
  Lett.} {\bfseries 120} no.~6, (2018) 061801},
  \href{http://arxiv.org/abs/1710.02867}{{\ttfamily arXiv:1710.02867
  [hep-ex]}}.

\bibitem{Archilli:2011zc}
{\bfseries KLOE-2} Collaboration, F.~Archilli {\em et~al.}, ``{Search for a
  vector gauge boson in $\phi$ meson decays with the KLOE detector},''
  \href{http://dx.doi.org/10.1016/j.physletb.2011.11.033}{{\em Phys. Lett. B}
  {\bfseries 706} (2012) 251--255},
  \href{http://arxiv.org/abs/1110.0411}{{\ttfamily arXiv:1110.0411 [hep-ex]}}.

\bibitem{Babusci:2012cr}
{\bfseries KLOE-2} Collaboration, D.~Babusci {\em et~al.}, ``{Limit on the
  production of a light vector gauge boson in $\phi$ meson decays with the KLOE
  detector},'' \href{http://dx.doi.org/10.1016/j.physletb.2013.01.067}{{\em
  Phys. Lett. B} {\bfseries 720} (2013) 111--115},
  \href{http://arxiv.org/abs/1210.3927}{{\ttfamily arXiv:1210.3927 [hep-ex]}}.

\bibitem{Anastasi:2015qla}
A.~Anastasi {\em et~al.}, ``{Limit on the production of a low-mass vector boson
  in $\mathrm{e}^{+}\mathrm{e}^{-} \to \mathrm{U}\gamma$, $\mathrm{U} \to
  \mathrm{e}^{+}\mathrm{e}^{-}$ with the KLOE experiment},''
  \href{http://dx.doi.org/10.1016/j.physletb.2015.10.003}{{\em Phys. Lett. B}
  {\bfseries 750} (2015) 633--637},
  \href{http://arxiv.org/abs/1509.00740}{{\ttfamily arXiv:1509.00740
  [hep-ex]}}.

\bibitem{Anastasi:2016ktq}
{\bfseries KLOE-2} Collaboration, A.~Anastasi {\em et~al.}, ``{Limit on the
  production of a new vector boson in $\mathrm{e^+ e^-}\rightarrow {\rm
  U}\gamma$, U$\rightarrow \pi^+\pi^-$ with the KLOE experiment},''
  \href{http://dx.doi.org/10.1016/j.physletb.2016.04.019}{{\em Phys. Lett. B}
  {\bfseries 757} (2016) 356--361},
  \href{http://arxiv.org/abs/1603.06086}{{\ttfamily arXiv:1603.06086
  [hep-ex]}}.

\bibitem{Abe:2010gxa}
{\bfseries Belle-II} Collaboration, T.~Abe {\em et~al.}, ``{Belle II Technical
  Design Report},'' \href{http://arxiv.org/abs/1011.0352}{{\ttfamily
  arXiv:1011.0352 [physics.ins-det]}}.

\bibitem{Kou:2018nap}
{\bfseries Belle-II} Collaboration, W.~Altmannshofer {\em et~al.}, ``{The Belle
  II Physics Book},'' \href{http://dx.doi.org/10.1093/ptep/ptz106}{{\em PTEP}
  {\bfseries 2019} no.~12, (2019) 123C01},
  \href{http://arxiv.org/abs/1808.10567}{{\ttfamily arXiv:1808.10567
  [hep-ex]}}. [Erratum: PTEP 2020, 029201 (2020)].

\bibitem{Ilten:2015hya}
P.~Ilten, J.~Thaler, M.~Williams, and W.~Xue, ``{Dark photons from charm mesons
  at LHCb},'' \href{http://dx.doi.org/10.1103/PhysRevD.92.115017}{{\em Phys.
  Rev. D} {\bfseries 92} no.~11, (2015) 115017},
  \href{http://arxiv.org/abs/1509.06765}{{\ttfamily arXiv:1509.06765
  [hep-ph]}}.

\bibitem{Ilten:2016tkc}
P.~Ilten, Y.~Soreq, J.~Thaler, M.~Williams, and W.~Xue, ``{Proposed Inclusive
  Dark Photon Search at LHCb},''
  \href{http://dx.doi.org/10.1103/PhysRevLett.116.251803}{{\em Phys. Rev.
  Lett.} {\bfseries 116} no.~25, (2016) 251803},
  \href{http://arxiv.org/abs/1603.08926}{{\ttfamily arXiv:1603.08926
  [hep-ph]}}.

\bibitem{Bergsma:1985is}
{\bfseries CHARM} Collaboration, F.~Bergsma {\em et~al.}, ``{A Search for
  Decays of Heavy Neutrinos in the Mass Range 0.5-2.8-{GeV}},''
  \href{http://dx.doi.org/10.1016/0370-2693(86)91601-1}{{\em Phys. Lett. B}
  {\bfseries 166} (1986) 473--478}.

\bibitem{Gninenko:2012eq}
S.~Gninenko, ``{Constraints on sub-GeV hidden sector gauge bosons from a search
  for heavy neutrino decays},''
  \href{http://dx.doi.org/10.1016/j.physletb.2012.06.002}{{\em Phys. Lett. B}
  {\bfseries 713} (2012) 244--248},
  \href{http://arxiv.org/abs/1204.3583}{{\ttfamily arXiv:1204.3583 [hep-ph]}}.

\bibitem{Athanassopoulos:1997er}
{\bfseries LSND} Collaboration, C.~Athanassopoulos {\em et~al.}, ``{Results on
  muon-neutrino ---> electron-neutrino oscillations from pion decay in flight
  neutrinos},'' \href{http://dx.doi.org/10.1103/PhysRevC.58.2489}{{\em Phys.
  Rev. C} {\bfseries 58} (1998) 2489--2511},
  \href{http://arxiv.org/abs/nucl-ex/9706006}{{\ttfamily
  arXiv:nucl-ex/9706006}}.

\bibitem{Batell:2009di}
B.~Batell, M.~Pospelov, and A.~Ritz, ``{Exploring Portals to a Hidden Sector
  Through Fixed Targets},''
  \href{http://dx.doi.org/10.1103/PhysRevD.80.095024}{{\em Phys. Rev. D}
  {\bfseries 80} (2009) 095024},
  \href{http://arxiv.org/abs/0906.5614}{{\ttfamily arXiv:0906.5614 [hep-ph]}}.

\bibitem{Blumlein:2011mv}
J.~Bl\"umlein and J.~Brunner, ``{New Exclusion Limits for Dark Gauge Forces
  from Beam-Dump Data},''
  \href{http://dx.doi.org/10.1016/j.physletb.2011.05.046}{{\em Phys. Lett. B}
  {\bfseries 701} (2011) 155--159},
  \href{http://arxiv.org/abs/1104.2747}{{\ttfamily arXiv:1104.2747 [hep-ex]}}.

\bibitem{Blumlein:2013cua}
J.~Bl{\"u}mlein and J.~Brunner, ``{New Exclusion Limits on Dark Gauge Forces
  from Proton Bremsstrahlung in Beam-Dump Data},''
  \href{http://dx.doi.org/10.1016/j.physletb.2014.02.029}{{\em Phys. Lett. B}
  {\bfseries 731} (2014) 320--326},
  \href{http://arxiv.org/abs/1311.3870}{{\ttfamily arXiv:1311.3870 [hep-ph]}}.

\bibitem{Alekhin:2015byh}
S.~Alekhin {\em et~al.}, ``{A facility to Search for Hidden Particles at the
  CERN SPS: the SHiP physics case},''
  \href{http://dx.doi.org/10.1088/0034-4885/79/12/124201}{{\em Rept. Prog.
  Phys.} {\bfseries 79} no.~12, (2016) 124201},
  \href{http://arxiv.org/abs/1504.04855}{{\ttfamily arXiv:1504.04855
  [hep-ph]}}.

\bibitem{Anelli:2015pba}
{\bfseries SHiP} Collaboration, M.~Anelli {\em et~al.}, ``{A facility to Search
  for Hidden Particles (SHiP) at the CERN SPS},''
  \href{http://arxiv.org/abs/1504.04956}{{\ttfamily arXiv:1504.04956
  [physics.ins-det]}}.

\bibitem{Feng:2017uoz}
J.~L. Feng, I.~Galon, F.~Kling, and S.~Trojanowski, ``{ForwArd Search
  ExpeRiment at the LHC},''
  \href{http://dx.doi.org/10.1103/PhysRevD.97.035001}{{\em Phys. Rev. D}
  {\bfseries 97} no.~3, (2018) 035001},
  \href{http://arxiv.org/abs/1708.09389}{{\ttfamily arXiv:1708.09389
  [hep-ph]}}.

\bibitem{Gardner:2015wea}
S.~Gardner, R.~Holt, and A.~Tadepalli, ``{New Prospects in Fixed Target
  Searches for Dark Forces with the SeaQuest Experiment at Fermilab},''
  \href{http://dx.doi.org/10.1103/PhysRevD.93.115015}{{\em Phys. Rev. D}
  {\bfseries 93} no.~11, (2016) 115015},
  \href{http://arxiv.org/abs/1509.00050}{{\ttfamily arXiv:1509.00050
  [hep-ph]}}.

\bibitem{Berlin:2018pwi}
A.~Berlin, S.~Gori, P.~Schuster, and N.~Toro, ``{Dark Sectors at the Fermilab
  SeaQuest Experiment},''
  \href{http://dx.doi.org/10.1103/PhysRevD.98.035011}{{\em Phys. Rev. D}
  {\bfseries 98} no.~3, (2018) 035011},
  \href{http://arxiv.org/abs/1804.00661}{{\ttfamily arXiv:1804.00661
  [hep-ph]}}.

\bibitem{Batell:2009yf}
B.~Batell, M.~Pospelov, and A.~Ritz, ``{Probing a Secluded U(1) at
  B-factories},'' \href{http://dx.doi.org/10.1103/PhysRevD.79.115008}{{\em
  Phys. Rev. D} {\bfseries 79} (2009) 115008},
  \href{http://arxiv.org/abs/0903.0363}{{\ttfamily arXiv:0903.0363 [hep-ph]}}.

\bibitem{DelNobile:2015uua}
E.~Del~Nobile, M.~Kaplinghat, and H.-B. Yu, ``{Direct Detection Signatures of
  Self-Interacting Dark Matter with a Light Mediator},''
  \href{http://dx.doi.org/10.1088/1475-7516/2015/10/055}{{\em JCAP} {\bfseries
  10} (2015) 055}, \href{http://arxiv.org/abs/1507.04007}{{\ttfamily
  arXiv:1507.04007 [hep-ph]}}.

\bibitem{Savage:2008er}
C.~Savage, G.~Gelmini, P.~Gondolo, and K.~Freese, ``{Compatibility of
  DAMA/LIBRA dark matter detection with other searches},''
  \href{http://dx.doi.org/10.1088/1475-7516/2009/04/010}{{\em JCAP} {\bfseries
  04} (2009) 010}, \href{http://arxiv.org/abs/0808.3607}{{\ttfamily
  arXiv:0808.3607 [astro-ph]}}.

\bibitem{Desai:2004pq}
{\bfseries Super-Kamiokande} Collaboration, S.~Desai {\em et~al.}, ``{Search
  for dark matter WIMPs using upward through-going muons in
  Super-Kamiokande},'' \href{http://dx.doi.org/10.1103/PhysRevD.70.083523}{{\em
  Phys. Rev. D} {\bfseries 70} (2004) 083523},
  \href{http://arxiv.org/abs/hep-ex/0404025}{{\ttfamily arXiv:hep-ex/0404025}}.
  [Erratum: Phys.Rev.D 70, 109901 (2004)].

\bibitem{Covi:2009xn}
L.~Covi, M.~Grefe, A.~Ibarra, and D.~Tran, ``{Neutrino Signals from Dark Matter
  Decay},'' \href{http://dx.doi.org/10.1088/1475-7516/2010/04/017}{{\em JCAP}
  {\bfseries 1004} (2010) 017},
\href{http://arxiv.org/abs/0912.3521}{{\ttfamily arXiv:0912.3521 [hep-ph]}}.

\bibitem{Aoki:2013yxa}
Y.~Aoki, E.~Shintani, and A.~Soni, ``{Proton decay matrix elements on the
  lattice},'' \href{http://dx.doi.org/10.1103/PhysRevD.89.014505}{{\em Phys.
  Rev. D} {\bfseries 89} no.~1, (2014) 014505},
  \href{http://arxiv.org/abs/1304.7424}{{\ttfamily arXiv:1304.7424 [hep-lat]}}.

\bibitem{Stone150}
E.~C. Stone, A.~C. Cummings, F.~B. McDonald, B.~C. Heikkila, N.~Lal, and W.~R.
  Webber, ``Voyager 1 observes low-energy galactic cosmic rays in a region
  depleted of heliospheric ions,''
  \href{http://dx.doi.org/10.1126/science.1236408}{{\em Science} {\bfseries
  341} no.~6142, (2013) 150--153},
  \href{http://arxiv.org/abs/https://science.sciencemag.org/content/341/6142/150.full.pdf}{{\ttfamily
  https://science.sciencemag.org/content/341/6142/150.full.pdf}}.
  \url{https://science.sciencemag.org/content/341/6142/150}.

\bibitem{Aguilar:2014mma}
{\bfseries AMS} Collaboration, M.~Aguilar {\em et~al.}, ``{Electron and
  Positron Fluxes in Primary Cosmic Rays Measured with the Alpha Magnetic
  Spectrometer on the International Space Station},''
  \href{http://dx.doi.org/10.1103/PhysRevLett.113.121102}{{\em Phys. Rev.
  Lett.} {\bfseries 113} (2014) 121102}.

\bibitem{Boudaud:2016mos}
M.~Boudaud, J.~Lavalle, and P.~Salati, ``{Novel cosmic-ray electron and
  positron constraints on MeV dark matter particles},''
  \href{http://dx.doi.org/10.1103/PhysRevLett.119.021103}{{\em Phys. Rev.
  Lett.} {\bfseries 119} no.~2, (2017) 021103},
  \href{http://arxiv.org/abs/1612.07698}{{\ttfamily arXiv:1612.07698
  [astro-ph.HE]}}.

\bibitem{Ackermann:2015zua}
{\bfseries Fermi-LAT} Collaboration, M.~Ackermann {\em et~al.}, ``{Searching
  for Dark Matter Annihilation from Milky Way Dwarf Spheroidal Galaxies with
  Six Years of Fermi Large Area Telescope Data},''
  \href{http://dx.doi.org/10.1103/PhysRevLett.115.231301}{{\em Phys. Rev.
  Lett.} {\bfseries 115} no.~23, (2015) 231301},
  \href{http://arxiv.org/abs/1503.02641}{{\ttfamily arXiv:1503.02641
  [astro-ph.HE]}}.

\bibitem{Hochberg:2015vrg}
Y.~Hochberg, E.~Kuflik, and H.~Murayama, ``{SIMP Spectroscopy},''
  \href{http://dx.doi.org/10.1007/JHEP05(2016)090}{{\em JHEP} {\bfseries 05}
  (2016) 090}, \href{http://arxiv.org/abs/1512.07917}{{\ttfamily
  arXiv:1512.07917 [hep-ph]}}.

\bibitem{Hochberg:2017khi}
Y.~Hochberg, E.~Kuflik, and H.~Murayama, ``{Dark spectroscopy at lepton
  colliders},'' \href{http://dx.doi.org/10.1103/PhysRevD.97.055030}{{\em Phys.
  Rev. D} {\bfseries 97} no.~5, (2018) 055030},
  \href{http://arxiv.org/abs/1706.05008}{{\ttfamily arXiv:1706.05008
  [hep-ph]}}.

\bibitem{Dror:2020fbh}
J.~A. Dror, ``{Discovering leptonic forces using nonconserved currents},''
  \href{http://dx.doi.org/10.1103/PhysRevD.101.095013}{{\em Phys. Rev. D}
  {\bfseries 101} no.~9, (2020) 095013},
  \href{http://arxiv.org/abs/2004.04750}{{\ttfamily arXiv:2004.04750
  [hep-ph]}}.

\bibitem{Pontecorvo:1957qd}
B.~Pontecorvo, ``{Inverse beta processes and nonconservation of lepton
  charge},'' {\em Sov. Phys. JETP} {\bfseries 7} (1958) 172--173.
[Zh. Eksp. Teor. Fiz.34,247(1957)].

\bibitem{Maki:1962mu}
Z.~Maki, M.~Nakagawa, and S.~Sakata, ``{Remarks on the unified model of
  elementary particles},'' \href{http://dx.doi.org/10.1143/PTP.28.870}{{\em
  Prog. Theor. Phys.} {\bfseries 28} (1962) 870--880}.
[,34(1962)].

\bibitem{Capozzi:2017ipn}
F.~Capozzi, E.~Di~Valentino, E.~Lisi, A.~Marrone, A.~Melchiorri, and
  A.~Palazzo, ``{Global constraints on absolute neutrino masses and their
  ordering},'' \href{http://dx.doi.org/10.1103/PhysRevD.95.096014}{{\em Phys.
  Rev.} {\bfseries D95} no.~9, (2017) 096014},
\href{http://arxiv.org/abs/1703.04471}{{\ttfamily arXiv:1703.04471 [hep-ph]}}.

\bibitem{Goto:1998qg}
T.~Goto and T.~Nihei, ``{Effect of an RRRR dimension 5 operator on proton decay
  in the minimal SU(5) SUGRA GUT model},''
  \href{http://dx.doi.org/10.1103/PhysRevD.59.115009}{{\em Phys. Rev. D}
  {\bfseries 59} (1999) 115009},
  \href{http://arxiv.org/abs/hep-ph/9808255}{{\ttfamily arXiv:hep-ph/9808255}}.

\bibitem{Murayama:2001ur}
H.~Murayama and A.~Pierce, ``{Not even decoupling can save the minimal
  supersymmetric SU(5) model},''
  \href{http://dx.doi.org/10.1103/PhysRevD.65.055009}{{\em Phys. Rev. D}
  {\bfseries 65} (2002) 055009},
  \href{http://arxiv.org/abs/hep-ph/0108104}{{\ttfamily arXiv:hep-ph/0108104}}.

\bibitem{Costa:1982uv}
G.~Costa, F.~Feruglio, and F.~Zwirner, ``{Baryon and Lepton Nonconservation in
  Supersymmetric Models},'' \href{http://dx.doi.org/10.1007/BF02814034}{{\em
  Nuovo Cim. A} {\bfseries 70} (1982) 201}.

\bibitem{ArkaniHamed:2004fb}
N.~Arkani-Hamed and S.~Dimopoulos, ``{Supersymmetric unification without low
  energy supersymmetry and signatures for fine-tuning at the LHC},''
  \href{http://dx.doi.org/10.1088/1126-6708/2005/06/073}{{\em JHEP} {\bfseries
  06} (2005) 073},
\href{http://arxiv.org/abs/hep-th/0405159}{{\ttfamily arXiv:hep-th/0405159
  [hep-th]}}.

\bibitem{Giudice:2004tc}
G.~F. Giudice and A.~Romanino, ``{Split supersymmetry},''
  \href{http://dx.doi.org/10.1016/j.nuclphysb.2004.11.048,
  10.1016/j.nuclphysb.2004.08.001}{{\em Nucl. Phys.} {\bfseries B699} (2004)
  65--89}, \href{http://arxiv.org/abs/hep-ph/0406088}{{\ttfamily
  arXiv:hep-ph/0406088 [hep-ph]}}.
[Erratum: Nucl. Phys.B706,487(2005)].

\bibitem{Wells:2004di}
J.~D. Wells, ``{PeV-scale supersymmetry},''
  \href{http://dx.doi.org/10.1103/PhysRevD.71.015013}{{\em Phys. Rev. D}
  {\bfseries 71} (2005) 015013},
  \href{http://arxiv.org/abs/hep-ph/0411041}{{\ttfamily arXiv:hep-ph/0411041}}.

\bibitem{Kawasaki:2008qe}
M.~Kawasaki, K.~Kohri, T.~Moroi, and A.~Yotsuyanagi, ``{Big-Bang
  Nucleosynthesis and Gravitinos},''
  \href{http://dx.doi.org/10.1103/PhysRevD.78.065011}{{\em Phys. Rev. D}
  {\bfseries 78} (2008) 065011},
  \href{http://arxiv.org/abs/0804.3745}{{\ttfamily arXiv:0804.3745 [hep-ph]}}.

\bibitem{Kawasaki:2017bqm}
M.~Kawasaki, K.~Kohri, T.~Moroi, and Y.~Takaesu, ``{Revisiting Big-Bang
  Nucleosynthesis Constraints on Long-Lived Decaying Particles},''
  \href{http://dx.doi.org/10.1103/PhysRevD.97.023502}{{\em Phys. Rev. D}
  {\bfseries 97} no.~2, (2018) 023502},
  \href{http://arxiv.org/abs/1709.01211}{{\ttfamily arXiv:1709.01211
  [hep-ph]}}.

\bibitem{Preskill:1979zi}
J.~Preskill, ``{Cosmological Production of Superheavy Magnetic Monopoles},''
  \href{http://dx.doi.org/10.1103/PhysRevLett.43.1365}{{\em Phys. Rev. Lett.}
  {\bfseries 43} (1979) 1365}.

\bibitem{Khoze:2014woa}
V.~V. Khoze and G.~Ro, ``{Dark matter monopoles, vectors and photons},''
  \href{http://dx.doi.org/10.1007/JHEP10(2014)061}{{\em JHEP} {\bfseries 10}
  (2014) 061}, \href{http://arxiv.org/abs/1406.2291}{{\ttfamily arXiv:1406.2291
  [hep-ph]}}.

\bibitem{Harigaya:2014waa}
K.~Harigaya, M.~Kawasaki, K.~Mukaida, and M.~Yamada, ``{Dark Matter Production
  in Late Time Reheating},''
  \href{http://dx.doi.org/10.1103/PhysRevD.89.083532}{{\em Phys. Rev. D}
  {\bfseries 89} no.~8, (2014) 083532},
  \href{http://arxiv.org/abs/1402.2846}{{\ttfamily arXiv:1402.2846 [hep-ph]}}.

\bibitem{DelNobile:2013sia}
M.~Cirelli, E.~Del~Nobile, and P.~Panci, ``{Tools for model-independent bounds
  in direct dark matter searches},''
  \href{http://dx.doi.org/10.1088/1475-7516/2013/10/019}{{\em JCAP} {\bfseries
  10} (2013) 019}, \href{http://arxiv.org/abs/1307.5955}{{\ttfamily
  arXiv:1307.5955 [hep-ph]}}.

\bibitem{Fitzpatrick:2012ib}
A.~Fitzpatrick, W.~Haxton, E.~Katz, N.~Lubbers, and Y.~Xu, ``{Model Independent
  Direct Detection Analyses},''
  \href{http://arxiv.org/abs/1211.2818}{{\ttfamily arXiv:1211.2818 [hep-ph]}}.

\end{thebibliography}\endgroup

\end{document}